\documentclass[a4paper,fleqn,usenatbib]{mnras}

\usepackage{newtxtext,newtxmath}

\usepackage[T1]{fontenc}
\usepackage{ae,aecompl}
\usepackage[normalem]{ulem}
\usepackage{cancel}
\usepackage{xcolor}

\usepackage{csquotes}

\usepackage{graphicx}	
\usepackage{amsmath}	

\def\gs{\mathrel{\raise0.35ex\hbox{$\scriptstyle >$}\kern-0.6em \lower0.40ex\hbox{{$\scriptstyle \sim$}}}}
\def\ls{\mathrel{\raise0.35ex\hbox{$\scriptstyle <$}\kern-0.6em \lower0.40ex\hbox{{$\scriptstyle \sim$}}}}
\newcommand{\Msolar}{\mbox{${\rm M_{\odot}}$}}

\newcommand{\HH}{H$_2$}

\newcommand{\kms}{${\rm km~s^{-1}}$}

\newcommand{\cpropstoo}{\texttt{CPROPS{\scriptsize TOO}}}
\newcommand{\cprops}{\texttt{CPROPS}}

\title[WISDOM Project -- IX.\ GMCs in NGC4429]{WISDOM Project -- IX.\ Giant
  Molecular Clouds in the Lenticular Galaxy NGC4429: Effects of Shear
  and Tidal Forces on Clouds}

\author[L.\ Liu et al.]{
Lijie Liu,$^{1}$\thanks{E-mail: ljliu.astro@gmail.com}
Martin Bureau,$^{1,2}$
Leo Blitz,$^{3}$
Timothy A. Davis,$^{4}$
Kyoko Onishi,$^{5}$
\newauthor
Mark Smith,$^{1}$
Eve North$^{3}$
and Satoru Iguchi$^{6,7}$
\\
\\
$^{1}$Sub-department of Astrophysics, Department of Physics,
University of Oxford, Keble Road, Oxford OX1 3RH, UK \\
$^{2}$Yonsei Frontier Lab and Department of Astronomy, Yonsei
University, 50 Yonsei-ro, Seodaemun-gu, Seoul 03722, Republic of Korea \\
$^{3}$Department of Astronomy and Radio Astronomy Laboratory,
University of California, Berkeley, CA 94720, USA \\
$^{4}$School of Physics \& Astronomy, Cardiff University, Queens
Buildings, The Parade, Cardiff, CF24 3AA, UK \\
$^{5}$Research Center for Space and Cosmic Evolution, Ehime
University, Matsuyama, Ehime, 790-8577, Japan \\
$^{6}$Department of Astronomical Science, SOKENDAI (The Graduate
University of Advanced Studies), Mitaka, Tokyo 181-8588, Japan \\
$^{7}$ National Astronomical Observatory of Japan, National Institutes of Natural Sciences, Mitaka, Tokyo
181-8588, Japan }

\date{Accepted XXX. Received YYY; in original form ZZZ}

\pubyear{2021}

\begin{document}
\label{firstpage}
\pagerange{\pageref{firstpage}--\pageref{lastpage}}
\maketitle

\begin{abstract}
  We present high spatial resolution ($\approx12$~pc) Atacama Large
  Millimeter/sub-millimeter Array $^{12}$CO$(J=3-2)$ observations of
  the nearby lenticular galaxy NGC4429. We identify $217$ giant
  molecular clouds within the $450$~pc radius molecular gas disc. The
  clouds generally have smaller sizes and masses but higher surface
  densities and observed linewidths than those of Milky Way disc
  clouds. An unusually steep size -- line width relation
  ($\sigma\propto R_{\rm c}^{0.8}$) and large cloud internal velocity
  gradients ($0.05$ -- $0.91$~km~s$^{-1}$~pc$^{-1}$) and observed
  Virial parameters ($\langle\alpha_{\rm obs,vir}\rangle\approx4.0$)
  are found, that appear due to internal rotation driven by the
  background galactic gravitational potential. Removing this rotation,
  an internal Virial equilibrium appears to be established between the
  self-gravitational ($U_{\rm sg}$) and turbulent kinetic
  ($E_{\rm turb}$) energies of each cloud, i.e.\
  $\langle\alpha_{\rm sg,vir}\equiv\frac{2E_{\rm turb}}{\vert U_{\rm
      sg}\vert}\rangle\approx1.3$. However, to properly account for
  both self and external gravity (shear and tidal forces), we
  formulate a modified Virial theorem and define an effective Virial
  parameter
  $\alpha_{\rm eff,vir}\equiv\alpha_{\rm sg,vir}+\frac{E_{\rm
      ext}}{\vert U_{\rm sg}\vert}$ (and associated effective velocity
  dispersion). The NGC4429 clouds then appear to be in a critical
  state in which the self-gravitational energy and the contribution of
  external gravity to the cloud's energy budget ($E_{\rm ext}$) are
  approximately equal, i.e.\
  $\frac{E_{\rm ext}}{\vert U_{\rm sg}\vert}\approx1$. As such,
  $\langle\alpha_{\rm eff,vir}\rangle\approx2.2$ and most clouds are
  not virialised but remain marginally gravitationally bound. We show
  this is consistent with the clouds having sizes similar to their
  tidal radii and being generally radially elongated. External gravity
  is thus as important as self-gravity to regulate the clouds of
  NGC4429.
\end{abstract}



\vspace*{0mm}


\section{Introduction}
\label{sec:introduction}
It is well-known that giant molecular clouds (GMCs) are the major gas
reservoirs for star formation (SF) and the sites where essentially all
stars are born. Understanding the properties of GMCs is thus key to
unraveling the interplay between gas and stars within galaxies.  Early
GMC studies were restricted to our own Milky Way (MW) and the
late-type galaxies (LTGs) in our Galactic neighbourhood
\citep[e.g.][]{engargiola2003, rosolowsky2005, rosolowsky2007_M31,
  rosolowsky2007_M33, gratier2012, colombo2014, wu2017, faesi2018},
where GMCs have relatively uniform properties and generally follow the
so-called Larson relations (between size, velocity dispersion and
luminosity; e.g.\ \citealt{blitz2007, bolatto2008}).  However, more
recent studies of other local galaxies have raised doubts on the
universality of cloud properties. The cloud properties in some LTGs
(such as M51 and NGC253) vary with galactic environment and do not
universally obey the usual scaling relations
\citep[e.g.][]{hughes2013, leroy2015, schruba2019}. The first study of
individual GMCs in an early-type galaxy (ETG; NGC4526) has also
clearly shown that the clouds in that galaxy do not follow the usual
size -- linewidth correlation and tend to be more luminous, denser and
to have larger velocity dispersions than the GMCs in the MW and other
Local Group galaxies \citep{utomo2015}. The differences in NGC4526 may
be due to a higher interstellar radiation field (and/or cloud
extinctions), a different external pressure relative to each cloud's
self-gravity, and/or different galactic dynamics. GMCs in ETGs seem to
have shorter orbital periods and be subjected to stronger shear/tidal
forces, analogous to the highly dynamic environment in the MW central
molecular zone (CMZ; e.g.\ \citealt{kruijssen2019, henshaw2019,
  dale2019}). Although we are entering an era of large surveys of GMC
populations \citep[e.g.][]{sun2018}, current samples of ETGs are still
very limited. More studies of GMCs in varied LTGs and ETGs are thus
required to provide a comprehensive census of GMC properties across
different galaxy environments.

A model introduced by \citet{meidt2018} suggests that gas motions at
the cloud scale combine the effects of gas self-gravity and the gas
response to the forces exerted by the background host galaxy. In the
ETG NGC4526, the gas motions at cloud scales appear to be driven by
the galactic potential. The measured line widths of the GMCs are much
larger than their Virial line widths (the line widths predicted by
assuming the clouds' Virial masses are equal to their gaseous masses),
an effect that appears to be due to dominant gas motions associated
with the background galactic potential. Cloud-scale velocity gradients
aligned with the large-scale velocity field indeed suggest a dominance
of rotational motions due to the galactic potential
\citep{utomo2015}. It is thus important to investigate whether
cloud-scale gas motions are generally dominated by motions due to
self-gravity (generally random) or motions due to the galactic
potential (generally circular), as this has implications for the
observed size -- linewidth relation, the Virial parameter, cloud
morphologies and the processes governing star formation
\citep{meidt2018}.

The dynamical state of a molecular cloud provides important insights
into its evolution. It also plays an important role to determine its
ability to form stars and stellar clusters
\citep[e.g.][]{hennebelle2013, padoan2017}.
In most Virial balance analyses of molecular clouds, the gravitational
term entering the Virial theorem includes only the cloud's own
self-gravitational energy. However, in some galactic environments
(e.g.\ in galactic nuclei), the external (i.e.\ galactic)
gravitational potential could also play an important role to regulate
the cloud dynamics \citep[e.g.][]{rosolowsky-blitz2005, thilliez2014,
  yusefzadeh2016}. To analyse the Virial balance of GMCs in such
environments, one thus needs to add another gravitational term related
to the background gravitational field
\citep[e.g.][]{ballesterosparedes2009, chen2016}.

The net effect of the external gravitational potential on the dynamics
of GMCs should however also include an additional kinetic energy term
related to the gas motions driven by the galactic potential, as they
provide another source of support against the cloud's self-gravity.
In this paper, we therefore revisit the Virial theorem by adding two
crucial terms that take into account the background galactic
gravitational potential: an external gravitational energy term and a
kinetic energy term associated with the gas motions due to galactic
potential. Although an extended Virial theorem including a background
tidal field has been formulated before \citep[see, e.g.,][]{chen2016},
our resulting Virial equation contains new terms that were previously
missing and is thus more general.

Early studies of GMCs suggested they are long-lived, quasi-equilibrium
entities, isolated from their interstellar environment
\citep[e.g.][]{solomon1987, elmegreen1989, blitz1993}.  However,
recent findings that the properties of GMCs vary with galactic
environment imply that the clouds are not decoupled from their
surroundings \citep[e.g.][]{hughes2013, colombo2014, faesi2018}.  The
main physical factors determining cloud properties include: (1) the
interstellar radiation field \citep[e.g.][]{mckee1989}; (2)
large-scale dynamics (e.g.\ galactic tides and shear due to
differential galactic rotation; \citealt{dib2012, meidt2015,
  melchior2017}); (3) interstellar gas pressure
\citep[e.g.][]{heyer2009, hughes2013, meidt2016}; and (4) the
large-scale atomic gas distribution and \ion{H}{i} column density
\citep[e.g.][]{engargiola2003, blitz2007, rosolowsky2007_M33}. In this
work, we will focus on the roles of galactic tide/shear to regulate
the properties of GMCs. One of our main purposes is indeed to
quantitatively investigate the effects of galactic tidal and shear
forces on the physical properties and dynamical states of the clouds.

We note an important conceptual point. We will not assume here that
the clouds are in dynamical equilibrium, to then infer the clouds'
gravitational motions due to the external (i.e.\ galactic)
potential. Instead, we will attempt to directly estimate the clouds'
gravitational motions due to the external potential, to then infer whether
the clouds are indeed in dynamical equilibrium or not. The question of
whether GMCs are in dynamical equilibrium (and thus long-lived) or out
of equilibrium (and thus transient) has remained unanswered for
decades. We thus believe this approach is not only well-justified and
worthwhile, but ultimately desirable.


The mm-Wave Interferometric Survey of Dark Object Masses (WISDOM) aims
to use the high angular resolution of the Atacama Large
Millimeter/sub-millimeter Array (ALMA) to study: (1) the masses and
properties of the supermassive black holes (SMBHs) lurking at the
centres of galaxies \citep[e.g.][]{onishi2017, davis2017, davis2018,
  smith2019, north2019, smith2020a, smith2020b}; (2) the physical
properties and dynamics of GMCs in the central parts of the same
galaxies. As part of WISDOM, we analyse here the properties and
dynamics of individual GMCs in the bulge of NGC4429, an SA0-type
galaxy located in the centre of the Virgo cluster. This paper is the
first of a series studying the GMCs in WISDOM galaxies, and it
introduces many of the methods and tools we will use to identify GMCs
and analyse their properties and dynamics. The paper is structured as
follows. In Section~\ref{sec:data_cloud_identification} we describe
the data and the methodology used to identify GMCs in NGC4429. We use
a modified version of the code {\cpropstoo}, that is more robust and
efficient at identifying clouds in complex and crowded
environments. The cloud properties, their probability distribution
functions and their mass distribution functions are reported in
Section~\ref{sec:cloud_properties}. Our analysis of the kinematics of
individual GMCs is presented in Section~\ref{sec:cloud_kinematics}. We
investigate the dynamical states of the GMCs utilising our modified
Virial theorem (taking into account the background galactic
gravitational potential) in
Section~\ref{sec:dynamical_state_clouds}. The shear motions within
clouds, the effects of self-gravity and the cloud morphologies are
discussed in Section~\ref{sec:discussion}. We conclude briefly in
Section~\ref{sec:conclusions}.



\section{Data and Cloud Identification}
\label{sec:data_cloud_identification}


\subsection{Target}
\begin{figure}
  \includegraphics[width=0.95\columnwidth]{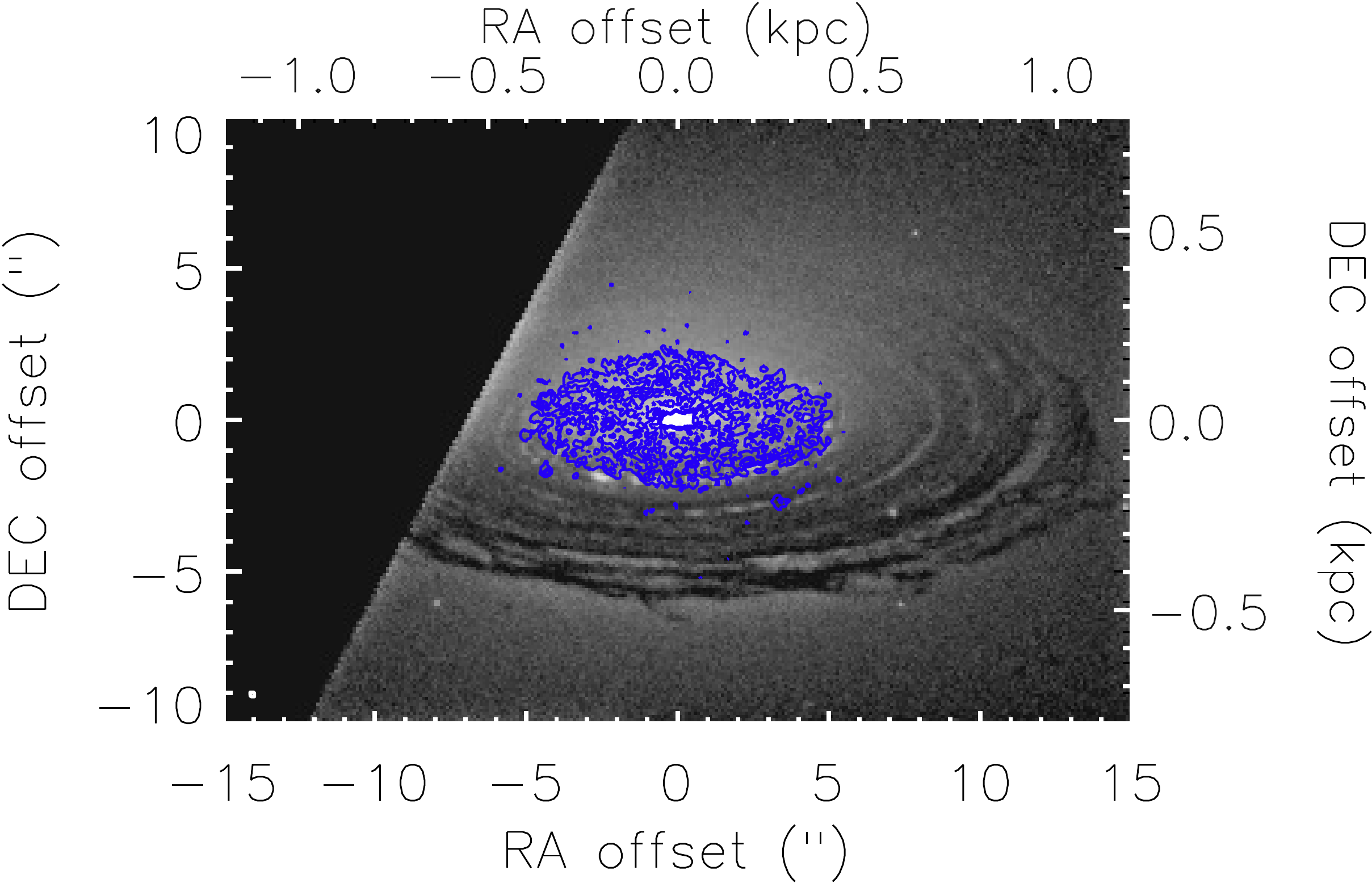}
  \caption{$^{12}$CO(3-2) molecular gas distribution of NGC4429 from
    our ALMA observations (blue contours; \citealt{davis2018}),
    overlaid on a {\it HST} Wide-Field Planetary Camera~2 (WFPC2)
    F606W image of a $2.8\times2.8$~kpc$^2$ region around its
    nucleus.}
  \label{fig:image_oplot}
\end{figure}

NGC4429 is a lenticular galaxy located in the centre of the Virgo
cluster, with a bar and stellar inner ring morphology
\citep{alatalo2013}. It contains a nuclear dust disc visible in
extinction against the stellar continuum in {\it Hubble Space
  Telescope} ({\it HST}) imaging (Fig.~\ref{fig:image_oplot} and
\citealt{davis2018}). NGC4429 has a total stellar mass of
$\approx1.5\times10^{11}$~{\Msolar}, a luminosity-weighted stellar
velocity dispersion within one effective radius
$\sigma_{\rm e}=177$~km~s$^{-1}$ \citep{cappellari2013}, and is a fast
rotator (specific angular momentum within one effective radius
$\lambda _{\rm R_e}=0.4$; \citealt{emsellem2011}).

The total molecular gas mass of NGC4429 detected via $^{12}$CO(1-0)
single-dish observations is $(1.1\pm0.08)\times10^8$~{\Msolar}
\citep{young2011}. The $^{12}$CO(1-0) Combined Array for Research in
Millimeter-wave Astronomy (CARMA) interferometric map shows the
molecular gas is co-spatial with the nuclear dust disc and regularly
rotates in the galaxy mid-plane \citep{davis2011, davis2013,
  davis2018}, with an inclination angle of $68^{\circ}$
\citep{davis2011, alatalo2013}. The $^{12}$CO(3-2) distribution is
more compact than that of $^{12}$CO(1-0), the $^{12}$CO(3-2) gas being
present only in the inner parts of the nuclear dust disc visible in
{\it HST} images (see Fig.~\ref{fig:image_oplot}). The star formation
rate (SFR) within this molecular gas disc has been estimated at
$0.1$~{\Msolar}~yr$^{-1}$ using mid-infrared and far-ultraviolet
emission \citep{davis2014}. The spatially-unresolved (sub-arcsecond)
radio continuum emission from the central regions of NGC4429 implies
the presence of a low-luminosity active galactic nucleus (LL-AGN;
\citealt{nyland2016}). The kinematics of the central CO gas, as probed
by the same dataset as used here, imply the presence of a
$(1.5\pm0.1)\times10^8$~{\Msolar} SMBH \citep{davis2018}. Throughout
this paper we assume a distance $D$ of $16.5\pm1.6$~Mpc for NGC4429
\citep{cappellari2011}. One arcsecond then corresponds to a physical
scale of $\approx80$~pc.


\subsection{Data}
NGC4429 was observed in the $^{12}$CO(3-2) line (345 GHz) using ALMA
as part of the WISDOM project. The data were calibrated and reduced in
a standard manner \citep{davis2018}, and the final $^{12}$CO(3-2) data
cube we adopt has a synthesised beam of $0\farcs18\times0\farcs14$
($14\times11$~pc$^2$) at a position angle of $311^{\circ}$ and a
channel width of $2$~{\kms}. It covers a region of
$17\farcs5\times17\farcs5$ ($1400\times1400$~pc$^2$), thus comprising
the entire nuclear dust and molecular gas disc. Pixels of $0\farcs05$
were chosen as a compromise between spatial sampling and cube size,
resulting in approximately $3.5\times2.8$~pixels$^2$ across the
synthesised beam \citep{davis2018}. Our spatial and spectral
resolutions allow for reliable estimates of the radii and velocity
dispersions of individual GMCs, that have a typical size of
$\approx50$~pc \citep{blitz1993} and a typical linewidth of several
{\kms} \citep[e.g.][]{solomon1987}. The root mean square (RMS) noise
in line-free channels of the cube is
$\sigma_{\rm rms}=1.34$~mJy~beam$^{-1}$ ($\approx0.5$~K) in $2$~{\kms}
channels. The integrated $^{12}$CO(3-2) spectrum of NGC4429 exhibits
the classic double-horn shape of a rotating disc, with a total flux of
$75.5\pm7.6$~Jy~km~s$^{-1}$.

As shown in \citet{davis2018}, the molecular gas disc of NGC4429 is
flocculent. Our ALMA observations reveal that the CO(3-2) gas surface
density does not decrease smoothly to our detection limit, but instead
appears to be truncated at an inner radius of $48\pm3$~pc and an outer
radius of $406\pm10$~pc \citep{davis2018}. As mentioned above, the
$^{12}$CO(3-2) disc thus lies only in the inner parts of the nuclear
dust disc visible in {\it HST} images (see
Fig.~\ref{fig:image_oplot}), and it has an extent smaller than that of
the $^{12}$CO(1-0) emission (that extends to the edge of the nuclear
dust disc; \citealt{davis2013}). As CO(3-2) is excited in denser and
warmer gas than CO(1-0) (with critical densities of
$\approx7\times10^4$ and $\approx1.4\times10^3$~cm$^{-3}$ and
excitation temperatures of $\approx15$ and $5.5$~K, respectively), we
are likely to identify a cloud population that is associated with
\ion{H}{ii} regions and thus ongoing star formation at the centre of
NGC4429 only. High-resolution observations of lower-$J$ CO transitions
may be required to conduct a study of the NGC4429 GMC population over
the entire molecular gas disc (if indeed additional clouds exist
beyond the CO(3-2) extent probed here).

Continuum $345$~GHz emission was also detected in NGC4429, with a
centre of ${\rm R.A.\,(J2000)}=12^{\rm h}27^{\rm m}26\fs504\pm0\fs013$
and ${\rm Dec.\,(J2000)}=11^{\circ}06\arcmin27\farcs57\pm0\farcs01$
derived by Gaussian fitting. This position is consistent with the
optical centre of NGC4429 \citep{adelmanmccarthy2008} and will be used
as the centre of the galaxy in this work.


\subsection{Cloud identification}
\label{sec:cloud_identification}

We use our own modified version of the {\cpropstoo} algorithms
\citep{leroy2015} to identify cloud structures. {\cpropstoo} is an
updated version of {\cprops} \citep{rosolowsky2006}, one of the cloud
identification algorithms most widely used in the literature. The key
modifications of {\cpropstoo} compared to {\cprops} were noted by
\citet{leroy2015}: {\cpropstoo} (1) deconvolves the beam in two
dimensions; (2) employs a larger suite of size and linewidth measures,
including measuring the area of and fitting an ellipse at the half
maximum flux level (in addition to measuring the second moment); and
(3) introduces additional extrapolation (aperture correction)
approaches, that essentially assume a Gaussian distribution to
extrapolate the ellipse fits. In this work we have further modified
{\cpropstoo}, to make it more robust when decomposing clouds in
complex and crowded environments.

The cloud identification algorithm first calculates a
spatially-varying estimate of the noise in the data cube, and then
uses the noise cube generated to create a three-dimensional (3D) mask
of bright emission. The mask initially includes only pixels where two
adjacent channels (at the same position) both have intensities above
$3\,\sigma_{\rm rms}$. It is then expanded to include all neighbouring
emission above a lower threshold -- two adjacent channels above
$2\,\sigma_{\rm rms}$. The regions thus identified are referred to as
\enquote{islands}. If an island has a projected area of less than two
synthesised beams, it is assumed to be a noise peak and is removed
from the mask. The resulting mask contains $\approx60\%$ of the
integrated flux of the galaxy, consistent with the fractions yielded
by {\cprops} in other studies of extragalactic clouds ($50$ -- $70\%$;
\citealt{wong2011, hughes2013, donovanmeyer2013, colombo2014,
  leroy2015, pan2017, miura2018, faesi2018, wong2019, imara2019}). We
checked the stringency of the mask by applying the same criteria to
the inverted data set (scaled by $-1$) and found no false positive, so
the masking criteria are likely robust.

Once regions of significant emission (i.e.\ islands) have been
identified, these islands are further decomposed into individual
\enquote{cloud} structures. Clouds are identified as local maxima
within a moving 3D box of area $3\times3$~spaxels$^{2}$
($\approx12\times12$~pc$^2$) and velocity width of $3$~channels
($6$~\kms). In our modified version of {\cpropstoo}, we add another
criterion to find local maxima, checking whether the
($3\times3\times3$~pixels$^3$) box centred on a local maximum also
represents a local maximum on a larger scale, as suggested by
\citet{yang2014}. This is to eliminate the impact of noisy pixels or
outliers, as a noise peak can easily become a local maximum within a
single box, but much less so on a larger scale. We thus consider a
($3\times3\times3$~pixels$^3$) box centred on each local maximum, and
require the sum of the flux densities in that box to be larger than
that in all eight spatially-adjacent ($3\times3\times3$~pixels$^3$)
boxes. The detection of local maxima in this way is much more robust
and efficient.

For each local maximum, the original {\cpropstoo} algorithm requires
all emission uniquely associated with that maximum (i.e.\ all emission
within the faintest intensity isosurface uniquely associated with that
maximum) to have a minimum area ($minarea$), minimum number of pixels
($minpix$) and minimum number of velocity channels ($minvchan$). It
also requires the local maximum's brightness temperature to lie at
least $\Delta T_{\rm max}$ above the merger level with any other
maximum (i.e.\ the brightest contour level enclosing another local
maximum). However, this decomposition algorithm often leads to cloud
size and velocity dispersion distributions that peak around the chosen
$minarea$, $minpix$ and $minvchan$. This is a well-known bias that
reflects the hierarchical structure of the ISM from parsec to
kiloparsec scales \citep[e.g.][]{verschuur1993, hughes2013,
  leroy2016}. It becomes especially problematic for complex and
crowded environments where the emission has low contrast and extends
over a range of scales (e.g.\ the centre of M51; \citealt{hughes2013,
  colombo2014}). Small $minarea$ and $minpix$ tend to identify the
sub-structures of a cloud (\enquote{over-decompositon}), whereas large
$minarea$ and $minpix$ tend to miss out small structures
(\enquote{under-decomposition}).

\begin{figure}
  \includegraphics[width=1.0\columnwidth]{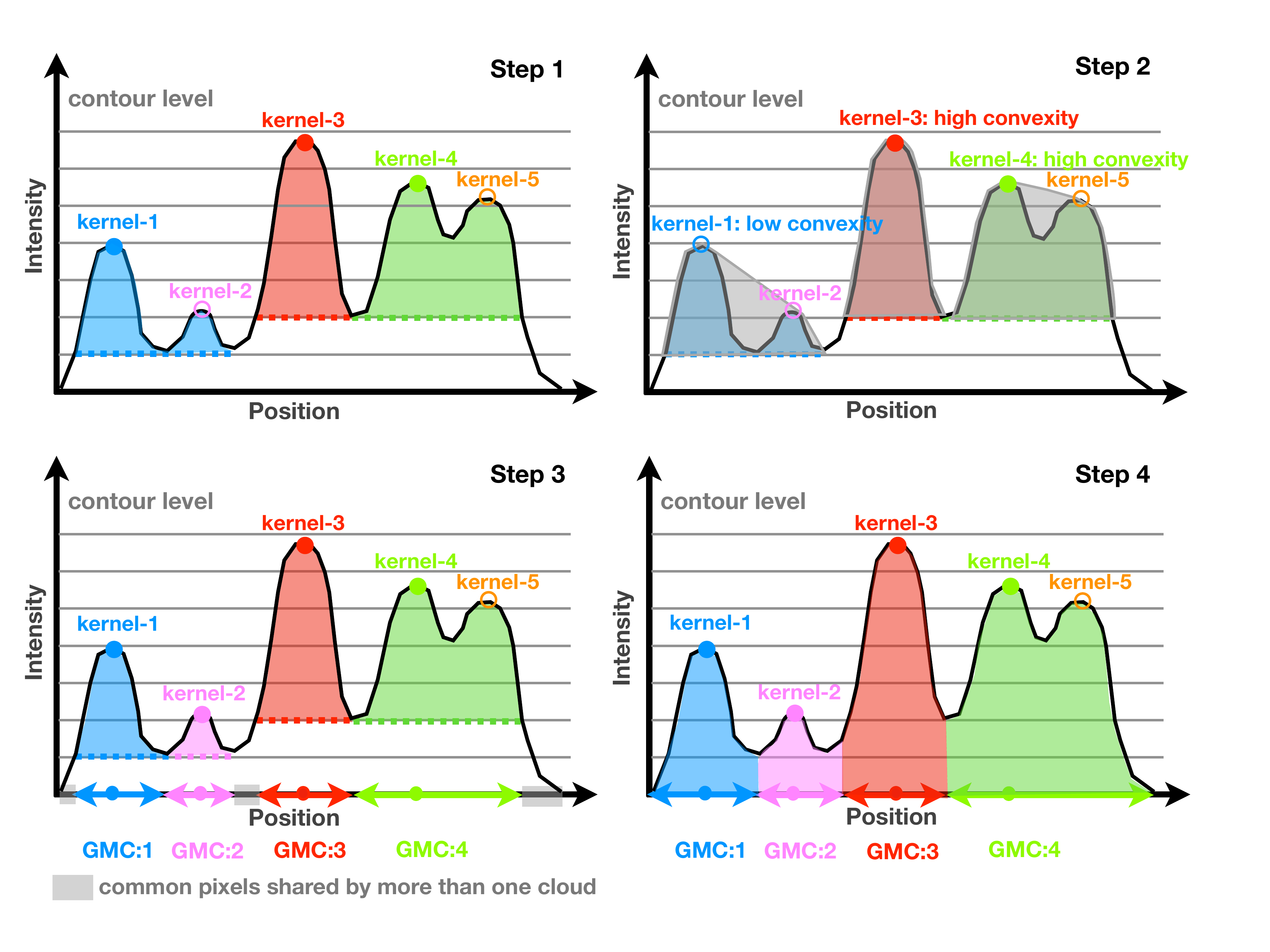}
  \caption{Schematic diagram of the cloud identification process using
    our modified {\cpropstoo} algorithm. Each panel shows a different
    step in the decomposition of a 1D line profile with five distinct
    kernels, each kernel corresponding to a local maximum and being
    identified by a different colour. Circles in matching colours
    indicate the kernels that are preserved or selected (solid
    circles) and rejected (open circles) at each step. Grey horizontal
    lines indicate characteristic brightness levels through the
    data. Each coloured dotted line indicates the unique level of the
    kernel in matching colour (i.e.\ the faintest level that is
    uniquely associated with that kernel), while each coloured region
    shows the emission uniquely associated with that kernel. {\bf Step
      1:} removal of kernels that do not meet the selection criteria
    given by $\Delta T_{\rm max}$, $minvchan$ and $minpix$/$minarea$
    (here kernel~2 and 5). {\bf Step 2:} removal of kernels that do
    not meet the selection criterion given by $minconvexity$ (here
    kernel~1). The $convexity$ parameter is defined as the ratio of
    the volume (or area in this 1D example) of the cloud (i.e.\ the
    coloured region of each kernel in matching colour) to the volume
    (or area) of the smallest convex hull encompassing the cloud
    (i.e.\ the associated grey regions). Only kernel~3 and 4 are
    preserved in this step. {\bf Step 3:} Repeat of steps~1 and 2
    adopting increasingly smaller $minpix$ and $minarea$ (here
    kernel~1 and 2 are re-selected due to the lower cloud size
    threshold; both have sufficient $convexity$). {\bf Step 4:}
    assigment of remaining emission (e.g.\ grey regions in the
    bottom-left panel) to the preserved kernels (using a
    friends-of-friends algorithm ensuring any pair of pixels in a
    kernel is connected by a continuous path). 
    }
  \label{fig:cprops_sketch}
\end{figure}

To remove this bias and identify cloud structures across multiple
scales, we modified {\cpropstoo} by setting each of $minarea$ and
$minpix$ to a range of values rather than a single value. In our work,
we assign $minarea$ a range of $100$ to $10$~spaxels (the synthesised
beam area) with a step of $5$~spaxels (half the beam area), similarly
in pixels for $minpix$. We start by searching for the largest cloud
structures using the largest $minarea$ ($100$~spaxels) and $minpix$
($100$~pixels), and then repeat the search process to identify
increasingly small clouds in the volume of the cube not yet assigned
to any cloud. We use a $minarea$ (resp.\ $minpix$) $5$~spaxels (resp.\
$5$~pixels) smaller than the previous one at each step, until all the
cloud structures larger than the beam size ($10$~spaxels) are
identified. As long as $minarea$ and $minpix$ cover large ranges, the
final results hardly depend on the specified ranges. We are therefore
able to remove two free parameters in the algorithm, making our
results less arbitrary and more robust. A schematic of our modified
{\cpropstoo} technique is shown in Figure~\ref{fig:cprops_sketch}
for a one-dimensional (1D) line profile.

The main concern about our newly-developed approach, however, is that
we may identify large clouds while ignoring potentially significant
sub-structures. To solve this problem, we introduce a new parameter,
$convexity$, inspired by an analogous quantity in studies of
biological structures \citep{lin2007}, that describes how significant
the sub-structure of a cloud is. The parameter $convexity$ is defined
as the ratio of the volume of the cloud (i.e.\ the volume of its 3D
intensity distribution) to the volume of the smallest convex hull
encompassing all of its flux (i.e.\ the volume of the smallest convex
envelope enclosing all of the cloud's 3D intensity distribution; see
the top-right panel of Fig.~\ref{fig:cprops_sketch} for an example
with a 1D line profile, i.e.\ a two-dimensional (2D) intensity
distribution). The $convexity$ of a cloud should thus be close to $1$
if the cloud has only one intensity peak and no sub-structure, and be
less than $1$ if the cloud has some sub-structures. The lower the
value of $convexity$, the more significant the sub-structure of a
cloud. Our modified {\cpropstoo} code requires all clouds to have a
minimum $convexity$ ($minconvexity$). Typical useful values are $0.5$
-- $0.7$, as determined by visual inspection, to ensure clouds are not
over- or under-decomposed. In this work, we set $minconvexity$ to
$0.55$. Overall, our new refinements allows {\cpropstoo} to identify
structures over multiple scales, with less arbitrariness than
previously.
 
We set the parameters $minvchan$ and $\Delta T_{\rm max}$ based on
physical priors described by \citet{rosolowsky2006}, that suggest a
cloud has a minimum
velocity dispersion $\Delta V_{\rm max}=2$~{\kms}
($minvchan=2\sqrt{2\ln2}\,\Delta V_{\rm max}\approx4$~{\kms}) and
$\Delta T_{\rm max}=1$~K, motivated by the properties of Galactic
GMCs. A factor of $2\sqrt{2\ln2}$ is applied to $\Delta V_{\rm max}$
to convert the velocity dispersion to a full width at half maximum
(FWHM). We set the parameters in physical units ({\kms} and K) rather
than data units (channel, $\sigma_{\rm rms}$) to reduce possible
biases when comparing cloud properties from different observations.
Our excellent spectral resolution (channel width of $2$~{\kms}) and
sensitivity ($\sigma_{\rm rms}\approx0.5$~K) allow us to reach and
thus use those physical parameters.

According to our algorithm, each surviving local maximum corresponds
to a cloud. {\cpropstoo} assigns the emission that is uniquely
associated with each local maximum (i.e.\ the emission within the
faintest intensity isosurface uniquely associated with that maximum)
to that cloud. The remaining emission shared among clouds is then
assigned to the \enquote{nearest} local maximum (i.e.\ the local
maximum with the shortest path through the data cube from a given
pixel). In our work, however, we apply a \enquote{friends-of-friends}
algorithm to assign all remaining emission, as for the ClumpFind
algorithm \citep{williams1994} and the original {\cprops} code
\citep{rosolowsky2006}. This friends-of-friends paradigm connects
pixels according to the brightnesses of neighbouring pixels, without
assuming a particular shape for the objects to decompose
\citep{rosolowsky2006}. This method conserves flux, so that all the
flux within the island regions is assigned to a particular cloud
\citep{tasker2009}. As each pair of pixels in a cloud can then be
connected by a continuous path through that cloud, we avoid assigning
disconnected pixels to the same cloud.

\begin{figure*}
  \includegraphics[width=0.9\textwidth]{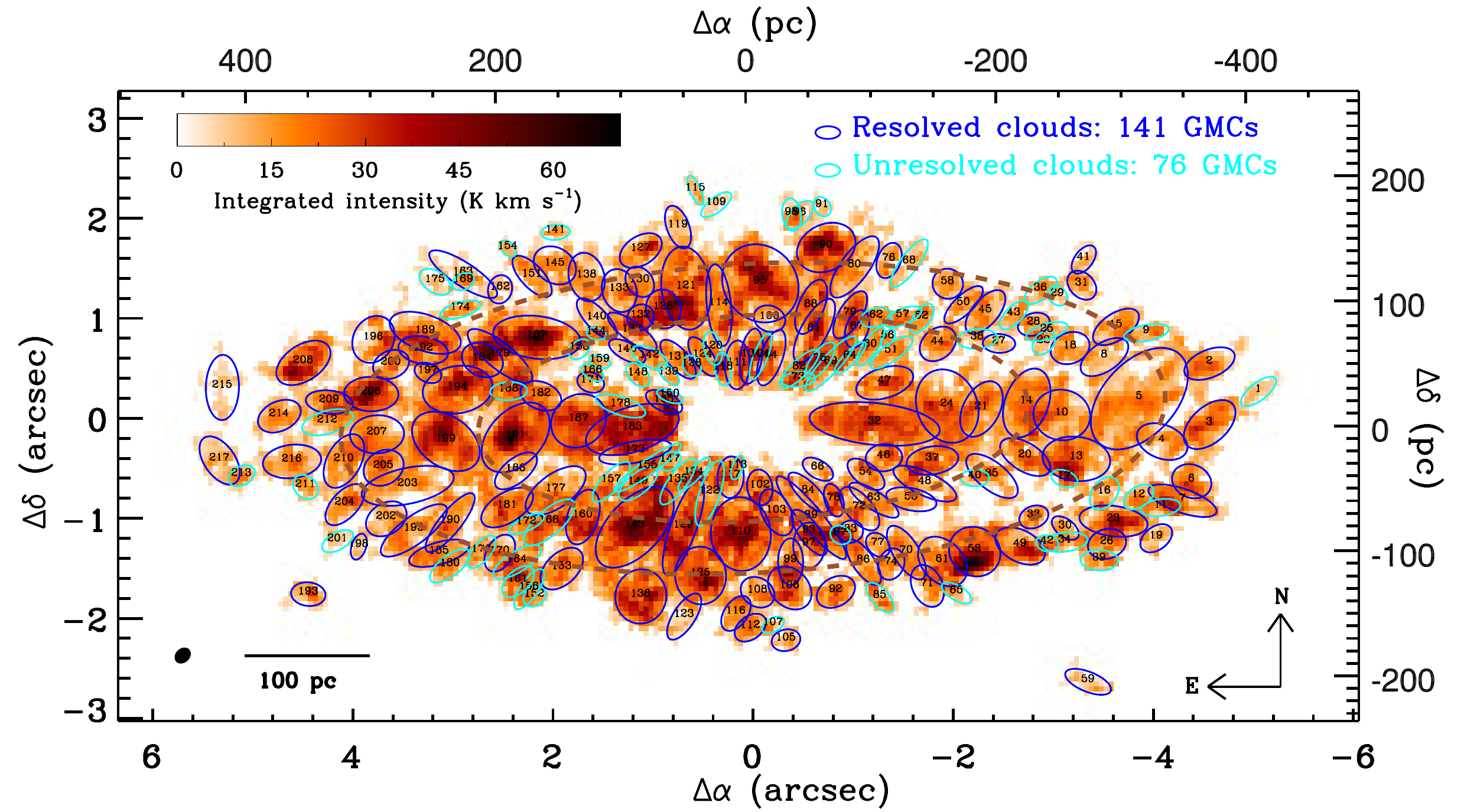}
  \caption{Molecular gas distribution of NGC4429 with GMCs identified.
    The integrated intensity map is shown (colour scale), blanking out
    non-signal areas using the mask generated by {\cpropstoo}. The
    mask covers pixels with connected emission above
    $2\,\sigma_{\rm rms}$ and at least two adjacent channels above
    $3\,\sigma_{\rm rms}$, where $\sigma_{\rm rms}$ is the RMS noise in
    the cube. The ellipses displayed, each corresponding to one
    labelled cloud, have been extrapolated to the limit of perfect
    sensitivity but have not been corrected for the finite spatial
    resolution. Dark blue (resp.\ cyan) ellipses indicate
    spatially-resolved (resp.\ unresolved) clouds. The two brown
    dashed ellipses at galactocentric distances of $220$ and $330$~pc
    define the three regions (inner, intermediate and outer) discussed
    in the text. The synthesised beam ($0\farcs18\times0\farcs14$ or
    $14\times11$~pc$^2$) is shown in the bottom-left corner along with
    a scale bar.}
  \label{fig:gmcs}
\end{figure*}

The resulting sample of GMCs in NGC4429 contains $217$ GMCs, $141$ of
which are spatially resolved, shown in Fig.~\ref{fig:gmcs}. The
majority of the resolved clouds have a single-peaked Gaussian-like
spatially-integrated line profile, although a few do reveal a
double-peaked line profile possibly indicating significant
rotation. Most line profiles are symmetric but a few are asymmetric,
with significant skewness (blue or red wing). The clouds identified
with our new refinements are $15\%$ fewer ($217$ versus $254$ clouds),
$18\%$ larger (median cloud size $\approx13$ versus $\approx11$~pc),
$18\%$ more massive (median gaseous mass $\approx2.0\times10^4$ versus
$1.7\times10^4$~M$_\odot$) and have velocity dispersions $30\%$ larger
(median velocity dispersion $5.2$ versus $4.0$~km~s$^{-1}$) than those
derived using the original {\cpropstoo} code. They also span a larger
range of sizes. A Gaussian fit to the size distribution yields a mean
of $16\pm0.5$~pc and a standard deviation of $\approx6$~pc for our
spatially-resolved clouds (see
Section~\ref{sec:probability_distribution_functions_gmc_properties}),
but $14\pm0.5$~pc and $\approx3.5$~pc, respectively, for those
identified using the original {\cpropstoo}. The resolved clouds
identified here also seem to have more regular morphologies, with a
mean $\langle convexity\rangle=0.57$ ($convexity>0.55$ by
construction) compared to $\langle convexity\rangle\approx0.45$ (and
$\approx54\%$ of resolved clouds with $convexity<0.55$) for
{\cpropstoo}-identified clouds. This confirms that our approach and
modified {\cpropstoo} code have great potential to identify clouds
over large spatial scales in crowded and complex environments (e.g.\
galactic centres and spiral arms).

\section{Cloud Properties} 
\label{sec:cloud_properties}


\subsection{Definition of GMC properties}
\label{sec:definition_gmc_properties}

Once all the pixels of every cloud have been identified, we calculate
the physical properties of the clouds by following the standard
{\cpropstoo}/{\cprops} definitions \citep{rosolowsky2006}. The
{\cpropstoo} algorithm applies moment methods to derive the size,
linewidth and flux of a cloud from its distribution within a
position-position-velocity data cube. One advantage of {\cpropstoo}
over other GMC identification algorithms is that it attempts to
correct the measured cloud properties for the finite sensitivity and
instrumental resolution \citep{rosolowsky2006}. To reduce the
sensitivity bias, the algorithm measures the size, velocity width and
luminosity as a function of the boundary intensity isosurface
($T_{\rm edge}$) and extrapolates them to the case of infinite
signal-to-noise ratio ($S/N$; i.e.\ $T_{\rm edge}=0$~K). The size and
linewidth are extrapolated linearly, while the luminosity is
extrapolated quadratically. To correct for the resolution bias,
{\cpropstoo} \enquote{deconvolves} the synthesised beam size from the
measured extrapolated cloud size in two
dimensions. \citet{rosolowsky2006} argued that moment measurements
combined with beam deconvolution and extrapolation represent a robust
way to compare heterogeneous observations of molecular clouds.

\textbf{\textit{Cloud centre.}} The central position ($x_{\rm c}$,
$y_{\rm c}$) and velocity ($v_{\rm c}$) of each cloud are obtained
directly from the intensity-weighted first spatial and velocity
moment,
\begin{equation}
  \left\{
    \begin{array}{lr}
  x_{\rm c}\equiv\frac{\sum_{i}^{\rm cloud}T_i x_i}{\sum_{i}^{\rm cloud} T_i}~,\\
  y_{\rm c}\equiv\frac{\sum_{i}^{\rm cloud}T_i y_i}{\sum_{i}^{\rm cloud} T_i}~,\\
  v_{\rm c}\equiv\frac{\sum_{i}^{\rm cloud}T_i v_i}{\sum_{i}^{\rm cloud} T_i}~,
  \end{array}
  \right.
\end{equation}
where ($x_i,y_i$) is the position of a given pixel, $v_i$ its velocity
and $T_i$ its flux (brightness temperature), and the sums are over all
pixels $i$ of each cloud.

\textbf{\textit{Cloud size.}} The radius $R_{\rm c}$ of each cloud is
calculated as the geometric mean of the second spatial moment of the
intensity distribution along the major and the minor axis:
\begin{equation}
  \label{eq:Rc}
  R_{\rm c}\equiv\eta\sqrt{\sigma_{\rm maj,dc}\,\,\sigma_{\rm min,dc}} =1.91\sqrt{\sigma_{\rm maj,dc}\,\,\sigma_{\rm min,dc}}~,
\end{equation}
where $\sigma_{\rm maj,dc}$ and $\sigma_{\rm min,dc}$ are the
deconvolved RMS spatial extent along the major and the minor axis,
respectively, extrapolated to the $T_{\rm edge}=0$~K isosurface, and
$\eta$ is a factor relating the one-dimensional RMS extent to the
radius of a cloud.  While $\eta$ formally depends on the shape and
density profile of the cloud, we follow \citet{solomon1987} and common
practice and adopt $\eta=1.91$ whenever we need to evaluate
expressions containing $R_{\rm c}$.
The major and minor axes are thus defined as the principal axes of the
moment of inertia tensor of the cloud (see Eq.~1 in
\citealt{rosolowsky2006}).

\textbf{\textit{Cloud velocity dispersion.}} The observed (i.e.\ 1D)
linewidth or velocity dispersion $\sigma_{\rm obs,los}$ of each cloud
is measured from the second moment of the intensity distribution along
the velocity axis, extrapolated to $T_{\rm edge}=0$~K. To account for
the potential bias toward a higher velocity dispersion due to the
finite spectral resolution, we perform a deconvolution as suggested by
\citet{rosolowsky2006}:
\begin{equation}
  \sigma_{\rm obs,los}\equiv\sqrt{\sigma^2_{\rm v}\,-\,\frac{\Delta V^2_{\rm chan}}{2\pi} }~,
\end{equation}
where $\sigma_{\rm v}$ is the extrapolated second moment along the
velocity axis, $\Delta V_{\rm chan}$ is the channel width and
$\frac{\Delta V_{\rm chan}}{\sqrt{2\pi}}$ is the standard deviation of
a Gaussian that has an integrated area equal to a spectral channel of
width $\Delta V_{\rm chan}$.

The observed velocity dispersion $\sigma_{\rm obs,los}$ includes the
effects of turbulent motions, intrinsic rotation of the cloud, and
shear motions due to the large-scale kinematics of the galactic disc
(such as galactic rotation and streaming motions).

In our work, we introduce another measured velocity dispersion,
$\sigma_{\rm gs,los}$, as defined by \citet{utomo2015}, although we
adopt the notation of \citet{henshaw2019}. We first calculate the
intensity-weighted mean velocity at each line of sight through a cloud
($\bar{v}(x_i,y_i)$), and measure its offset with respect to the mean
velocity at the cloud centre ($\bar{v}(x_0,y_0)$). We assume that this
offset ($\bar{v}(x_i,y_j)-\bar{v}(x_0,y_0)$) is produced by both
intrinsic motions within the cloud and/or large-scale galactic disc
motions, and thereby shift the velocities at each line of sight to
match their mean velocity to that of the cloud centre
($\bar{v}(x_0,y_0)$). We then measure the second moment of the shifted
emission distribution along the velocity axis and extrapolate it to
$T_{\rm edge}=0$~K. The final derived gradient-subtracted velocity
dispersion, $\sigma_{\rm gs,los}$, is also deconvolved for the channel
width as above. We thus obtain a measure of the turbulent (random)
motions within the cloud only, free of any bulk motion.

\textbf{\textit{Cloud luminosity.}} The CO(3-2) luminosity of each
cloud is given by
\begin{equation}
  \label{eq:LCO3-2}
  \frac{L_{\rm CO(3-2)}}{\rm K~km~s^{-1}~pc^2}=\frac{F_{\rm
      CO(3-2)}}{\rm K~km~s^{-1}~arcsec^2}\,\left(\frac{D}{\rm
      pc}\right)^2\left(\frac{\pi}{180\times3600}\right)^2~,
\end{equation}
where $F_{\rm CO(3-2)}$
is the zeroth moment (total flux) of the cloud extrapolated to
$T_{\rm edge}=0$~K using a quadratic extrapolation and $D$ is the
distance to NGC4429.

\textbf{\textit{Cloud gaseous mass.}} The CO luminosity-based mass of
each cloud is obtained from $L_{\rm CO(3-2)}$ using
\begin{equation}
  \frac{M_{\rm gas}}{\Msolar}=4.4~\frac{L_{\rm CO}}{\rm K~km~s^{-1}~pc^2}\,\,\frac{X_{\rm CO}}{2\times10^{20}~{\rm cm^{-2}~(K~km~s^{-1})^{-1}}}~,
\end{equation}
where $L_{\rm CO}$ is the cloud's CO(1-0) luminosity (see
Eq.~\ref{eq:LCO3-2} above) and $X_{\rm CO}$ is the assumed
CO-to-{\HH} conversion factor. The CO(3-2)/CO(1-0) intensity ratio was
measured to be $1.06\pm0.15$ (in beam temperature units) overall in
NGC4429 \citep{davis2018}, and we assume that value for all the clouds
here. We further adopt a standard Galactic conversion factor
$X_{\rm CO}=2.3\times10^{20}$~cm$^{-2}$~(K~km~s$^{-1}$)$^{-1}$
(including the mass contribution from helium; \citealt{strong1988,
bolatto2013}), commonly used in
previous extragalactic studies \citep[e.g.][]{hughes2013, colombo2014,
  utomo2015, sun2018}, although it has been suggested that this
conversion factor depends on the environment of each molecular cloud,
e.g.\ metallicity and radiation field (see \citealt{bolatto2013} for a
review). The final gaseous mass of each cloud is thus obtained from
\begin{equation}
  \frac{M_{\rm gas}}{\Msolar}=4.7~\frac{L_{\rm CO(3-2)}}{\rm K~km~s^{-1}~pc^2}~.
\end{equation}

\textbf{\textit{Cloud Virial mass.}} The Virial (i.e.\ dynamical) mass
of each cloud is calculated with the formula
\begin{equation}
  \label{eq:Virial_mass}
  M_{\rm vir}=\frac{\sigma^2R_{\rm c}}{b_{\rm s}G}
  =\frac{5\sigma^2R_{\rm c}}{G}
\end{equation}
\citep{maclaren1988}, where $G$ is the gravitational constant,
$\sigma$ the observed (i.e.\ 1D) cloud velocity dispersion,
$R_{\rm c}$ the cloud radius (see Eq.~\ref{eq:Rc}) and $b_{\rm s}$ is
a geometrical factor that quantifies the effects of inhomogeneities
and/or non-sphericity of the cloud mass distribution on its
self-gravitational energy.  For a cloud in which the isodensity
contours are homoeoidal ellipsoids,
$b_{\rm s}=b_{\rm s_1}b_{\rm s_2}$, where $b_{\rm s_1}$ quantifies the
effects of the inhomogeneities and $b_{\rm s_2}$ those of the
ellipticity (see Appendix~\ref{app:modified_Virial_theorem} for more
details on $b_{\rm s_1}$ and $b_{\rm s_2}$). We adopt
$b_{\rm s}=\frac{1}{5}$ for a spherical homogeneous (i.e.\ constant
density) cloud whenever we need to evaluate $M_{\rm vir}$. The Virial
mass obtained from Eq.~\ref{eq:Virial_mass} assumes that each cloud is
spherical and virialised (with isotropic velocity dispersions), with
no magnetic support or pressure confinement. We note that, to
investigate the dynamical state of each cloud in the presence of
strong tidal/shear forces, in the sections that follow we will define
different $M_{\rm vir}$ using velocity dispersions $\sigma$ calculated
in different ways. These will be clearly labeled when used to avoid
confusion.

\textbf{\textit{Cloud distance from the centre.}} The deprojected
distance ($R_{\rm gal}$) of a cloud from the centre of the galaxy
(${\rm R.A.\,(J2000)}=12^{\rm h}27^{\rm m}26\fs504\pm0\fs013$ and
${\rm Dec.\,(J2000)}=11^{\circ}06\arcmin27\farcs57\pm0\farcs01$ is
calculated assuming the clouds are located in an infinitelly thin
molecular gas disc with a position angle of $93^\circ$ and an
inclination angle of $68^\circ$ (i.e.\ an axis ratio of $0.37$; see
\citealt{davis2018}).

\textbf{\textit{Uncertainties.}} The uncertainties of our measured
cloud properties are estimated via a bootstrapping technique. For each
cloud, we generate $1000$ realisations of the data by randomly
sampling the initial distribution, with repetition allowed, to reach
the same number of cloud pixels. The cloud properties are measured for
each sampled structure, and the median absolute deviation is used to
estimate the fractional uncertainty of each property. The final
uncertainties are scaled by the square root of the number of spaxels
per synthesised beam area to account for the fact that not all of the
pixels are independent. Our bootstrap approach assumes the boundary of
each cloud is fixed, and therefore does not take into account the
uncertainties in defining the cloud themselves. Nevertheless, we have
compared the uncertainties produced by our bootstrapping method to
those derived from other techniques \citep[e.g.][]{rosolowsky2006,
  faesi2016}, demonstrating that they are similar and thus reliable.We
note that the uncertainty of the gradient-subtracted velocity
dispersion $\sigma_{\rm gs,los}$ is derived via the same bootstrapping
technique, and thus includes the uncertainty of the adopted mean
velocity at the cloud centre.

The uncertainty of the adopted distance $D$ to NGC4429 was not
propagated through the uncertainties of the measured quantities. This
is because an error on the distance to NGC4429 translates to a
systematic (rather than random) scaling of some of the measured
quantities (no effect on the others), i.e.\ $R_{\rm c}\propto D$,
$L_{\rm CO(3-2)}\propto D^2$, $M_{\rm gas}\propto D^2$,
$\omega\propto D^{-1}$ and $R_{\rm gal}\propto D$.


\subsection{Table of GMC properties}
\label{sec:GMCproperties}

Table~\ref{tab:gmcs} lists the positions and
properties of the $217$ GMCs identified in our work. Around $65\%$
($141/217$) of the GMCs identified are resolved spatially, i.e.\ with
a deconvolved diameter larger than or equal to the synthesised beam
size.  All are resolved spectrally, i.e.\ with a deconvolved velocity
width at least half of one (Hanning smoothed) velocity channel
\citep{donovanmeyer2013}. All masked CO flux has been assigned to a
cloud, and the total flux of all clouds ($\approx43$~Jy~km~s$^{-1}$)
is about $60\%$ of the integrated flux of the galaxy
($75$~Jy~km~s$^{-1}$). The diffuse emission below the adopted
threshold of $2$ times the RMS noise is not included in our
analysis. As our primary beam covers all the CO emission in NGC4429,
our derived GMC catalogue is complete at $^{12}$CO(3-2).


\begin{table*}
  \centering
  \caption{Observed properties of the clouds in NGC4429.}
  \label{tab:gmcs}
  \resizebox{\textwidth}{!}{%
    \begin{tabular}{rcccccccccccc}
      \hline\hline
      ID & RA(2000) & Dec(2000) & $V_{\rm LSR}$ & $R_{\rm c}$ & $\sigma_{\rm obs,los}$ & $\sigma_{\rm gs,los}$ & $L_{\rm CO(3-2)}$ & $M_{\rm gas}$ & $T_{\rm max}$ & $\omega$ & $\phi_{\rm rot}$ & $R_{\rm gal}$ \\
         & (h:m:s)  & ($^\circ:^\prime:^{\prime \prime}$) & (${\rm km~s^{-1}}$) & (pc) & (${\rm km~s^{-1}}$) & (${\rm km~s^{-1}}$) & ($10^4~{\rm K~km~s^{-1}~pc^2}$) & ($10^5$ M$_\odot$) & (K) & (${\rm km~s^{-1}~pc^{-1}}$) & ($^\circ$) & (pc) \\
      \hline\hline
  1& 12:27:26.2& 11:06:27.9& \phantom{1}853.8& $\dots$ & $\phantom{1}1.50\pm1.06$ & $1.25\pm1.12$ & $\phantom{1}0.92\pm0.26$ & $0.43\pm0.12$ & 3.8& $\dots$  & $\dots$  & 404 \\
  2& 12:27:26.2& 11:06:28.2& \phantom{1}864.4& $16.69\pm5.27$ & $\phantom{1}4.81\pm1.45$ & $2.81\pm1.02$ & $\phantom{1}3.01\pm0.56$ & $1.40\pm0.26$ & 3.3& $0.43\pm0.14$ & $139\pm\phantom{1}25$ & 372 \\
  3& 12:27:26.2& 11:06:27.6& \phantom{1}864.6& $23.45\pm3.63$ & $\phantom{1}6.69\pm1.00$ & $3.79\pm0.74$ & $\phantom{1}4.99\pm0.71$ & $2.32\pm0.33$ & 3.8& $0.27\pm0.04$ & $175\pm\phantom{1}14$ & 370 \\
  4& 12:27:26.2& 11:06:27.4& \phantom{1}872.4& $20.18\pm4.12$ & $\phantom{1}4.47\pm0.92$ & $2.95\pm0.85$ & $\phantom{1}2.59\pm0.44$ & $1.21\pm0.21$ & 3.4& $0.23\pm0.09$ & $290\pm\phantom{1}19$ & 339 \\
  5& 12:27:26.2& 11:06:27.8& \phantom{1}875.9& $46.89\pm3.03$ & $\phantom{1}5.53\pm0.54$ & $2.35\pm0.33$ & $11.49\pm1.01$ & $5.35\pm0.47$ & 4.2& $0.20\pm0.01$ & $163\pm\phantom{1}\phantom{1}3$ & 309 \\
  6& 12:27:26.2& 11:06:27.0& \phantom{1}876.9& $17.79\pm3.15$ & $\phantom{1}4.51\pm0.99$ & $2.86\pm0.90$ & $\phantom{1}2.34\pm0.50$ & $1.09\pm0.23$ & 3.4& $0.28\pm0.09$ & $222\pm\phantom{1}24$ & 394 \\
  7& 12:27:26.2& 11:06:26.8& \phantom{1}883.7& $17.76\pm4.90$ & $\phantom{1}5.10\pm1.05$ & $2.93\pm0.80$ & $\phantom{1}2.96\pm0.50$ & $1.38\pm0.23$ & 3.9& $0.13\pm0.04$ & $149\pm\phantom{1}25$ & 407 \\
  8& 12:27:26.3& 11:06:28.3& \phantom{1}887.3& $18.99\pm4.82$ & $\phantom{1}3.58\pm1.18$ & $1.83\pm1.07$ & $\phantom{1}1.50\pm0.28$ & $0.70\pm0.13$ & 3.8& $0.16\pm0.10$ & $138\pm\phantom{1}49$ & 296 \\
  9& 12:27:26.2& 11:06:28.5& \phantom{1}885.2& $\dots$ & $\phantom{1}3.62\pm1.11$ & $3.08\pm1.27$ & $\phantom{1}1.27\pm0.30$ & $0.59\pm0.14$ & 3.7& $\dots$  & $\dots$  & 345 \\
 10& 12:27:26.3& 11:06:27.7& \phantom{1}888.2& $29.40\pm3.14$ & $\phantom{1}4.81\pm0.75$ & $2.97\pm0.70$ & $\phantom{1}4.69\pm0.59$ & $2.18\pm0.28$ & 3.8& $0.16\pm0.03$ & $230\pm\phantom{1}13$ & 248 \\
 11& 12:27:26.2& 11:06:26.7& \phantom{1}892.8& $\dots$ & $\phantom{1}2.84\pm0.74$ & $2.52\pm0.81$ & $\phantom{1}0.99\pm0.27$ & $0.46\pm0.13$ & 4.2& $\dots$  & $\dots$  & 400 \\
 12& 12:27:26.2& 11:06:26.9& \phantom{1}894.8& $\dots$ & $\phantom{1}5.49\pm1.42$ & $3.11\pm1.24$ & $\phantom{1}1.38\pm0.37$ & $0.64\pm0.17$ & 3.5& $\dots$  & $\dots$  & 371 \\
 13& 12:27:26.3& 11:06:27.2& \phantom{1}892.9& $26.18\pm2.65$ & $\phantom{1}6.34\pm0.86$ & $3.22\pm0.58$ & $\phantom{1}6.12\pm0.60$ & $2.85\pm0.28$ & 4.8& $0.33\pm0.03$ & $213\pm\phantom{1}\phantom{1}7$ & 285 \\
 14& 12:27:26.3& 11:06:27.8& \phantom{1}898.0& $26.25\pm2.73$ & $\phantom{1}5.14\pm0.62$ & $2.52\pm0.42$ & $\phantom{1}5.33\pm0.56$ & $2.49\pm0.26$ & 4.4& $0.27\pm0.02$ & $202\pm\phantom{1}\phantom{1}6$ & 219 \\
 15& 12:27:26.2& 11:06:28.6& \phantom{1}898.4& $14.93\pm4.44$ & $\phantom{1}3.36\pm0.87$ & $2.46\pm0.68$ & $\phantom{1}2.37\pm0.52$ & $1.11\pm0.24$ & 3.5& $0.17\pm0.07$ & $\phantom{1}72\pm\phantom{1}39$ & 330 \\
 16& 12:27:26.3& 11:06:26.9& \phantom{1}900.3& $\dots$ & $\phantom{1}5.08\pm1.80$ & $2.57\pm1.55$ & $\phantom{1}1.37\pm0.41$ & $0.64\pm0.19$ & 2.7& $\dots$  & $\dots$  & 342 \\
 17& 12:27:26.3& 11:06:27.0& \phantom{1}902.0& $\dots$ & $\phantom{1}4.76\pm1.75$ & $4.20\pm1.95$ & $\phantom{1}0.82\pm0.19$ & $0.38\pm0.09$ & 3.8& $\dots$  & $\dots$  & 296 \\
 18& 12:27:26.3& 11:06:28.3& \phantom{1}901.2& $16.35\pm3.67$ & $\phantom{1}5.70\pm1.23$ & $2.18\pm0.81$ & $\phantom{1}1.75\pm0.40$ & $0.81\pm0.19$ & 3.3& $0.52\pm0.19$ & $148\pm\phantom{1}28$ & 279 \\
 19& 12:27:26.2& 11:06:26.4& \phantom{1}906.3& $14.16\pm5.81$ & $\phantom{1}6.04\pm2.54$ & $4.16\pm2.67$ & $\phantom{1}1.48\pm0.46$ & $0.69\pm0.21$ & 2.8& $0.34\pm0.92$ & $260\pm121$ & 438 \\
 20& 12:27:26.3& 11:06:27.3& \phantom{1}908.3& $20.56\pm4.42$ & $\phantom{1}3.33\pm0.84$ & $2.51\pm0.82$ & $\phantom{1}2.57\pm0.50$ & $1.20\pm0.23$ & 3.6& $0.10\pm0.04$ & $181\pm\phantom{1}55$ & 244 \\
      $ \dots$& $ \dots$& $ \dots$& $ \dots$& $ \dots$& $ \dots$& $ \dots$& $ \dots$& $ \dots$ & $ \dots$ & $ \dots$ & $ \dots$ & $ \dots$ \\
     217& 12:27:26.9& 11:06:27.2& 1344.3& $20.42\pm3.92$ & $\phantom{1}2.23\pm0.68$ & $1.58\pm0.89$ & $\phantom{1}1.94\pm0.34$ & $0.90\pm0.16$ & 3.8& $0.01\pm0.04$ & $ 85\pm138$ & 428 \\
      \hline\hline 
    \end{tabular}
  } \raggedright Notes.\ -- Measurements of $M_{\rm gas}$ assume a
  CO(3-2)/CO(1-0) line ratio of $1.06\pm0.15$ (in beam temperature
  units; \citealt{davis2018}) and a standard Galactic conversion
  factor $X_{\rm CO}=2\times10^{20}$~cm$^{-2}$~(K~km~s$^{-1}$)$^{-1}$
  (including the mass contribution from helium). All uncertainties are
  quoted at the $1\,\sigma$ level. As noted in the text, the
  uncertainty of the adopted distance $D$ to NGC4429 was not
  propagated through the tabulated uncertainties of the measured
  quantities. This is because an error on the distance to NGC4429
  translates to a systematic (rather than random) scaling of some of
  the measured quantities (no effect on the others), i.e.\
  $R_{\rm c}\propto D$, $L_{\rm CO(3-2)}\propto D^2$,
  $M_{\rm gas}\propto D^2$, $\omega\propto D^{-1}$ and
  $R_{\rm gal}\propto D$. Table~\ref{tab:gmcs} is available in its
  entirety in machine-readable form in the electronic edition.
\end{table*}

Table~\ref{tab:gmcs} lists each cloud's identification number, central
position in both R.A.\ and Dec., local standard of rest velocity
$V_{\rm LSR}$, radius $R_{\rm c}$, observed velocity dispersion
$\sigma_{\rm obs,los}$ and gradient-subtracted velocity dispersion
$\sigma_{\rm gs,los}$, total CO(3-2) luminosity $L_{\rm CO(3-2)}$,
gaseous mass $M_{\rm gas}$, peak intensity $T_{\rm max}$, angular
velocity $\omega$ and position angle of the rotation axis
$\phi_{\rm rot}$ (see
Section~\ref{sec:velocity_gradient_cloud}), and deprojected
distance from the centre of the galaxy $R_{\rm gal}$.


\subsection{Probability distribution functions of GMC properties}
\label{sec:probability_distribution_functions_gmc_properties}

\begin{figure*}
  \includegraphics[width=0.9\textwidth]{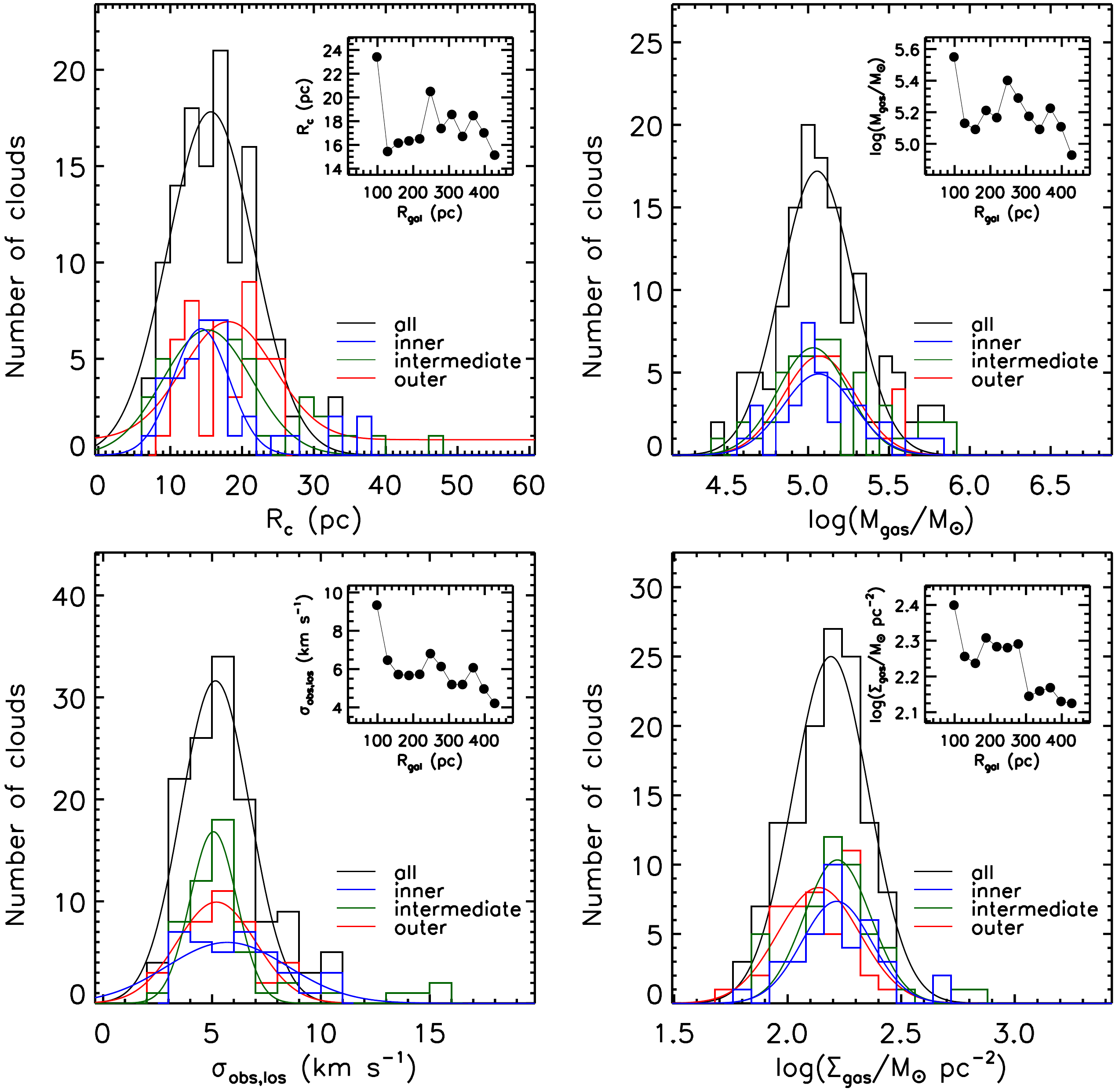}
  \caption{Distributions of $R_{\rm c}$,
    $\log(M_{\rm gas}/{\rm M_\odot})$, $\sigma_{\rm obs,los}$ and
    $\log(\Sigma_{\rm gas}/{\rm M_\odot~pc^{-2}})$ with their Gaussian
    fits for the $141$ spatially-resolved clouds identified in NGC4429
    (black histograms), and for only the clouds in the inner (blue
    histograms), intermediate (green histograms) and outer (red
    histograms) region of the galaxy, respectively. The insets show
    the median $R_{\rm c}$, $\log(M_{\rm gas}/{\rm M_\odot})$,
    $\sigma_{\rm obs,los}$ and
    $\log(\Sigma_{\rm gas}/{\rm M_\odot~pc^{-2}})$ in elliptical
    annuli of constant $R_{\rm gal}$ (and equal width
    $\Delta R_{\rm gal}=30$~pc).}
  \label{fig:gmc_properties}
\end{figure*}

The number distributions of $R_{\rm c}$,
$\log(M_{\rm gas}/{\rm M_\odot})$, $\sigma_{\rm obs,los}$ and
$\log(\Sigma_{\rm gas}/{\rm M_\odot~pc^{-2}})$ (where
$\Sigma_{\rm gas}$ is the characteristic gaseous mass surface density
of each cloud,
$\Sigma_{\rm gas}\equiv\frac{M_{\rm gas}}{\pi R_{\rm c}^2}$) for the
$141$ spatially-resolved clouds of NGC4429 are shown in
Fig.~\ref{fig:gmc_properties}. We divide the
galaxy into three distinct regions (separated by the two brown dashed
ellipses in Fig.~\ref{fig:gmcs}): inner ($R_{\rm gal}\le220$~pc),
intermediate ($220<R_{\rm gal}\le330$~pc) and outer
($R_{\rm gal}>330$~pc) region. In each panel, the black histogram
(data) and curve (Gaussian fit) show the full sample, while the blue,
green and red colours show only the clouds in the inner, intermediate
and outer region, respectively. The insets show the median
$R_{\rm c}$, $\log(M_{\rm gas}/{\rm M_\odot})$, $\sigma_{\rm obs,los}$
and $\log(\Sigma_{\rm gas}/{\rm M_\odot~pc^{-2}})$ as functions of the
galactocentric distance $R_{\rm gal}$.

The spatially-resolved clouds of NGC4429 have sizes $R_{\rm c}$
ranging from $7$ to about $50$~pc (see Fig.~\ref{fig:gmc_properties},
top-left panel).
A Gaussian fit to the size distribution yields a mean of $16\pm0.5$~pc
and a standard deviation of $\approx6$~pc. The clouds in NGC4429
appear to have sizes smaller than those of clouds in the MW disc
(typical sizes $\approx30$ -- $50$~pc;
\citealt{mivilledeschenes2017}), Local Group galaxies (typical sizes
$\approx20$ -- $70$~pc; \citealt{rosolowsky2003, rosolowsky2007_M31,
  rosolowsky2007_M33, hirota2011}) and most late-type galaxies
(typical sizes $\approx20$ -- $200$~pc; \citealt{donovanmeyer2012,
  hughes2013, rebolledo2015}), but slightly larger than those of
clouds in the Galactic Centre (typical sizes $\approx5$ -- $15$~pc;
\citealt{oka2001, kauffmann2017}) and the ETG NGC4526 (typical sizes
$\approx5$ -- $30$~pc; \citealt{utomo2015}). We note however that the
CO $J=3-2$ transition used in our work traces the warm molecular
medium ($10-50$~K) around active SF regions, and has a higher
characteristic density than the $J=1-0$ transition
($\approx7\times10^4$ versus $\approx1.4\times10^3$~cm$^{-3}$). The
CO(3-2) line could therefore potentially trace more compact structures
than CO(1-0) \citep{miville-deschenes2017,colombo2019}. The inset in
the top-left panel presents the median cloud size as a function of
galactocentric distance. We note that the three innermost resolved
clouds (clouds No.\ 32, 165 and 183; $R_{\rm gal}\le100$~pc), that all
lie along the major axis, have exceptionally large masses and/or
surface densities. Except for these three innermost resolved clouds,
the clouds in the inner region generally have slightly smaller sizes
than the clouds at larger radii (i.e.\ in the intermediate and outer
regions). The sizes of the clouds appear to slightly increase with
galactocentric distance but drop at the outer edge of the molecular
disc ($R_{\rm gal}\gs375$~pc).

The gaseous masses $M_{\rm gas}$ of the spatially-resolved clouds of
NGC4429 range from $2.8\times10^4$ to $8\times10^5$~{\Msolar} (see
Fig.~\ref{fig:gmc_properties}, top-right panel).  The median cloud
gaseous mass of the sample is $\approx1.6\times10^5$~{\Msolar}. More
than one third ($54/141$) of the resolved clouds are light
($M_{\rm gas}\le10^5$~{\Msolar}), but they overall contribute only
$\approx16\%$ of the total molecular gas mass in clouds. There is no
cloud more massive than $M_{\rm gas}=10^6$~{\Msolar} in NGC4429. The
clouds in NGC4429 have gaseous masses slightly smaller than those of
clouds in the MW disc ($\approx10^{4.5}$ -- $10^{7.0}$~{\Msolar};
\citealt{rice2016}), NGC4826 ($\approx10^{6.0}$ --
$10^{7.2}$~{\Msolar}; \citealt{rosolowsky-blitz2005}), NGC1068
($\approx10^{4.2}$ -- $10^{7.6}$~{\Msolar}; \citealt{tosaki2016}), M51
($\approx10^{5.0}$ -- $10^{7.5}$~{\Msolar}; \citealt{colombo2014}),
NGC253 ($\approx10^{6.3}$ -- $10^{7.8}$~{\Msolar};
\citealt{leroy2015}) and the LMC ($\approx10^{4.2}$ --
$10^{6.8}$~{\Msolar}; \citealt{hughes2010}), but similar to those of
clouds in M31 ($\approx10^4$ -- $10^6$~{\Msolar};
\citealt{rosolowsky2007_M31}), M33 ($\approx10^4$ -- $10^6$~{\Msolar};
\citealt{rosolowsky2003, rosolowsky2007_M33}), the SMC ($\approx10^4$
-- $10^6$~{\Msolar}; \citealt{muller2010}) and the ETG NGC4526
($\approx10^{4.7}$ -- $10^{6.6}$~{\Msolar}; \citealt{utomo2015}). The
clouds in the intermediate region tend to be more massive than the
clouds in the inner and outer regions (see the inset in the top-right
panel). The median cloud mass also appears to drop abruptly in the
outermost region of the molecular disc ($R_{\rm gal}\gs375$~pc).


The spatially-resolved clouds of NGC4429 have observed velocity
dispersions (linewidths) $\sigma_{\rm obs,los}$ between $2$ and
$16$~{\kms} (see Fig.~\ref{fig:gmc_properties}, bottom-left panel).  A
Gaussian fit to the velocity dispersion distribution yields a mean of
$5.2\pm0.2$~{\kms}. The clouds in NGC4429 have observed velocity
dispersions higher than those of clouds with the same sizes in the MW
and Local Group galaxies (where $\sigma_{\rm obs,los}$ is typically
$2$ -- $3$~{\kms}; \citealt{rosolowsky2003, rosolowsky2007_M31,
  rosolowsky2007_M33, fukui2008, muller2010}), but similar to those of
the clouds in the ETG NGC4526 ($\sigma_{\rm obs,los}\approx5$ --
$16$~{\kms}; \citealt{utomo2015}). Almost all clouds with high
velocity dispersions ($\sigma_{\rm obs,los}\ge10$~{\kms}) are located
in the inner and intermediate regions. We find a general trend of
slightly decreasing velocity dispersion with galatocentric radius (see
the inset in the bottom-left panel).

The gaseous mass surface densities $\Sigma_{\rm gas}$ of
spatially-resolved clouds in NGC4429 have a range of $\approx40$ --
$650$~{\Msolar}~pc$^{-2}$ (see Fig.~\ref{fig:gmc_properties},
bottom-right panel). A Gaussian fit to the distribution of
$\log(\Sigma_{\rm gas}/{\rm M_\odot~pc^{-2}})$ yields a mean of
$2.2\pm0.17$~dex. The clouds in NGC4429 have an average gaseous mass
surface density that is lower than that of the clouds in the ETG
NGC4526
($\langle\Sigma_{\rm gas}\rangle\approx1000$~{\Msolar}~pc$^{-2}$;
\citealt{utomo2015}), but is comparable to that of the clouds in M33
and M64
($\langle\Sigma_{\rm gas}\rangle\approx100$~{\Msolar}~pc$^{-2}$;
\citealt{rosolowsky2003, rosolowsky-blitz2005}) and is larger than
that of the clouds in the MW disc and the LMC
($\langle\Sigma_{\rm gas}\rangle\approx50$~{\Msolar}~pc$^{-2}$;
\citealt{lombardi2010, heyer2009, hughes2010,
  mivilledeschenes2017}). The gaseous mass surface densities of
individual clouds in NGC4429 vary by more than an order of
magnitude. We find that the clouds in the inner region tend to have a
slightly larger minimum gaseous mass surface density
($\Sigma_{\rm gas}\ge70$~{\Msolar}~pc$^{-2}$) than the clouds in the
intermediate ($\Sigma_{\rm gas}\ge60$~{\Msolar}~pc$^{-2}$) and outer
($\Sigma_{\rm gas}\ge40$~{\Msolar}~pc$^{-2}$) region. The general
trend is that the clouds at smaller radii have higher gaseous mass
surface densities (see the inset in the bottom-right panel).


\subsection{GMC mass spectra}
\label{sec:gmc_mass_spectra}

\begin{figure}
  \includegraphics[width=0.95\columnwidth]{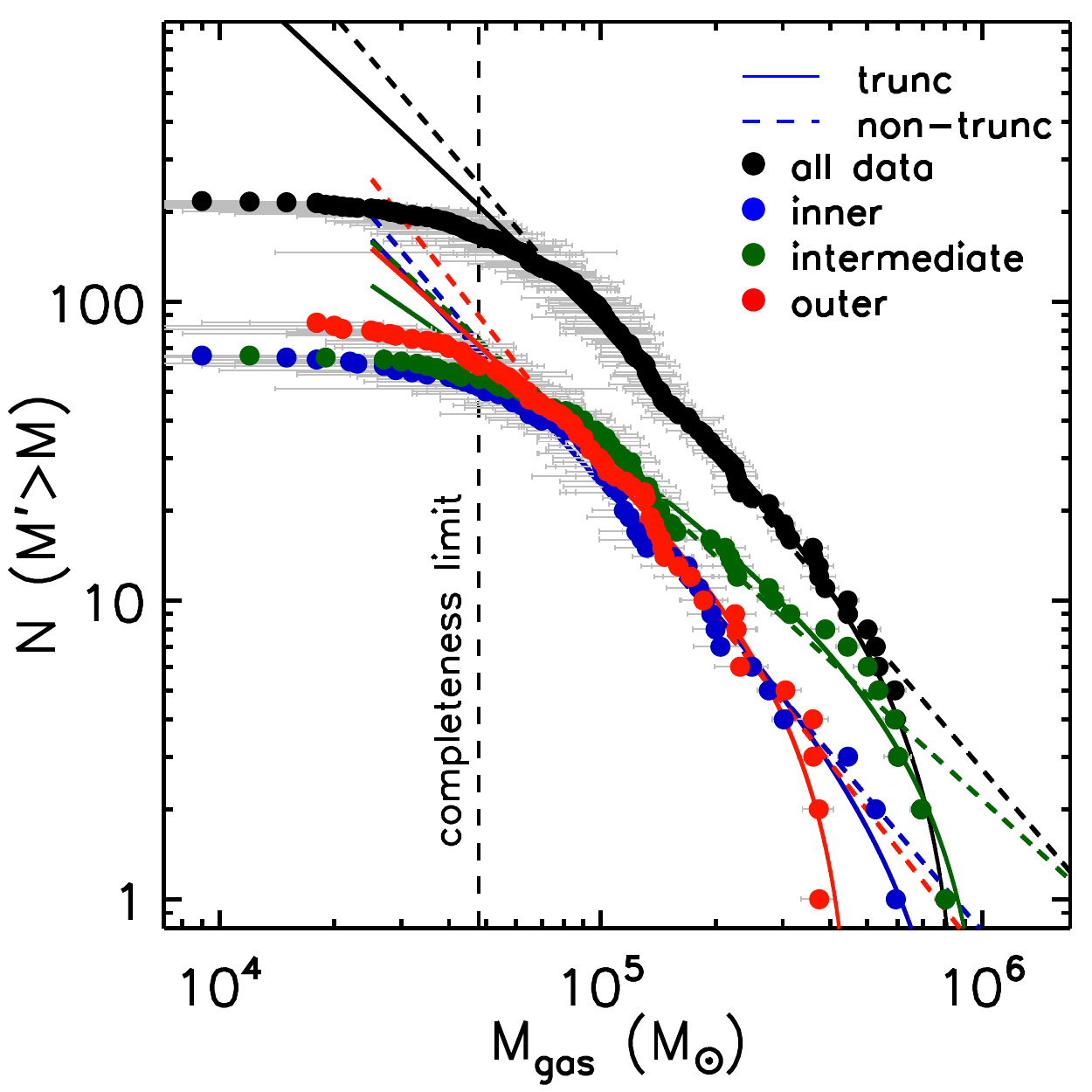}
  \caption{Cumulative gaseous mass distribution of all the clouds of
    NGC4429 (black data points) and only the clouds in the inner (blue
    data points), intermediate (green data points) and outer (red data
    points) region, respectively. Truncated (solid curve) and
    non-truncated (dashed curves) power-law fits are overlaid in
    matching colours. Our mass completeness limit is indicated by the
    black vertical dashed line.}
  \label{fig:mass_function}
\end{figure}

The distribution of GMCs by mass is a critical diagnostic of a GMC
population and provides important clues to GMC formation and
destruction \citep{rosolowsky2005, colombo2014}. We choose the gaseous
mass over the Viral mass to determine the mass function, because gas
mass does not require assumptions about the dynamical state of the
GMCs and is well defined even for spatially-unresolved clouds. We fit
the cumulative mass distribution (see Fig.~\ref{fig:mass_function})
instead of the differential mass distribution, as
\citet{rosolowsky2005} argues that the former is more reliable than
the latter as it is not affected by biases related to binning and it
can account for uncertainties of the cloud masses.

Cumulative mass distribution functions can be characterised
quantitatively by a power-law function
\begin{equation}
  N(M^\prime>M)=\left(\frac{M}{M_0}\right)^{\gamma+1}~,
\end{equation}
where $N(M^\prime>M)$ is the number of clouds with a mass greater than
$M$, $M_0$ sets the normalisation, and $\gamma$ is the power-law
index. Alternatively, a truncated power-law function can be used,
\begin{equation}
  N(M^\prime>M)=N_0\left[\left(\frac{M}{M_0}\right)^{\gamma+1}-1\right]~,
\end{equation}
where $M_0$ is now the cut-off mass of the distribution and $N_0$ is
the number of clouds with a mass $M>2^{1/(\gamma+1)}M_0$, the cut-off
point of the distribution (for a meaningful truncation to exist, one
expects $N_0\gg1$).

We fit the cumulative mass spectra by applying the \enquote{error in
  variables} method developed by \citet{rosolowsky2005}, that adopts
an iterative maximum-likelihood approach to estimate the best-fitting
parameters and account for uncertainties of both the cloud mass and
the number distribution. Fitting is only performed above the
completeness limit of $M_{\rm com}=4\times10^4$~{\Msolar}, shown as a
black vertical dashed line in Fig.~\ref{fig:mass_function}. We
estimate the mass completeness limit based on the minimum
spatially-resolved cloud (gaseous) mass ($M_{\rm min}$) and the
observational sensitivity, i.e.\
$M_{\rm com}\equiv M_{\rm min}+10\delta_{\rm M}$, where the
contribution to the mass due to noise, $\delta_{\rm M}$, is estimated
by multiplying our RMS column density sensitivity limit of
$10$~{\Msolar}~pc$^{-2}$ by the synthesised beam area of
$\approx180$~pc$^2$. The parameters of the best-fitting truncated
power laws to the cumulative (gaseous) mass distributions of the
clouds in NGC4429 are listed in Table~\ref{tab:mass_function}.


\begin{table}
  \centering
  \caption{Parameters of the truncated power laws best fitting the
    cumulative gaseous mass distributions of the clouds in NGC4429.}
  \label{tab:mass_function}
  \resizebox{\columnwidth}{!}{%
    \begin{tabular}{lcccc}
      \hline\hline
      Region & Distance & $\gamma$ & $M_0$             & $N_0$ \\
             & (pc)     &          & ($10^5$~M$_\odot$) & \\
      \hline\hline
      All & $\phantom{00}0<R_{\rm gal}\le450$ & $-2.18\pm0.21$ & $\phantom{0}8.8\pm1.3$ & $6.9\pm4.4$ \\
      Inner & $\phantom{00}0<R_{\rm gal}\le220$ & $-2.32\pm0.24$ & $\phantom{0}9.2\pm2.5$ & $1.4\pm1.9$  \\
      Intermediate & $220<R_{\rm gal}\le330$ & $-1.83\pm0.33$ & $10.6\pm1.6$ & $5.2\pm5.5$ \\
      Outer & $330 <R_{\rm gal}\le450$ & $-2.08\pm0.32$ & $\phantom{0}4.6\pm0.4$ & $6.6\pm5.4$ \\
      \hline\hline 
    \end{tabular}
  }
  \raggedright Notes.\ -- All uncertainties are quoted at the $1\,\sigma$ level.
\end{table}

We find strong evidence for a curtailment of very massive GMCs in
NGC4429, as a truncated power-law function (black solid line in
Fig.~\ref{fig:mass_function}) with a high value of $N_0$ ($6.9\pm4.4$)
fits the gaseous mass distribution much better than a pure power-law
function (black dashed line). This implies that NGC4429 lacks the
processes that actively accumulate molecular gas clumps into high-mass
GMCs.
The best truncated fit yields a slope $\gamma=-2.18\pm0.21$, a slope
steeper than $-2$ implying that most of the molecular gas mass of
NGC4429 is in low-mass clouds and there should thus be a significant
amount of gas below our completeness limit. This is consistent with
the fact that only $\approx60\%$ of the emission is decomposed into
clouds at our resolution (see
Section~\ref{sec:GMCproperties}). However, there must also be a lower
gaseous mass limit for the molecular clouds or a turnover at low mass
for the total mass to remain finite.

Our derived slope $\gamma$ is similar to that measured for the clouds
in in the outer Galaxy ($-2.2\pm 0.1$; \citealt{rice2016}), the ETG
NGC4526 ($-2.39\pm0.03$; \citealt{utomo2015}), M51 ($-2.3\pm1$;
\citealt{colombo2014}) and the outer regions of M33 ($-2.1\pm1$;
\citealt{rosolowsky2007_M33}), but is steeper than that for the clouds
in the inner Galaxy ($-1.6\pm0.1$; \citealt{rice2016}), the spiral
arms of M51 ($-1.79\pm0.09$; \citealt{colombo2014}), NGC1068
($-1.25\pm0.07$; \citealt{tosaki2016}), the inner regions of M33
($-1.8\pm1$; \citealt{rosolowsky2007_M33}), NGC300 ($-1.80\pm0.07$;
\citealt{faesi2016}) and the overall mass spectrum of Local Group
galaxies ($\approx-1.7$; \citealt{blitz2007}).

The best-fitting cut-off gaseous mass $M_0$ of our truncated
distribution ($(8.8\pm1.3)\times10^5$~{\Msolar}) is comparable to that
for the clouds in the outer Galaxy ($(1.5\pm0.5)\times10^6$~{\Msolar};
\citealt{rice2016}) and the inner regions of M33
($(7.4\pm0.5)\times10^5$~{\Msolar}; \citealt{rosolowsky2007_M33}), but
is much lower than that for the clouds in most other galaxies such as
the inner Galaxy ($(1.0\pm0.2)\times10^7$~{\Msolar};
\citealt{rice2016}), the ETG NGC4526
($(4.12\pm0.08)\times10^6$~{\Msolar}; \citealt{utomo2015}), M51
($(1.8\pm0.3)\times10^6$~{\Msolar}; \citealt{colombo2014}), NGC1068
($(5.9\pm0.6)\times10^7$~{\Msolar}; \citealt{tosaki2016}) and the
outer regions of M33 ($(3.4\pm1.2)\times10^6$~{\Msolar};
\citealt{rosolowsky2007_M33}).

Variations of the GMC gaseous mass distribution as a function of
galactocentric distance can also be quantified. We find the cloud
cumulative gaseous mass functions of the three regions to be slightly
different, with a best-fitting truncated slope $\gamma$ of
$-2.32\pm0.24$, $-1.83\pm0.33$ and $-2.08\pm0.32$ and a cut-off
gaseous mass $M_0$ of $(9.2\pm2.5)\times10^5$,
$(10.6\pm1.6)\times10^5$ and $(4.6\pm0.4)\times10^5~\Msolar$ in the
inner, intermediate and outer region, respectively. The distributions
of the clouds in the inner and outer regions appear to be similar at
gaseous masses below $2\times10^5$~{\Msolar}, but the latter shows a
truncation while the former seems to be better fit by a pure power law
even at the high-mass end. Massive clouds appear to be suppressed at
the galaxy centre and especially in the outer regions of the
disc. Indeed, the distribution of clouds with gaseous masses greater
than the completeness limit cuts off abruptly inside $40$~pc and
beyond $450$~pc (see Fig.~\ref{fig:gmcs}). More than half of the most
massive clouds ($>2.5\times10^5$~{\Msolar}) are located in the
intermediate region, implying that the survival of massive clouds is
more favoured in this region. Overall, the environmental dependence of
the gaseous mass spectrum indicates that the formation and destruction
mechanisms of GMCs are (slightly) different at different
galactocentric distances.

\section{Cloud Kinematics}
\label{sec:cloud_kinematics}


\subsection{Velocity gradients of individual clouds}
\label{sec:velocity_gradient_cloud}

We observe strong velocity gradients within individual GMCs. Many
authors argue that these gradients are the signature of cloud rotation
\citep[e.g.][]{blitz1993, phillips1999, rosolowsky2003,
  rosolowsky2007_M31, utomo2015}. The observed velocity gradient of
each cloud can be quantified by fitting a plane to its
intensity-weighted first moment (i.e.\ mean line-of-sight velocity)
map $\bar{v}(x,y)$:
\begin{equation}
  \label{eq:cloud_vfield}
  \bar{v}(x,y)=ax+by+c~,
\end{equation}
where $a$ and $b$ are the projected velocity gradient along
respectively the $x$- and the $y$-axis on the sky (selected here in
the standard/intuitive manner, i.e.\ respectively reversely
proportional to the right ascension and proportional to the
declination). We adopt the code \texttt{lts\_planefit} to perform the
fits. This code combines least-trimmed-squares robust techniques
\citep{rousseeuw2006} into a least-squares fitting algorithm, and
allows for intrinsic scatter, uncertainties, possible large outliers
and weighting of each pixel by its flux (i.e.\ gaseous mass surface
density). The projected angular velocity $\omega_{\rm obs}$ (i.e.\ the
magnitude of the projected velocity gradient) and position angle of
the rotation axis $\phi_{\rm rot}$ are then given by the best-fitting
coefficients:
\begin{gather}
  \omega_{\rm obs}=\sqrt{a^2+b^2}~,\\
  \phi_{\rm rot}=\tan^{-1}(b/a)~.
\end{gather}
The uncertainties of $\omega_{\rm obs}$ and $\phi_{\rm rot}$ are
estimated from the uncertainties of the parameters $a$ and $b$ using
standard error propagation rules. We note that these derived projected
angular velocities $\omega_{\rm obs}$ are underestimated by a factor
$1-\cos(\varphi)$ compared to the intrinsic ones (i.e.\
$\omega_{\rm obs}=\cos(\varphi)\omega_{\rm int}$), where $\varphi$ is
the angle between the cloud rotation axis and the plane of the
sky. This is however inconsequential for all following analyses and
discussions, as all modelled quantities will themselves be projected
onto the sky (according to the model assumptions) before comparison.

Fitting a plane to the mean line-of-sight velocity map of each cloud
implicitly assumes cloud solid-body rotation. While this may not be
intrinsically true (i.e.\ the angular velocity may depend on the
radius within each cloud), because our clouds are generally relatively
poorly spatially resolved, $\omega_{\rm obs}$ as defined above
nevertheless provides a useful single quantity to quantify the bulk
(projected) rotation of each cloud.
  
\begin{figure}
  \includegraphics[width=0.95\columnwidth]{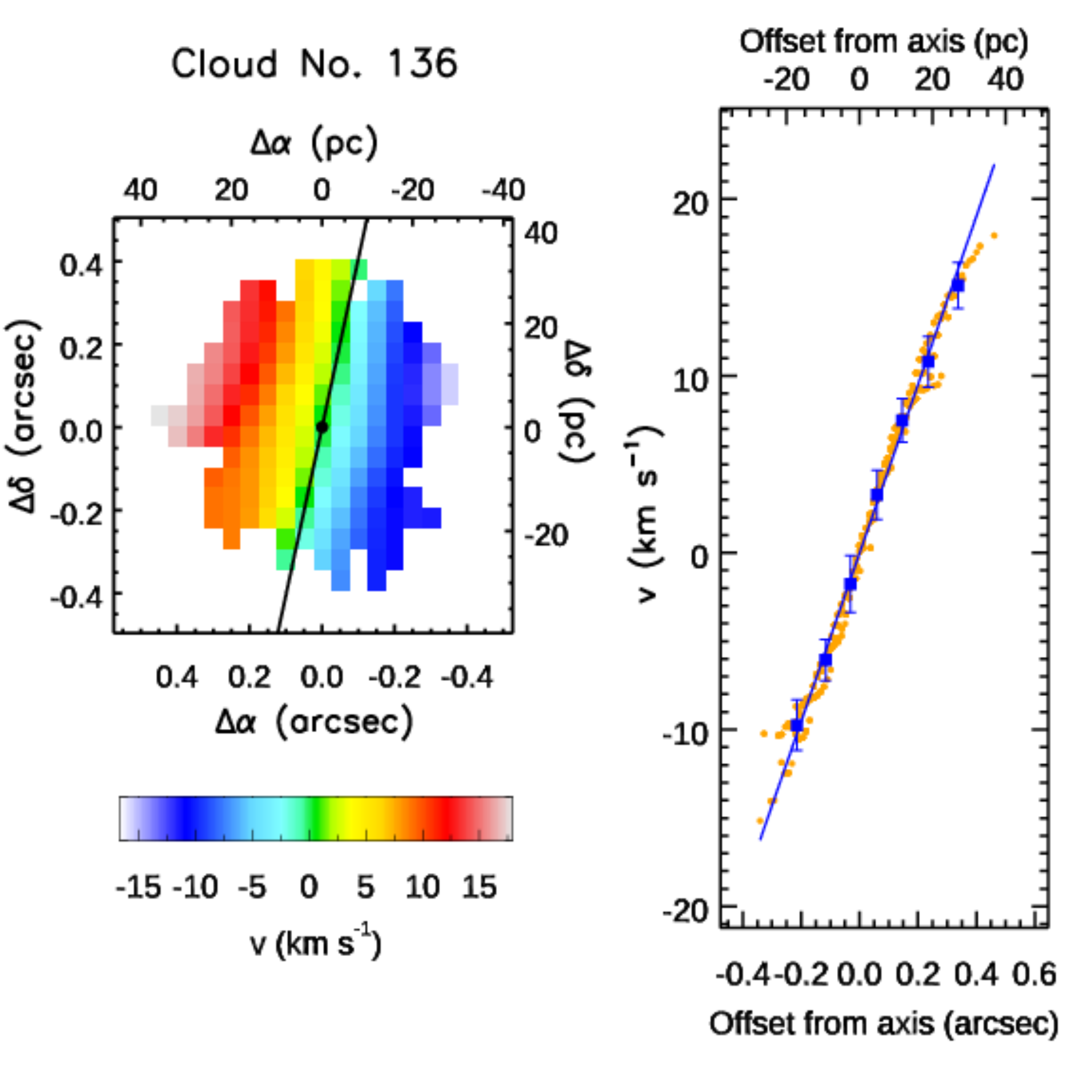}
  \caption{One example of plane fitting to the intensity-weighted
    first moment (i.e.\ mean line-of-sight velocity) map of a cloud of
    NGC4429 (here cloud No.\ 136). The left panel shows the cloud's
    mean velocity map with the best-fitting rotation axis (black line)
    and centre (black solid circle) overplotted. The right panel shows
    the mean line-of-sight velocity of each pixel within the cloud
    against the perpendicular distance of the pixel from the
    best-fitting rotation axis (orange data points). Blue squares are
    means of the velocity in bins of perpendicular distance from the
    rotation axis. For illustrative purposes only, the blue line shows
    the best-fitting straight line to the data, indicating that
    solid-body rotation is a good description of the cloud's
    kinematics.}
  \label{fig:cloud_kinematics}
\end{figure}

Figure~\ref{fig:cloud_kinematics} provides one example of our plane
fitting to the mean line-of-sight velocity map of a cloud of
NGC4429. The left panel shows the intensity-weighted first moment map
with the best-fitting rotation axis (black line) and centre (black
solid circle) overplotted. For illustrative purposes only, the right
panel shows the mean velocity of each pixel within the cloud
($\bar{v}(x,y)$) against the perpendicular distance of the pixel from
the best-fitting cloud rotation axis. A cloud with solid-body rotation
should have all its data points well fit by a straight line, as is the
case here. Overall, we find that planes are reasonable fits to the
velocity maps of most of the clouds in NGC4429, and the median value
of the reduced $\chi^2$ for the $141$ spatially-resolved clouds is
$\chi_{\rm r}^2=0.8$. More than half ($82$) of the resolved clouds are
well-fit by solid-body rotation ($\chi_{\rm r}^2\le1$).

The best-fitting results are listed in Table~\ref{tab:gmcs}. The
projected velocity gradients $\omega_{\rm obs}$ of the $141$
spatially-resolved clouds range from $0.05$ to
$0.91$~{\kms}~pc$^{-1}$, with an average of
$\approx0.33$~{\kms}~pc$^{-1}$. Our derived velocity gradients are
significantly larger than those inferred for MW clouds
($\sim0.1$~{\kms}~pc$^{-1}$; \citealt{blitz1993, phillips1999,
  imara2011a}), M33 ($\approx0.15$~{\kms}~pc$^{-1}$;
\citealt{rosolowsky2003, imara2011b}) and M31 ($0$ --
$0.2$~{\kms}~pc$^{-1}$; \citealt{rosolowsky2007_M31}), but they are
comparable to those inferred for the clouds of the ETG NGC4526 ($0$ --
$1.0$~{\kms}~pc$^{-1}$; \citealt{utomo2015}).


\subsection{Origin of the clouds' velocity gradients} 
\label{sec:origin_velocity_gradients}

The observed velocity gradients of the clouds can arise from turbulent
motions, the clouds' intrinsic rotation and/or galaxy rotation.
\citet{burkert2000} suggested that turbulent velocity fields can
produce observed linear gradients, that were estimated to be of order
$0.08$~{\kms}~${\rm pc^{-1}}$ for their median cloud radius of
$20$~pc. As our measured (i.e.\ projected) velocity gradients are
generally much larger than this, we suggest turbulence is not
important to account for them.

\begin{figure*}
  \includegraphics[width=0.9\textwidth]{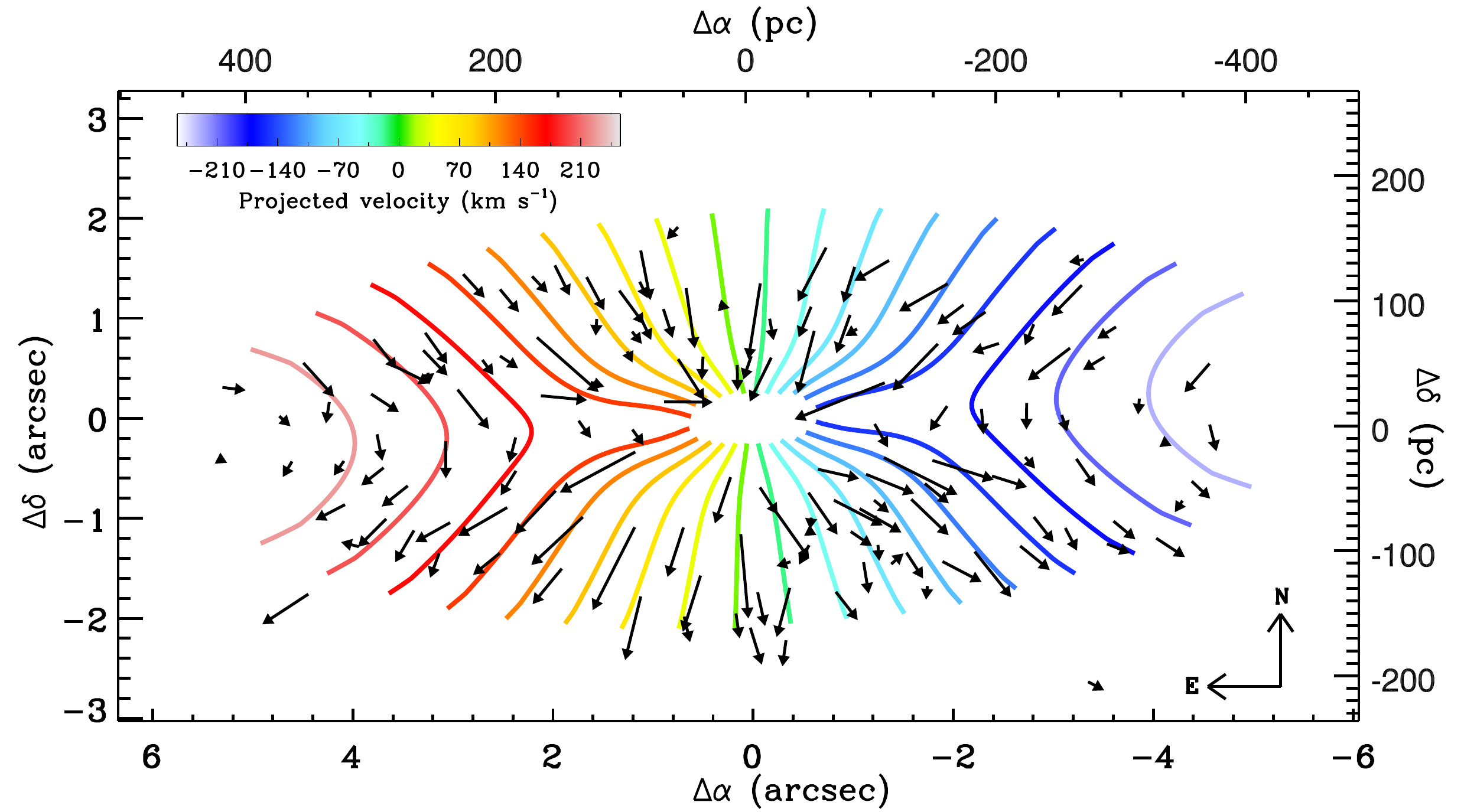}
  \caption{Projected directions of the angular momentum vectors of
    individual spatially-resolved GMCs in NGC4429 (black arrows),
    overplotted on the isovelocity contours of the molecular gas
    (colour coded by the projected velocities). The length of the
    arrows represents the magnitudes of the velocity gradients (i.e.\
    $\omega_{\rm obs}$). The projected velocities are derived from
    our gas dynamical model assuming pure rotation (see text).}
  \label{fig:gmc_vfield}
\end{figure*}

The observed velocity gradients of the clouds in NGC4429 are more
likely produced by the intrinsic rotation of the clouds and/or galaxy
rotation. Galaxy rotation can produce velocity gradients across the
small areas occupied by GMCs, especially at small galactocentric
distances corresponding to the steep part of the rotation curve. To
identify the origin of the observed velocity gradients of the clouds
of NGC4429, we overplot the rotation axes of the individual clouds
(i.e.\ the projected directions of their angular momentum vectors) on
the isovelocity contours of the galaxy in
Fig.~\ref{fig:gmc_vfield}. If the velocity gradients of the clouds are
produced by the clouds' intrinsic rotation, their rotation axes should
be randomly distributed. On the other hand, if the velocity gradients
of the clouds are produced by the galaxy rotation, their rotation axes
should show a strong alignment with the galaxy isovelocity contours.

As shown in Fig.~\ref{fig:gmc_vfield}, we do find a strong
tendency for the projected angular momentum vectors of the clouds to
be tangential to the isovelocity contours of NGC4429, implying that
the observed velocity gradients of the clouds are primarily a
consequence of galactic rotation. This is similar to the trend in
NGC4526 \citep{utomo2015}, but different from that in the MW
\citep{koda2006} and M31 \citep{rosolowsky2007_M31}, where the
distributions of position angles are random.

Here the isovelocity contours due to the galaxy rotation were derived
by creating a gas dynamical model using the \texttt{Kinematic
  Molecular Simulation} (\texttt{KinMS}) package of
\citet{davis2013}. Inputs to the model include the stellar mass
distribution, stellar mass-to-light ratio, SMBH mass, as well as the
disc orientation (position angle and inclination) and position
(spatially and spectrally). The stellar mass distribution is
parametrised by a multi-Gaussian expansion (MGE;
\citealt{emsellem1994}) fit to a $V$-band image from {\it HST}
\citep{davis2018}. The free parameters are derived by fitting to the
observed gas kinematics, assuming the object is axisymmetric (in the
central parts where CO is located) and the gas in circular rotation
(see \citealt{davis2018} for details of the fitting procedures and the
best-fitting parameters). The dark matter and gas masses are not
included in our model, as they are small compared to those of the SMBH
and stars. We note that a variable mass-to-light ratio has been
adopted, as required by the data, with a piecewise linear form as a
function of radius. An inclination angle of $68^{\circ}$ and a
kinematic position angle of $93^{\circ}$ (as measured in that work)
are adopted to calculate the line-of-sight projection of the gas
circular velocities.

To further quantify the effects of the galaxy rotation on our observed
velocity gradients, we compare the measured angular velocities and
position angles of the rotation axes of the clouds in NGC4429 to those
expected from a pure galaxy rotation model. We measure the projected
angular velocities and rotation axes of the model over the same areas
as for the observed clouds, using the methods described in
Section~\ref{sec:velocity_gradient_cloud}. We assume that the motion
of the gas within each cloud (i.e.\ each fluid element of each cloud)
follows perfectly circular orbits defined by our kinetic model above.
We find a strong correlation between the modelled and observed
position angles (with a median angle difference of
$\approx 19^{\circ}$), supporting the idea that the observed
cloud-scale velocity gradients are aligned with the large-scale
velocity field, as suggested by Fig.~\ref{fig:gmc_vfield}.

\begin{figure*}
  \includegraphics[width=0.8\textwidth]{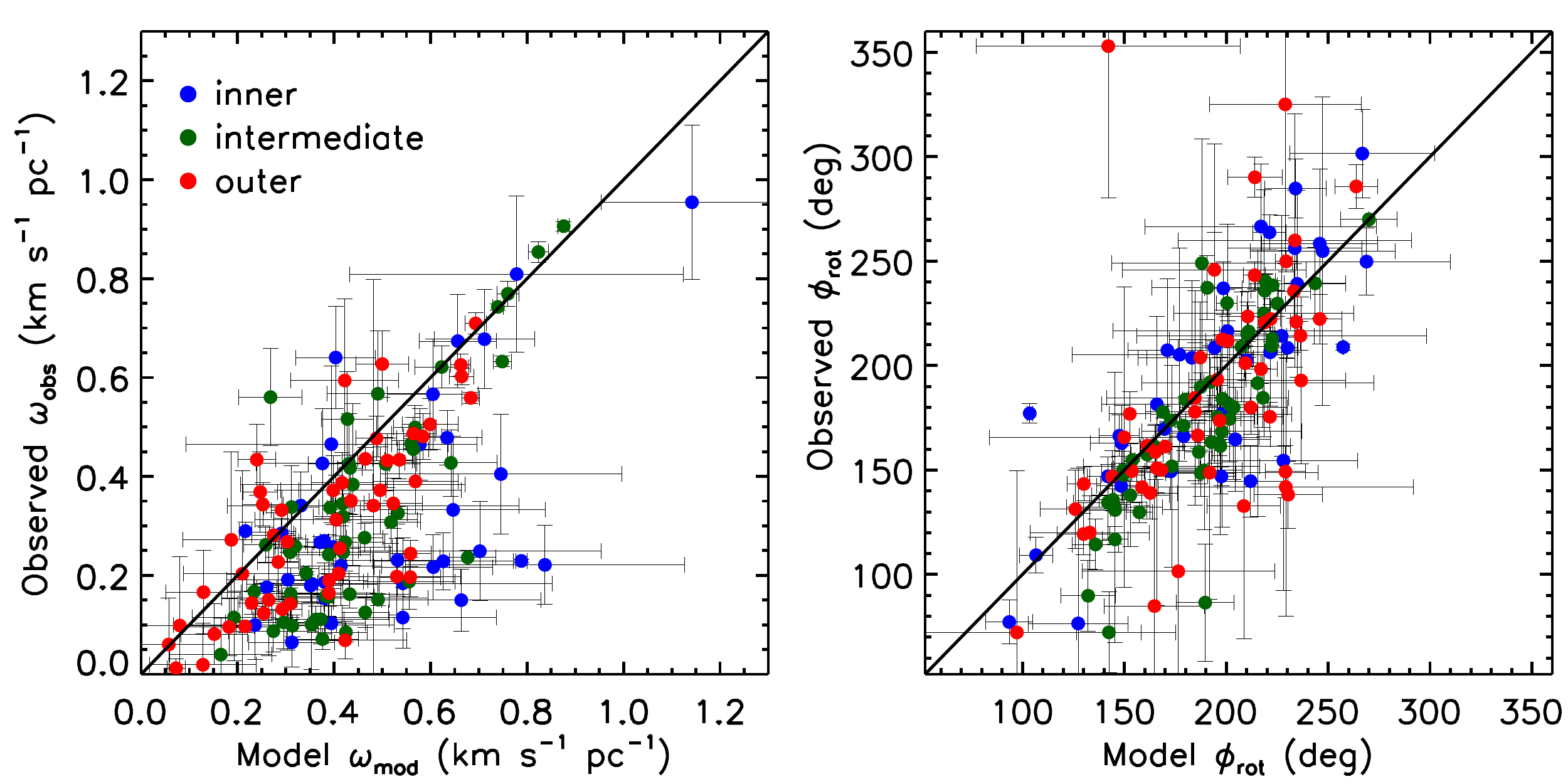}
  \caption{Correlations between the modelled and observed projected
    angular velocities $\omega_{\rm obs}$ (left panel) and position
    angles of the rotation axes $\phi_{\rm rot}$ (right panel) for the
    $141$ spatially-resolved clouds of NGC4429. The data points are
    colour-coded by region and the black solid lines show the $1:1$
    relations.}
  \label{fig:angmom_comparison}
\end{figure*}

A general correlation between the modelled and observed angular
velocities is also found. Our model overestimates the observed angular
velocities $\omega_{\rm obs}$ by a median factor of $2$, much smaller
than the $\omega_{\rm mod}/\omega_{\rm obs}$ ratios found for clouds
in WISDOM late-type galaxies
($\omega_{\rm mod}/\omega_{\rm obs}\gtrsim10$; Shu et al., in prep;
Choi et al., in prep). This discrepancy between the amplitudes of the
observed and modelled angular velocities is unlikely to be due to the
clouds' own rotations, as the observed position angles
$\phi_{\rm rot}$ of the clouds would then be expected to deviate from
the modelled ones randomly. A possible explanation is that the
self-gravity of the clouds is also important, so that the clouds do
not follow pure galaxy rotation (see
Section~\ref{sec:equilibrium_between_self-gravity_external-gravity}
for more discussion of this). The discrepancy could also partly be due
to the limitation of {\cprops} to isolate individual clouds in
highly-crowded environments. To reduce the ambiguities due to cloud
blending, we fit both the data and model again without including the
outermost boundary pixels of each cloud. In this case, a strong
correlation between the modelled and observed position angles is again
present (see the right panel of Fig.~\ref{fig:angmom_comparison}),
with a median angle difference of $\approx16^{\circ}$, but the model
overestimates the observed angular velocities by a reduced median
factor of $1.5$ only (left panel of Fig.~\ref{fig:angmom_comparison}).
In the inner region, where the clouds are more blended in both space
and velocity, the discrepancies between the modelled and observed
angular velocities is worse (with a median factor of $2$), and the
angle difference is larger (with a median value of
$\approx20^{\circ}$). In the outer region, where clouds are less
blended, the model shows a much better agreement with the observations
(with a median angular velocity discrepancy factor of only $1.2$ and a
median angle difference of only $\approx14^{\circ}$)

In summary, a comparison of the observed and modelled projected
angular velocities and rotation axes of individual clouds suggests
that the observed velocity gradients of the clouds in NGC4429 are
primarily caused by the local circular orbital motions, themselves due
to the galaxy potential. We note that the good match between our
observations and model suggests that the motion of the gas within each
cloud of NGC4429 mainly follows gravitational orbital (and thus shear)
motions rather than epicyclic motions (see
Section~\ref{sec:linear_shear_flow_non-zero_e_ext} for more discussion
of this issue).


\section{Dynamical State of Clouds}
\label{sec:dynamical_state_clouds}


\subsection{Cloud scaling relations using the observed velocity
  dispersion}
\label{sec:larson_sigma_obs}

The scaling relations between the physical properties of molecular
clouds have become a standard tool for assessing the clouds' physical
states and dynamical conditions \citep[e.g.][]{blitz2007, hughes2013}.
The most fundamental relation is the size -- linewidth relation,
a.k.a.\ Larson's first relation \citep[e.g.][]{larson1981,
  solomon1987}, that has become the yardstick for GMC studies in the
MW and external galaxies \citep[e.g.][]{bolatto2008}. The size --
linewidth relationship is usually interpreted as a signature of the
turbulent motions within clouds \citep[e.g.][]{falgarone1991,
  elmegreen1996, lequeux2005}, and it provides a unique probe of the
dynamical state of the turbulent molecular gas in extragalactic
star-forming systems.

Another important scaling relation providing crucial insights is the
correlation between the clouds' dynamical (i.e.\ Virial) masses
$M_{\rm vir}$ and their true masses $M$ (here taken to be the gaseous
masses $M_{\rm gas}$). The comparison of the Virial and gaseous masses
provides an important clue to the dynamical state of the clouds
according to the Virial theorem. Indeed, the Virial
parameter
\begin{equation}
  \label{eq:alpha_vir}
  \begin{split}
    \alpha_{\rm vir} & \equiv\frac{M_{\rm vir}}{M} \\
    & = \frac{\sigma^2R_{\rm c}}{b_{\rm s}GM} \\
    & = \frac{2\,\frac{1}{2}M\sigma^2}{b_{\rm s}GM^2/R_{\rm c}}
  \end{split}
\end{equation}
(see Eq.~\ref{eq:Virial_mass}) is equal to the ratio of two times the
turbulent kinetic energy to the (absolute value of the)
self-gravitational energy of a cloud, quantifying the degree of
gravitational boundedness of the cloud. If the Virial parameter of a
cloud $\alpha_{\rm vir}\approx1$, the cloud is gravitationally bound
and in Virial equilibrium. If its Virial mass is much larger than its
gaseous mass ($\alpha_{\rm vir}\gg1$), the cloud has to be confined by
external pressure (it would otherwise disperse) and it is unlikely to
be bound (i.e.\ it is a transient feature of the ISM). If
$\alpha_{\rm vir}\lesssim1$, the molecular cloud is likely unstable to
gravitational collapse. We note that a critical parameter
$\alpha_{\rm crit}\approx2$ is often regarded as the threshold between
gravitationally-bound and unbound objects \citep{kauffmann2013,
  kauffmann2017}.

\begin{figure*}
  \includegraphics[width=0.95\textwidth]{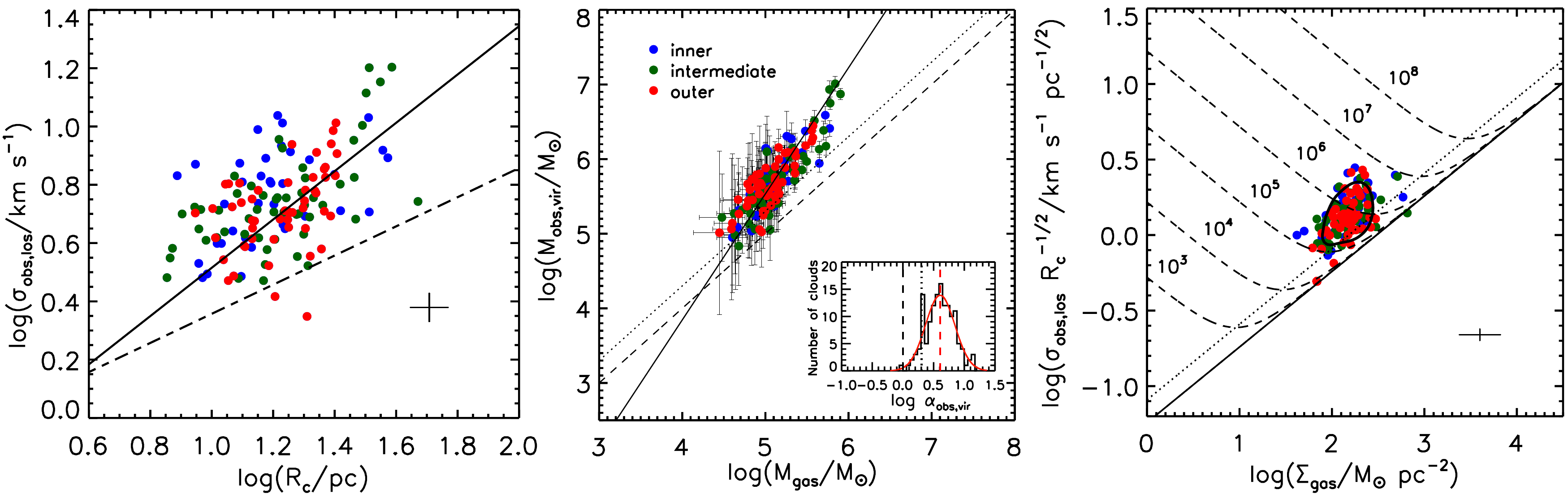}
  \caption{Left: Size -- linewidth relation of the $141$
    spatially-resolved clouds of NGC4429, using the observed velocity
    dispersion $\sigma_{\rm obs,los}$. The black solid line shows the
    best-fitting relation, while the black dashed line shows Larson's
    relation for the Milky Way disc \citep{solomon1987}.  Middle:
    Correlation between Virial mass and gaseous mass for the same
    spatially-resolved clouds. The black solid line shows the
    best-fitting relation, while the black dashed and dotted diagonal
    lines show the $1:1$ and $2:1$ relations, respectively. The
    distribution of $\log(\alpha_{\rm obs,vir}$) (black histogram)
    with a log-normal fit overlaid (red solid line) is shown in an
    inset. The red dashed line in the inset indicates the mean of the
    log-normal fit, while the black dashed and dotted lines indicate
    $\alpha_{\rm vir}=1$ and $\alpha_{\rm vir}=2$,
    respectively. Right: Correlation between
    $\sigma_{\rm obs,los}R^{-1/2}$ and gaseous mass surface density
    ($\Sigma_{\rm gas}$) for the same spatially-resolved clouds. The
    black solid contour encloses $68\%$ of the data points. The black
    solid and dotted diagonal lines show the solution for simple
    (i.e.\ $\alpha_{\rm vir}=1$) and marginal (i.e.\
    $\alpha_{\rm vir}=2$) Virial equilibria, respectively. The
    V-shaped black dashed curves show solutions for pressure-bound
    clouds at different pressures ($P_{\rm ext}/k_{\rm B}=10^3$,
    $10^4$, $\cdots$, $10^{8}$~K~cm$^{-3}$). Data points are
    colour-coded by region in all three panels. Typical uncertainties
    are shown as a black cross in the bottom-right corner of the left
    and right panels.}
  \label{fig:dynamics_obs}
\end{figure*}

A third important scaling relation is the correlation between the
clouds' mass surface densities $\Sigma$ (again taken here to be the
gaseous mass surface densities $\Sigma_{\rm gas}$) and the quantities
$\sigma R_{\rm c}^{-1/2}$ (where as before $\sigma$ and $R_{\rm c}$
are a measure of the observed/1D velocity dispersion and size of each
cloud, respectively). The $\sigma R_{\rm c}^{-1/2}$ --
$\Sigma_{\rm gas}$ plot provides a necessary modification to Larson's
scaling relations. It implies an additional constraint to the velocity
dispersion, whereby the velocity dispersion of a cloud depends on both
its spatial extent and its gaseous mass surface density
\citep{field2011}. If clouds are virialised (and do not necessarily
obey Larson's first relation), observations should cluster around the
line $\sigma R_{\rm c}^{-1/2}=\sqrt{\pi Gb_{\rm s}\Sigma_{\rm gas}}$
($b_{\rm s}=1/5$ for a homogeneous spherical cloud; see the black
solid diagonal line in the right panel of e.g.\
Fig.~\ref{fig:dynamics_obs}). If clouds are not virialised but are
marginally gravitationally bound (i.e.\
$\alpha_{\rm vir}\approx\alpha_{\rm vir,crit}=2$), the data points
should cluster around the line
$\sigma R_{\rm c}^{-1/2}=\sqrt{2\pi Gb_{\rm s}\Sigma_{\rm gas}}$ (see
the black dotted diagonal line in the right panel of e.g.\
Fig.~\ref{fig:dynamics_obs}). If clouds are not gravitationally bound,
external pressure ($P_{\rm ext}$) must play an important role to
confine the clouds, and the clouds should be distributed along the
black V-shaped dashed curves in the right panel of
Fig.~\ref{fig:dynamics_obs}:
$\sigma R_{\rm c}^{-1/2}=\sqrt{\pi Gb_{\rm s}\Sigma_{\rm
    gas}+\frac{4P_{\rm ext}}{3\Sigma_{\rm gas}}}$
\citep{field2011}. We note that for the largest $\Sigma_{\rm gas}$ of
each V-shaped curve, the clouds are dominated by self-gravity and the
equilibrium curve is asymptotic to the solution of the simple Virial
equilibrium (SVE, i.e.\ the black solid diagonal line;
\citealt{field2011}).

For consistency with GMC studies in the MW and external galaxies in
the literature, we first adopt the observed velocity dispersion
$\sigma_{\rm obs,los}$ (see
Section~\ref{sec:definition_gmc_properties}) to explore the above
three scaling relations. As seen in the left panel of
Fig.~\ref{fig:dynamics_obs} (data points and black solid line), there
is a strong correlation between size and linewidth (with a Spearman
rank correlation coefficient of $0.5$) for the $141$ clouds of NGC4429
that are spatially resolved, the only clouds where a reliable
measurement of the size $R_{\rm c}$ is possible (see
Table~\ref{tab:gmcs}). However, the relation departs from the
traditional one derived for clouds in the MW disc (black dashed line
in the left panel of Fig.~\ref{fig:dynamics_obs};
\citealt{solomon1987, dame2001, rice2016}). The observed tendency is
for clouds to exhibit a higher velocity dispersion at a given
size. Our results also reveal a steep size -- linewidth relation,
\begin{equation}
  \log\left(\frac{\sigma_{\rm obs,los}}{\rm km~s^{-1}}\right)=(-0.30\pm0.17)+(0.82\pm0.13)\,\log\left(\frac{R_{\rm c}}{\rm pc}\right)~,
\end{equation}
steeper than that of clouds in the MW disc ($0.5\pm0.05$;
\citealt{solomon1987}). The slope is also marginally steeper than that
derived for CMZ clouds ($0.66\pm0.18$; \citealt{kauffmann2017}), but
the zero-point is much smaller ($5.5\pm1.0$ for CMZ clouds;
\citealt{kauffmann2017}), and the velocity dispersions of CMZ clouds
are indeed higher than those of the NGC4429 clouds at any given size.

The Virial masses of the spatially-resolved clouds of NGC4429
calculated from their observed velocity dispersions,
\begin{equation}
  \label{eq:M_obs,vir}
  M_{\rm obs,vir}\equiv\frac{\sigma_{\rm obs,los}^2R_{\rm c}}{b_{\rm s}G}
\end{equation}
(see Eq.~\ref{eq:Virial_mass}), are compared to their gaseous
masses $M_{\rm gas}$ in the middle panel of
Fig.~\ref{fig:dynamics_obs}, where as always we have assumed
$b_{\rm s}=\frac{1}{5}$ (spherical homogeneous clouds). We find Virial
masses significantly larger than the gaseous masses. A linear fit
yields (black solid line in the middle panel of
Fig.~\ref{fig:dynamics_obs})
\begin{equation}
  \log\left(\frac{M_{\rm obs,vir}}{\rm M_\odot}\right)=(-2.91\pm0.43)+(1.69\pm0.08)\,\log\left(\frac{M_{\rm gas}}{\rm M_\odot}\right)~.
\end{equation}
A log-normal fit to the distribution of the resulting Virial
parameters,
\begin{equation}
  \label{eq:alpha_obs,vir}
  \alpha_{\rm obs,vir}\equiv\frac{M_{\rm obs,vir}}{M}=\frac{M_{\rm obs,vir}}{M_{\rm gas}}~,
\end{equation}
shown as an inset in the middle panel of Fig.~\ref{fig:dynamics_obs},
yields a mean $\langle\alpha_{\rm obs,vir}\rangle=4.04\pm0.22$ and a
standard deviation of $0.24$~dex. In particular, all resolved clouds
have $\alpha_{\rm obs,vir}>1$.

The derived $\sigma_{\rm obs,los}R_{\rm c}^{-1/2}-\Sigma_{\rm gas}$
relation is presented in the right panel of
Fig.~\ref{fig:dynamics_obs} for the spatially-resolved clouds of
NGC4429. The gaseous mass surface densities $\Sigma_{\rm gas}$ of the
GMCs vary by one order of magnitude, and the size -- linewidth
coefficient ($\sigma_{\rm obs,los}R_{\rm c}^{-1/2}$) increases with
increasing $\Sigma_{\rm gas}$. The data points do not lie along the
solid diagonal line of the SVE, but are instead offset from it and
distributed across the ${\rm V}$-shaped curves. If pressure is
important to the dynamical state of the clouds, the clouds in NGC4429
seem to experience a wide range of considerable external pressures
($P_{\rm ext}/k_{\rm B}\approx10^5$ -- $10^{7}$~K~cm$^{-3}$, where
$k_{\rm B}$ is Boltzmann's constant).  Overall,
Fig.~\ref{fig:dynamics_obs} thus seems to suggest that the kinetic
energy of the clouds in NGC4429 is more important than their
gravitational energy, hence the clouds are either not bound or tend
toward pressure equilibria.

However, a major concern about the use of the above relations to
assess the dynamical states of clouds in NGC4429 is the applicability
of the observed velocity dispersion $\sigma_{\rm obs,los}$. The
difference of the derived size -- linewidth relation with respect to
the \citet{solomon1987} trend seems to imply that the measured
linewidths of the clouds are not set purely by their internal
virialised motions and/or turbulence \citep{meidt2013,
  kauffmann2017}. Recent works suggest that, in the centre of galaxies
where strong shear and tidal forces are present, a considerable part
of the cloud-scale gas motions is due to these external galactic
forces \citep[e.g.][]{meidt2018, utreras2020}. We have already
demonstrated that the observed strong velocity gradients of the clouds
in NGC4429, that reflect the velocity gradients in the plane of the
galaxy, are mainly a consequence of local orbital motions defined by
the background galactic gravitational potential (i.e.\ the galaxy
circular velocity curve; see
Section~\ref{sec:origin_velocity_gradients}). In this case, the steep
slope of the size -- linewidth relation (see the left panel of
Fig.~\ref{fig:dynamics_obs}) can be explained as resulting from the
decay of fast orbit-induced large-scale motions to transonic
conditions on small spatial scales \citep{kauffmann2017}.

The question then is whether gas motions associated with the
background galactic potential should also be involved in assessing the
dynamical states and stability of the clouds. Intuitively, gas motions
due to external galactic forces should be considered when calculating
a cloud's kinetic energy that is meant to balance its
self-gravitational energy \citep{chen2016, meidt2018}. Conversely, in
the presence of strong galactic forces, self-gravity is no longer the
only force binding a cloud. Therefore, to verify whether clouds are
virialised in a galactic environment where tidal/shear forces are
strong, one needs to modify the conventional Virial theorem to include
(1) external forces arising from the background galactic potential and
(2) the gas motions induced by these forces. We do exactly that in the
next sub-sections.


\subsection{Basic framework}
\label{sec:basic_framework}

We recall here a key conceptual point emphasised in
Section~\ref{sec:introduction}. We will not assume here that the
clouds of NGC4429 are in dynamical equilibrium, and then deduce the
clouds' gravitational motions due to the external (i.e.\ galactic)
potential. Rather, we will measure and quantify the clouds' gravitational
motions due to the external potential, and then deduce whether the
clouds are indeed in dynamical equilibrium. This is the only way to
reliably assess whether GMCs are in dynamical equilibrium (and thus
long-lived) or out of equilibrium (and thus transient), arguably the
most important question in the field.

As described in detail in Appendix~\ref{app:modified_Virial_theorem},
we envision each cloud as a continuous structure with well-defined
borders in position- and velocity-space, located in a rotating gas
disc with a circular velocity determined by the shape of the
background galactic gravitational potential. Each cloud's centre of
mass (CoM) is assumed to be in the mid-plane of the disc. We assume
that each fluid element of a cloud experiences two kinds of motions:
(1) random turbulent motions arising from self-gravity (cloud
gravitational potential $\Phi_{\rm sg}$), that have a velocity
dispersion $\sigma_{\rm sg}$, and (2) bulk gravitational motions
associated with the external (i.e.\ galactic) potential
($\Phi_{\rm gal}$), that have a RMS velocity $\sigma_{\rm gal}$
($\sigma_{\rm gal}\equiv\frac{\int{v_{\rm gal}^2\,dm}}{M}$, where
$v_{\rm gal}$ is the velocity of each fluid element due to
gravitational motions relative to the CoM, the integral is over all
fluid elements $dm$, and $\int{dm}=M$). Thermal motions are ignored,
as they are often small compared to turbulent motions in a cold gas
cloud \citep[e.g.][]{fleck1980}.  We assume the motions due to
self-gravity ($\sigma_{\rm sg}$) and the background galactic potential
($\sigma_{\rm gal}$) to be uncorrelated, and the cloud's own
gravitational potential $\Phi_{\rm sg}$ to be (statistically)
independent of the local external gravitational potential defined by
the galaxy $\Phi_{\rm gal}$.  The turbulent motions due to
self-gravity are expected to be quasi-isotropic in three dimensions
\citep{field2008, ballesterosparedes2011}, while the gas motions
induced by the external gravitational potential are often
non-isotropic \citep{meidt2018}. Gravitational motions in the plane
are assumed to be separable from those in the vertical direction. We
consider only the effects of gravitational forces and ignore external
pressure and magnetic fields.

With those considerations, the resulting modified Virial theorem (MVT)
can be split into two independent parts:
\begin{gather}
  \label{eq:modified_Virial_theorem_1}
  \begin{split}
    \frac{\ddot{I}}{2}\approx & \underbrace{\left[3M\sigma_{\rm sg,los}^2-3b_{\rm s}GM^2/R_{\rm c}\right]}_{\rm self~gravity} \\
    + & \left[\underbrace{M\left(\sigma_{\rm gal,z}^2-b_{\rm e}\nu_0^2Z_{\rm c}^2\right)}_{\rm external,~vertical~direction}+\underbrace{M\left(\sigma_{\rm
            gal,r}^2+\sigma_{\rm gal,t}^2+b_{\rm e}(T_0-2\Omega_0^2)R_{\rm c}^2\right)}_{\rm external,~plane}\right]~,
  \end{split}
\end{gather}
where $I$, $M$, $R_{\rm c}$ and $Z_{\rm c}$ are respectively the
cloud's moment of inertia, mass, radius and scale height,
$\nu_0^2\equiv4\pi G\rho_{\ast,0}$ (formally the total mass volume
density evaluated at the cloud's CoM, but we use here $\rho_{\ast,0}$,
the stellar mass volume density $\rho_\ast$ evaluated at the cloud's
CoM using our MGE model, as it is accurately constrained; see
Appendix~\ref{app:stellar_density_calculation}), $b_{\rm s}$ is the
aforementioned geometrical factor that quantifies the effects of
inhomogeneities and/or non-sphericity associated with self-gravity,
$b_{\rm e}$ is a geometrical factor that quantifies the effects of
inhomogeneities (only) associated with external gravity
($b_{\rm e}=\frac{(1-\psi/3)}{(5-\psi)}$ for a spherical cloud with a
radial mass volume density profile $\rho(r)\propto r^{-\psi}$, thus
$b_{\rm e}=b_{\rm s}=\frac{1}{5}$ for a spherical homogeneous cloud as
before; see Appendix~\ref{app:modified_Virial_theorem} for more
details on $b_{\rm e}$), $\sigma_{\rm sg,los}$ is the cloud's 1D
turbulent velocity dispersion due to self-gravity,
$\sigma_{\rm gal,r}$, $\sigma_{\rm gal,t}$ and $\sigma_{\rm gal,z}$
are the RMS velocity of gas motions due to external gravity in
respectively the radial (i.e.\ the direction pointing from the galaxy
centre to the cloud's CoM), azimuthal (i.e.\ the direction along the
orbital rotation) and vertical (i.e.\ the direction perpendicular to
the cloud's orbital plane) direction (as measured in an inertial
frame, i.e.\ by a distant observer; see
Appendix~\ref{app:modified_Virial_theorem} for a more detailed
discussion of $\sigma_{\rm gal,r}$ and $\sigma_{\rm gal,t}$),
$\Omega_0$ is the circular orbital angular velocity $\Omega$ at the
cloud's CoM, and $T_0\equiv-R\frac{d\Omega^2(R)}{dR}\vert_{R=R_0}$ is
the tidal acceleration per unit length in the radial direction $T$
\citep[e.g.][]{stark1978} evaluated at the cloud's CoM ($R$ is the
galactocentric distance in the plane of the disc and $R_0$ that of the
cloud's CoM). We note that here and throughout, $\Omega(R)$ is the
theoretical quantity
$\Omega(R)\equiv\sqrt{\frac{1}{R}\frac{d\Phi_{\rm gal}(R)}{dR}}$
defined by the galaxy potential $\Phi_{\rm gal}$, i.e.\ it is the
angular velocity of a fluid element moving in perfect circular motion
($\Omega(R)=V_{\rm circ}(R)/R$, where $V_{\rm circ}(R)$ is the
circular velocity curve) rather than the observed angular velocity of
the fluid element ($V_{\rm rot}(R)/R$, where $V_{\rm rot}$ is the
observed rotation curve). The first term in square brackets on the
right-hand side (RHS) of Eq.~\ref{eq:modified_Virial_theorem_1}
comprises the energy terms regulated by self-gravity, while the second
term in square brackets contains the contributions of external gravity
to the cloud's energy budget ($E_{\rm ext}$) in respectively the
vertical direction ($E_{\rm ext,z}$) and the plane
($E_{\rm ext,plane}$). The detailed derivation of
Eq.~\ref{eq:modified_Virial_theorem_1} and its more general form for a
homogeneous ellipsoidal cloud
(Eq.~\ref{eq:modified_Virial_theorem_appendix}) is provided in
Appendix~\ref{app:modified_Virial_theorem}.

\begin{figure}
  \includegraphics[width=0.95\columnwidth]{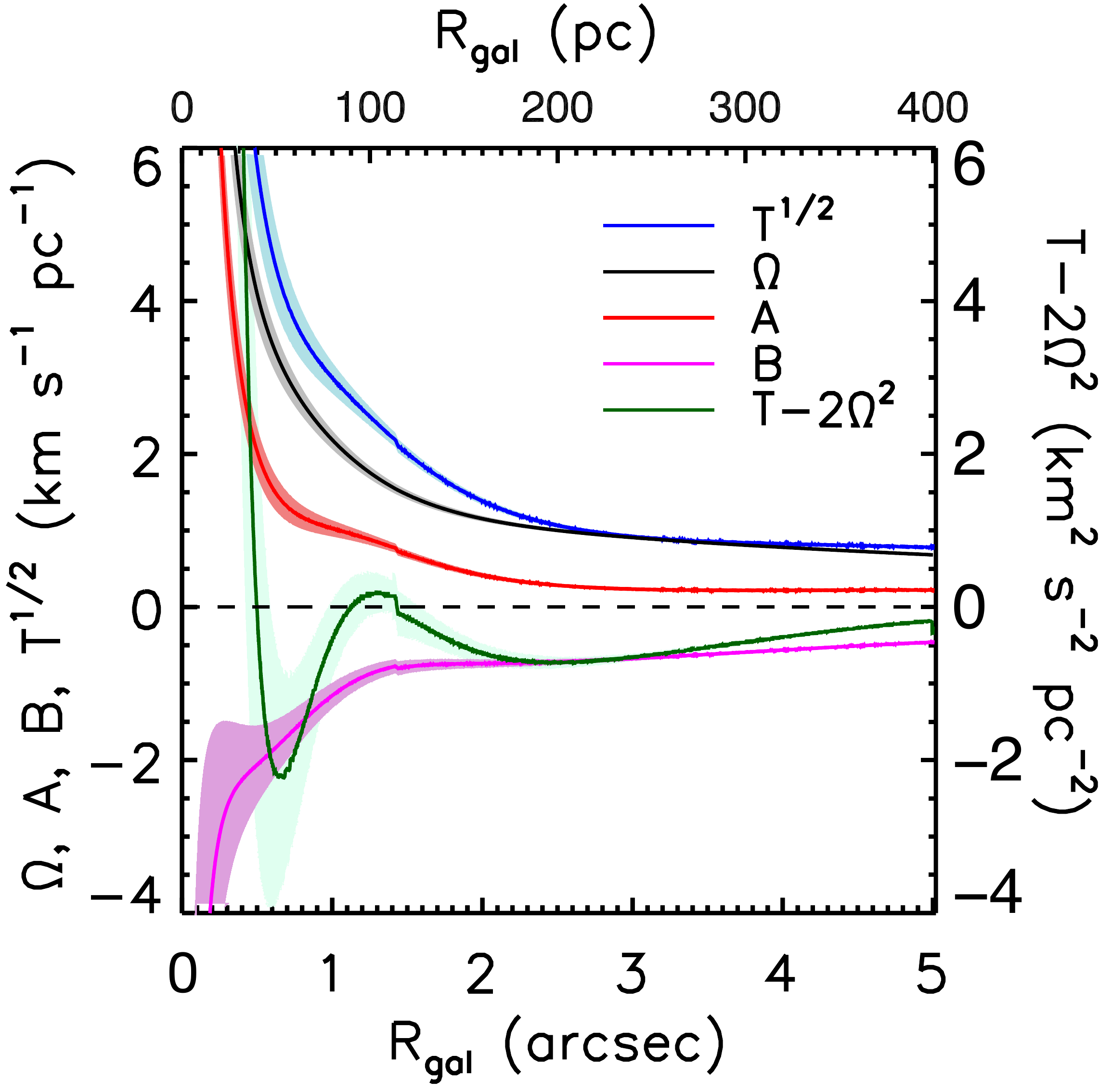}
  \caption{Galactocentric distance ($R_{\rm gal}$) dependence of the
    orbital angular velocity $\Omega$, Oort's constants $A$ and $B$,
    the tidal acceleration per unit length in the radial direction
    $T$, and the function $T-2\Omega^2$ in NGC4429, as calculated from
    our gas dynamical model. The black dashed horizontal line
    indicates an ordinate of $0$. The coloured envelopes around each
    curve indicate the $\pm1\,\sigma$ uncertainties. We note that the
    slight discontinuity in the radial profiles of $A$, $B$, $T$ and
    $T-2\Omega^2$ at $R_{\rm gal}\approx1\farcs4$ is caused by our
    adopted piecewise linear mass-to-light ratio radial profile
    $M/L(R)$ \citep[see][]{davis2018}, so that while $M/L(R)$ is
    continuous $\frac{d\,M/L(R)}{dR}$ is not.}
  \label{fig:a_vs_r}
\end{figure}

For reference, we show in Fig.~\ref{fig:a_vs_r} the dependence of
$\Omega$, Oort's constants $A$ and $B$, $T$ and $T-2\Omega^2$ on
galactocentric distance $R$ in NGC4429. The functions $\Omega$, $A$
and $T$ are always positive, $B$ is always negative, while
$T-2\Omega^2$ is generally negative except in the very centre.  We
note that $T=4A\Omega=4\Omega\,(B+\Omega)$. The rotational shear
(i.e.\ Oort's constant $A$) in NGC4429 is much larger
($\ge0.2~{\rm km~s^{-1}~pc^{-1}}$ at galactocentric distances
$R\ls450$~pc, where the clouds are located) than that in the bulk of
the Galactic disc ($\approx0.02~{\rm km~s^{-1}~pc^{-1}}$ at
$R\ge3$~kpc; \citealt{dib2012}) and the LMC
($\approx0.018~{\rm km~s^{-1}~pc^{-1}}$ at $R\ge1$~kpc;
\citealt{thilliez2014}).

\subsection{Role of self-gravity}
\label{sec:role_self_gravity}

The first term in square brackets on the RHS of
Eq.~\ref{eq:modified_Virial_theorem_1} describes an internal
equilibrium regulated by self-gravity. For a cloud that attains Virial
balance between its internal turbulent kinetic energy
($\frac{3}{2}M\sigma^2_{\rm sg,los}$) and its self-gravitational
energy ($U_{\rm sg}\equiv-3 b_{\rm s}GM^2/R_{\rm c}$), such as an
isolated self-gravitating cloud, these two terms should cancel out. To
investigate the role of self-gravity, one thus needs to measure the
cloud's turbulent velocity dispersion due to self-gravity only
($\sigma_{\rm sg,los}$). However, the observed velocity dispersion
$\sigma_{\rm obs,los}$ is not necessarily equal to
$\sigma_{\rm sg,los}$, as there are potentially significant
contributions from bulk (galaxy-driven) gravitational motions.
Indeed, the observed velocity dispersion $\sigma_{\rm obs,los}$ of a
cloud can be expressed as
\begin{equation}
  \label{eq:sigma_obslos}
  \sigma_{\rm obs,los}^2\approx\sigma_{\rm sg,los}^2\,+\,\left(\sigma_{\rm gal,r}^2\sin^2\theta+\sigma_{\rm gal,t}^2\cos^2\theta\right)\sin^2i\,+\,\sigma_{\rm gal,z}^2\cos^2i~,
\end{equation}
where $i$ is the inclination of the galactic disc with respect to the
line of sight, and $\theta$ is the (deprojected) azimuthal angle of
the cloud's CoM with respect to the kinematic major axis of the disc
(see Eq.~32 of \citealt{meidt2018}).

\begin{figure*}
  \includegraphics[width=0.33\textwidth]{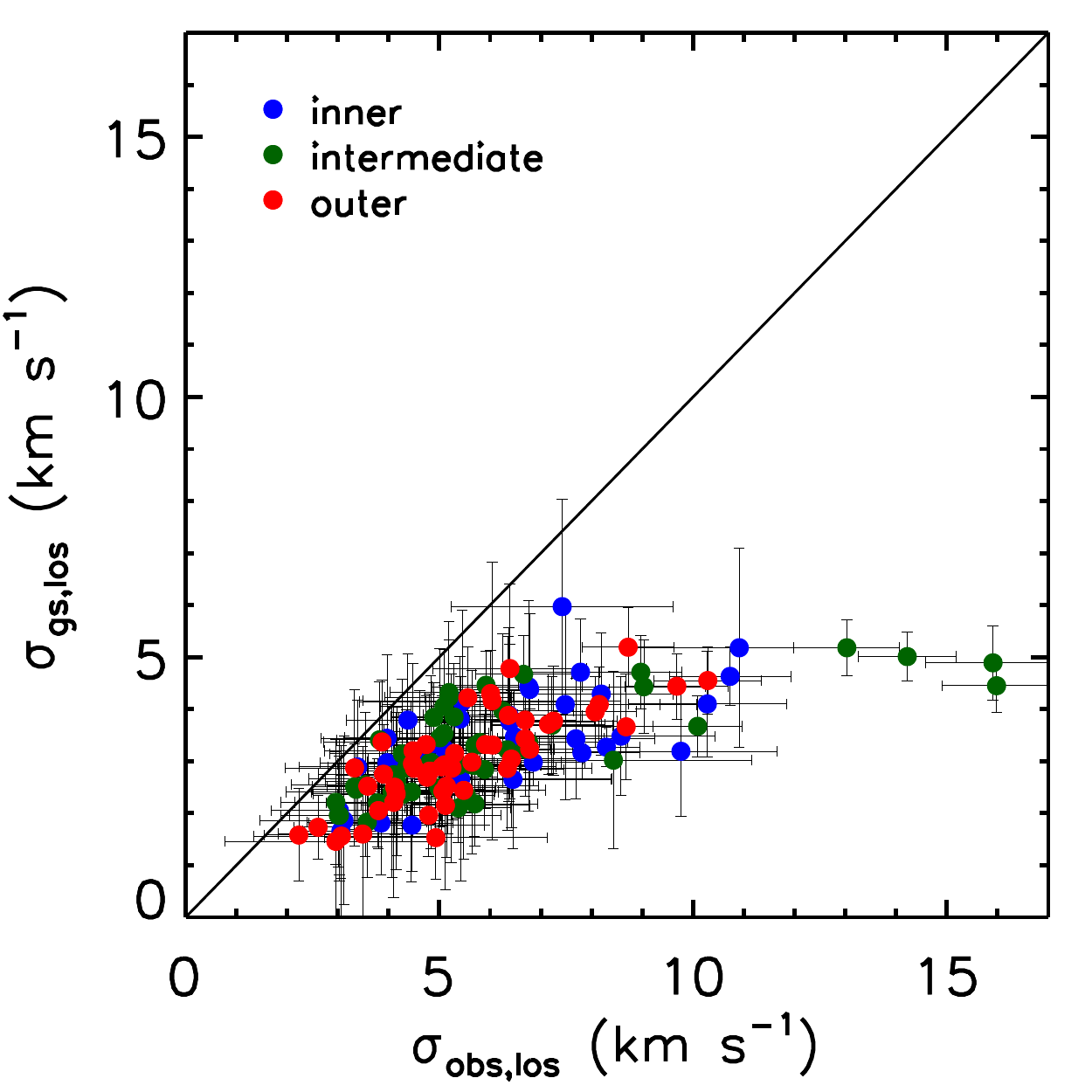}
  \includegraphics[width=0.33\textwidth]{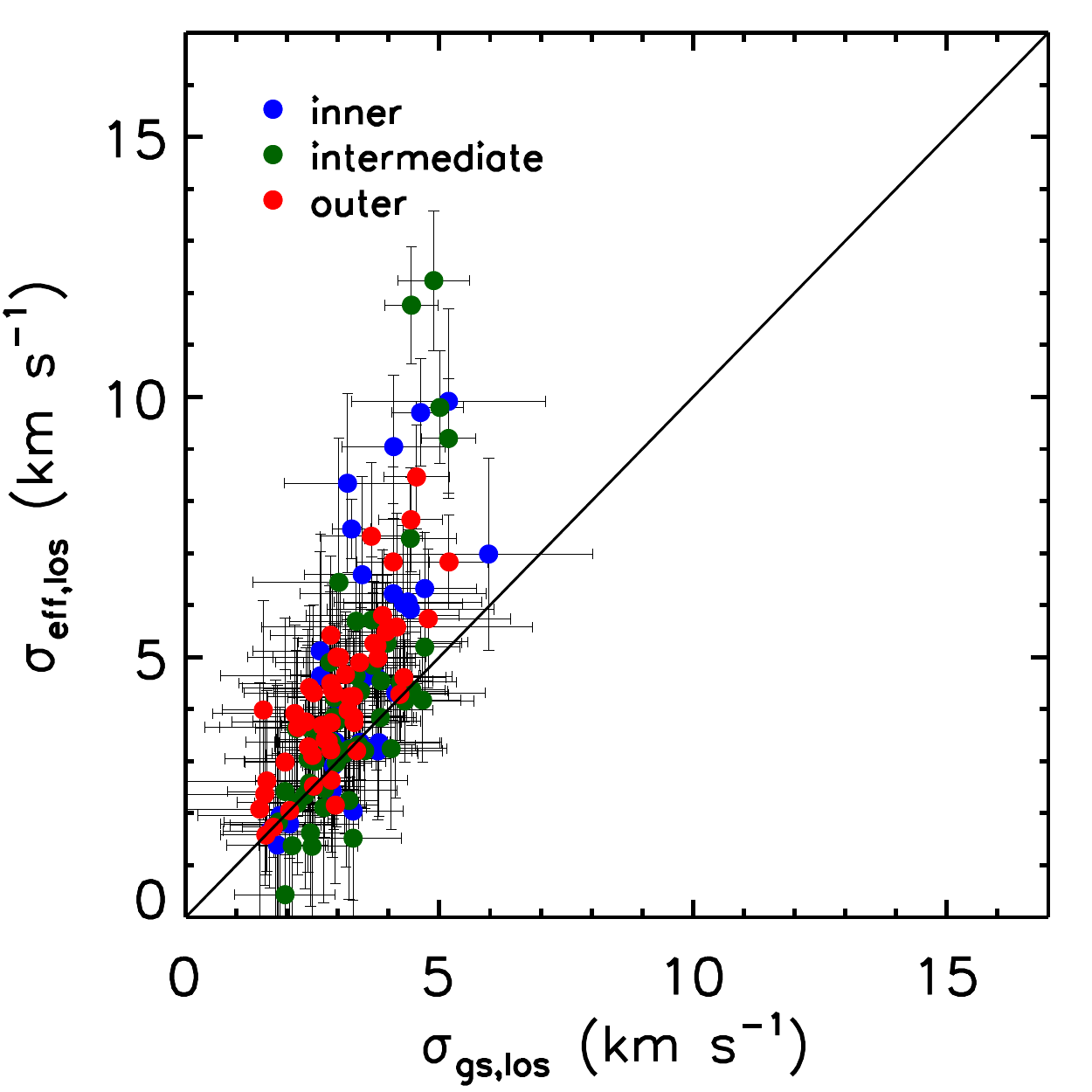}
  \includegraphics[width=0.33\textwidth]{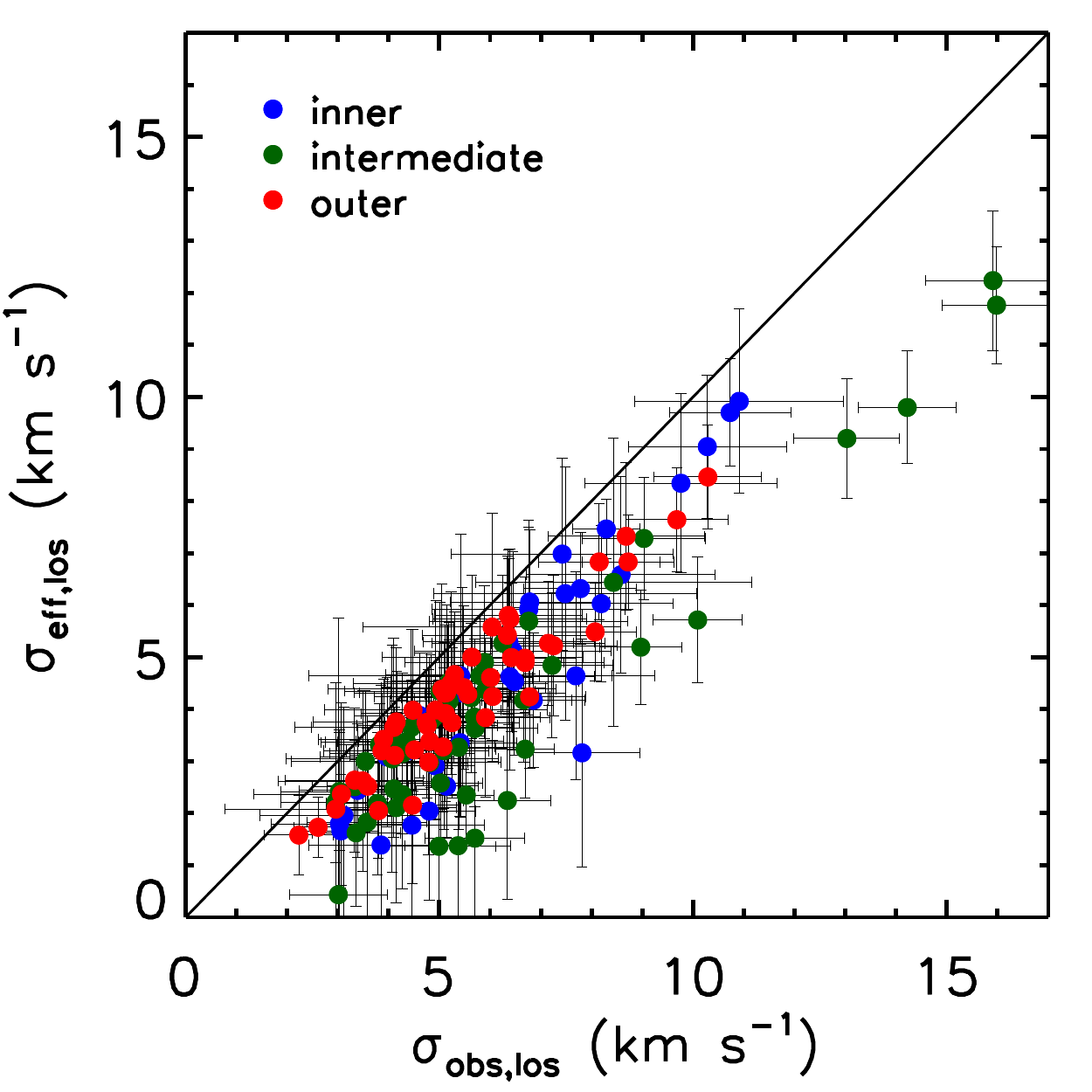}
  \caption{Comparisons of our observed ($\sigma_{\rm obs,los}$),
    gradient-subtracted ($\sigma_{\rm gs,los}$) and effective
    ($\sigma_{\rm eff,los}$) cloud velocity dispersion measures for
    the $141$ spatially-resolved clouds of NGC4429. Data points are
    colour-coded by region in all three panels.}
  \label{fig:sigma_comp}
\end{figure*}

We therefore need to reduce the contamination of our measured velocity
dispersions by bulk gravitational motions. This is why we introduced a
new measure of the velocity dispersion, $\sigma_{\rm gs,los}$, in
Section~\ref{sec:definition_gmc_properties}, where we first shifted
each line-of-sight velocity spectrum to match its centroid velocity
($\bar{v}(x,y)$) to that of the cloud's CoM ($\bar{v}(0,0)$), and then
measured the velocity dispersion (i.e.\ the second moment along the
velocity axis) of the shifted emission distribution and extrapolated
it to $T_{\rm edge}=0$~K. The derived gradient-subtracted velocity
dispersion $\sigma_{\rm gs,los}$ was then deconvolved by the channel
width ($\Delta V_{\rm chan}/\sqrt{2\pi}$), yielding our final adopted
measure. Table~\ref{tab:gmcs} lists the derived $\sigma_{\rm gs,los}$
of all spatially-resolved clouds and the left panel of
Fig.~\ref{fig:sigma_comp} shows a comparison of $\sigma_{\rm gs,los}$
and $\sigma_{\rm obs,los}$. As expected,
$\sigma_{\rm gs,los}<\sigma_{\rm obs,los}$, and all particularly large
$\sigma_{\rm obs,los}$ measurements have been corrected to
$\ls5$~{\kms}.

The observed velocity gradient of a cloud is due to bulk motions
within the cloud only. Assuming that the vertical gravitational
motions can be treated as random motions that balance the weight of
the disc (i.e.\ no bulk motion in the vertical direction), analogously
to turbulent motions due to self-gravity, the only bulk motions will
originate from in-plane gravitational motions. Our newly-derived
gradient-subtracted velocity dispersion $\sigma_{\rm gs,los}$ can
therefore be written as
\begin{equation}
  \label{eq:vrms_gs_los}
  \sigma_{\rm gs,los}^2\approx\sigma_{\rm sg,los}^2\,+\,\sigma_{\rm gal,z}^2\cos^2i~,
\end{equation}
minimising contamination from bulk gas motions in the plane. Our
gradient-subtracted velocity dispersion $\sigma_{\rm gs,los}$ thus
removed the second term (in-plane bulk gravitational motions) but kept
the first term (turbulent self-gravitational motions) and last term
(vertical random gravitational motions) on the RHS of
Eq.~\ref{eq:sigma_obslos}.
However, as we will demonstrate below, the
$\sigma^2_{\rm gal,z}\cos^2i$ term is negligible compared to
$\sigma^2_{\rm sg,los}$ in NGC4429 and can thus safely be ignored, so
that $\sigma_{\rm gs,los}\approx\sigma_{\rm sg,los}$ in NGC4429. Using
our newly derived $\sigma_{\rm gs,los}$ measure, we thus revisit the
scaling relations of Fig.~\ref{fig:dynamics_obs} in
Fig.~\ref{fig:dynamics_gs}.


\subsection{Cloud scaling relations using the gradient-subtracted velocity
  dispersion}
\label{sec:larson_sigma_gs}

\begin{figure*}
  \includegraphics[width=0.95\textwidth]{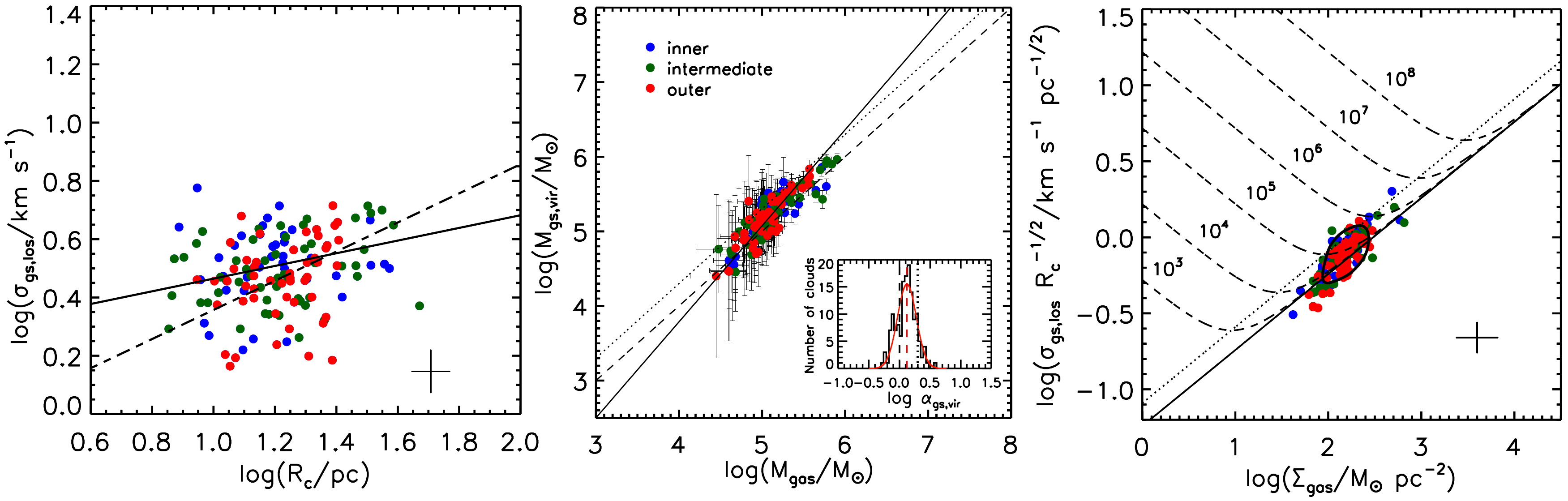}
  \caption{Same as Fig.~\ref{fig:dynamics_obs}, but using our
    gradient-subtracted measure of velocity dispersion
    $\sigma_{\rm gs,los}$.}
  \label{fig:dynamics_gs}
\end{figure*}

The left panel of Fig.~\ref{fig:dynamics_gs} (data points and black
solid line) presents the size -- linewidth relation based on our
$\sigma_{\rm gs,los}$ measure for the $141$ spatially-resolved clouds
of NGC4429. We now find the size -- $\sigma_{\rm gs,los}$ correlation
to be rather weak, with a Spearman rank coefficient of
$0.25$. However, compared with the size -- linewidth relation using
$\sigma_{\rm obs,los}$, it appears to better follow the relation of
the MW disc clouds (black dashed line in the left panel of
Fig.~\ref{fig:dynamics_gs}). Indeed, the data points seem to cluster
around the MW disc scaling law \citep{solomon1987}, although there is
a large scatter. A weak size -- linewidth relation has also been
inferred in other galaxies \citep[e.g.][]{hughes2013, utomo2015}, but
a small dynamic range and the relatively large uncertainties of our
$\sigma_{\rm gs,los}$ measurements probably at least partially explain
the poor correlation.

We find a nearly linear correlation between the
$\sigma_{\rm gs,los}$-derived Virial masses,
\begin{equation}
  \label{eq:M_gs,vir}
  M_{\rm gs,vir}\equiv\frac{\sigma_{\rm gs,los}^2R_{\rm c}}{b_{\rm s}G}
\end{equation}
(see Eq.~\ref{eq:Virial_mass}), and the CO-derived gaseous
masses $M_{\rm gas}$ of the spatially-resolved clouds (black solid
line in the middle panel of Fig.~\ref{fig:dynamics_gs}), where we have
again assumed $b_{\rm s}=\frac{1}{5}$ (spherical homogeneous clouds):
\begin{equation}
  \log\left(\frac{M_{\rm gs,vir}}{\rm M_\odot}\right)=(-1.36\pm0.28)+(1.28\pm0.06)\,\log\left(\frac{M_{\rm gas}}{\rm M_\odot}\right)~.
\end{equation} 
A log-normal fit to the distribution of the resulting Virial
parameters,
\begin{equation}
  \label{eq:alpha_gs,vir}
  \alpha_{\rm gs,vir}\equiv\frac{M_{\rm gs,vir}}{M}=\frac{M_{\rm gs,vir}}{M_{\rm gas}}~,
\end{equation}
shown as an inset in the middle panel of Fig.~\ref{fig:dynamics_gs},
yields a mean $\langle\alpha_{\rm gs,vir}\rangle=1.28\pm0.04$ and a
standard deviation of $0.15$~dex. No systematic variation is observed
in the Virial parameter $\alpha_{\rm gs,vir}$ for clouds over a wide
range of galactocentric distances.

The right panel of Fig.~\ref{fig:dynamics_gs} shows the comparison
between $\sigma_{\rm gs,los}R^{-1/2}_{\rm c}$ and the gaseous mass
surface density $\Sigma_{\rm gas}$ for the spatially-resolved
clouds. The data points are distributed along the black solid diagonal
line, suggesting a simple Virial equilibrium. Therefore, when the
contamination of in-plane bulk motions is removed, the clouds in
NGC4429 do seem to reach a state of Virial equilibrium.

A full determination of the internal equilibrium state of clouds
regulated by self-gravity (i.e.\ the first term in brackets on the RHS
of Eq.~\ref{eq:modified_Virial_theorem_1}) requires a knowledge
of $\sigma_{\rm sg,los}$ rather than $\sigma_{\rm gs,los}$. However,
we can still gain important insights from
Fig.~\ref{fig:dynamics_gs}. First, our measured
$\sigma^2_{\rm gs,los}$ should be strongly dominated by
$\sigma^2_{\rm sg,los}$, i.e.\
$\sigma^2_{\rm sg,los}\gg\sigma^2_{\rm gal,z}\cos^2i$ and thus
$\sigma^2_{\rm gs,los}\approx\sigma^2_{\rm sg,los}$ (see
Eq.~\ref{eq:vrms_gs_los}), otherwise (if
$\sigma^2_{\rm gal,z}\cos^2i$ were to contribute significantly to
$\sigma^2_{\rm gs,los}$) the scaling relations presented in
Fig.~\ref{fig:dynamics_gs} would depend on the galaxy's inclination
angle and the trend seen in Fig.~\ref{fig:dynamics_gs} (suggesting a
state of gravitational equilibrium) would turn out to be merely a
coincidence. But we note that a similar result was obtained in another
ETG. Indeed, NGC4526 revealed a good agreement between the
$\sigma_{\rm gs,los}$-derived Virial masses and the CO-derived gaseous
masses ($\langle\alpha_{\rm gs,vir}\rangle=0.99\pm0.02$), and
similarly a $\sigma_{\rm gs,los}R^{-1/2}_{\rm c}$ --
$\Sigma_{\rm gas}$ correlation as expected from Virial equilibrium
\citep{utomo2015}. We thereby consider that the most likely
explanation of our results in Fig.~\ref{fig:dynamics_gs} (and the
results of \citealt{utomo2015}) is that $\sigma^2_{\rm gs,los}$ is
dominated by $\sigma^2_{\rm sg,los}$ (that is assumed isotropic and
thus independent of the galaxy inclination angle) and that an internal
gravitational equilibrium has been achieved through self-gravity. This
assumption is particularly reasonable in NGC4429, as in any case only
a very small part of $\sigma^2_{\rm gal,z}$ can contribute to
$\sigma^2_{\rm gs,los}$ considering its high disc inclination
($i=68^{\circ}$ so $\cos^2i\approx0.1$).

If $\sigma^2_{\rm gs,los}\approx\sigma^2_{\rm sg,los}$, then the left
panel of Fig.~\ref{fig:dynamics_gs} seems to suggest that the clouds'
internal turbulent motions associated with self-gravity follow the
classical size -- linewidth relation of MW clouds, despite a large
scatter. This supports the traditional interpretation of turbulent
motions as the origin of the size -- line width relation
\citep[e.g.][and references therein]{maclow2004,
  ballesterosparedes2006a, ballesterosparedes2007}, that emerges
entirely as a consequence of the gas self-gravity \citep{camacho2016,
  ibanezmejia2016}. The middle panel of Fig.~\ref{fig:dynamics_gs}
then implies that $M_{\rm gas}\approx M_{\rm sg,vir}$, i.e.\ that GMCs
attain approximate Virial balance between their internal turbulent
kinetic energies and their self-gravitational energies. The fact that
the mean $ \alpha_{\rm gs,vir}$ is slightly larger than one
($\langle\alpha_{\rm gs,vir}\rangle=1.28\pm0.04$) may be due to
contamination of $\sigma^2_{\rm gs,los}$ by the
$\sigma^2_{\rm gal,z}\cos^2i$ term. Indeed, in
Section~\ref{sec:cloud_scale_height} we will show that the motions
induced by (external) gravity contribute around $20\%$ of the total
$\sigma^2_{\rm gs,los}$. The right panel of Fig.~\ref{fig:dynamics_gs}
then further indicates that an internal virialisation has been roughly
achieved by self-gravity. In other words, the gravitational potential
defined by the mass of a cloud is matched by the kinetic energy
induced by self-gravity. In this case
($M_{\rm gas}\approx M_{\rm sg,vir}$), we have
\begin{equation}
  \label{eq:internal_turbulent_velocity_dispersion}
  \sigma_{\rm sg,los}^2\approx\pi b_{\rm s}R_{\rm c}G\Sigma_{\rm gas}
\end{equation}
(see Eq.~\ref{eq:Virial_mass}), and
$\alpha_{\rm sg,vir}\equiv\frac{M_{\rm sg,vir}}{M_{\rm gas}}\approx1$
(where $\alpha_{\rm sg,vir}$ is the Virial parameter set purely by a
cloud's self-gravity), as has been suggested by many previous studies
of self-gravitating clouds (e.g.\ Eq.~10 in \citealt{heyer2009}).

We note that this internal Virial equilibrium is established by
self-gravity despite the presence of an external galactic potential,
which seems to support our previous assumption that the motions due to
self-gravity emerge independently of the background galactic
potential. For more discussion of how a Virial equilibrium is
established through the balance of turbulent kinetic and
self-gravitational energy, see \citet{meidt2018}.


\subsection{Role of external gravity}
\label{sec:role_external_gravity}

The contribution of external gravity to a cloud's gravitational energy
budget ($E_{\rm ext}$) is given by the second term in brackets on the
RHS of Eq.~\ref{eq:modified_Virial_theorem_1}:
\begin{equation}
  \label{eq:contribution_external_gravity}
  E_{\rm ext}\approx\underbrace{M\left(\sigma^2_{\rm gal,z}-b_{\rm e}\nu_0^2Z^2_{\rm c}\right)}_{\rm external,~vertical~direction}+\,\underbrace{M\left(\sigma^2_{\rm gal,r}+\sigma^2_{\rm gal,t}+b_{\rm e}(T_0-2\Omega_0^2)R^2_{\rm c}\right)}_{\rm external,~plane}~.
\end{equation}
If $E_{\rm ext}>0$, external gravity acts against self-gravity and
makes the cloud less bound. If $E_{\rm ext}<0$, external gravity acts
with self-gravity and contributes to the collapse of the cloud. If
$E_{\rm ext}\approx0$, the effect of external gravity can be
ignored. A more general form of $E_{\rm ext}$ for a homogeneous
ellipsoidal cloud is provided in
Appendix~\ref{app:modified_Virial_theorem} (Eq.~\ref{eq:eext}). We can
split $E_{\rm ext}$ into two independent parts, one in the vertical
direction ($E_{\rm ext,z}$) and one in the plane
($E_{\rm ext,plane}$), and consider them in turn.

\textbf{\textit{Vertical direction.}} The contribution of the external
potential to the gravitational energy budget of the cloud in the
vertical direction is
\begin{equation}
  \label{eq:eext_vertical}
  E_{\rm ext,z}\approx M\left(\sigma^2_{\rm gal,z}-b_{\rm e}\nu_0^2Z^2_{\rm c}\right)~.
\end{equation}
It is similar to the vertical hydrostatic
equilibrium equation of a gaseous disc \citep[e.g.\ Eq.~3
in][]{koyama2009}. If it is positive, the cloud will be disrupted in
the vertical direction, but if it is negative, the cloud will collapse
in the vertical direction. However, as neither $\sigma_{\rm gal,z}$
nor $Z_{\rm c}$ can be measured directly from our observations, we can
not really assess the vertical equilibrium state of the clouds. But if
we make the common assumption that vertical equilibrium is satisfied
on a cloud scale, i.e.\ that the vertical contribution of external
gravity to the net energy budget of a cloud is negligible (i.e.\
$M(\sigma^2_{\rm gal,z}- b_{\rm e}\nu_0^2Z^2_{\rm c})\approx0$), then
we can derive a relation between $\sigma_{\rm gal,z}$ and $Z_{\rm c}$:
\begin{equation}
  \label{eq:vertical_rms_velocity}
  \sigma^2_{\rm gal,z}\approx  b_{\rm e}\nu_0^2Z_{\rm c}^2~.
\end{equation}

The measured scale heights $Z_{\rm c}$ of clouds in edge-on disc
galaxies can thus be used to determine their unobservable vertical
velocity dispersions $\sigma_{\rm gal,z}$, or conversely the measured
line-of-sight velocity dispersions $\sigma_{\rm gal,z}$ of clouds in
face-on galaxies can be used to determine the unobservable scale
heights $Z_{\rm c}$, as suggested by \citet{koyama2009}. In our work,
we can estimate the value of $\sigma_{\rm gal,z}$ from the deviation
of $\sigma_{\rm gs,los}$ from $\sigma_{\rm sg,los}$, and then infer a
cloud's scale height (combining Eqs.~\ref{eq:vrms_gs_los} and
\ref{eq:vertical_rms_velocity}; see
Section~\ref{sec:cloud_scale_height}). We note that our derived
$\sigma_{\rm gal,z}$ -- $Z_{\rm c}$ correlation is different from the
one derived via the epicyclic approximation by \citet{meidt2018}, by a
factor of $b_{\rm e}$
\citep[$\sigma^2_{\rm gal,z}\approx\nu_0^2Z_{\rm c}^2$
in][]{meidt2018}.  This is because we assumed a spherical cloud with a
radial mass volume density distribution (i.e.\
$\rho(r)\propto r^{-\psi}$) while \citet{meidt2018} assumed a cloud
with an exponential vertical mass volume density distribution (i.e.\
$\rho(z)\propto\exp(-z)$). Overall, to retain vertical hydrostatic
equilibrium on a cloud scale, the gravitationally-induced vertical
motions ($\sigma_{\rm gal,z}$) need to balance the vertical weight of
the background galaxy.

We note here that assuming vertical equilibrium for the clouds goes
against our stated aim of {\em inferring} whether the clouds are
indeed in equilibrium directly from measurements. However, galaxies
are highly symmetric vertically and there is no bulk motion in the
vertical direction, and we will show below that we do not need to
assume the clouds are in equilibrium in the plane. We therefore keep
moving forward with our plan, even if it can only be partially
achieved.

\textbf{\textit{Plane.}} The contribution of the external potential to
the gravitational energy budget of a cloud in the plane is
\begin{equation}
  \label{eq:eext_plane}
  E_{\rm ext,plane}\approx M\left(\sigma^2_{\rm gal,r}+\sigma^2_{\rm gal,t}+  b_{\rm e}(T_0-2\Omega_0^2)R^2_{\rm c}\right)~.
\end{equation} 
The orbital angular velocity $\Omega_0$ and the tidal acceleration
parameter $T_0$ at the cloud's CoM can be derived from our gas
dynamical model (see Section~\ref{sec:origin_velocity_gradients} and
Fig.~\ref{fig:a_vs_r}) and they are listed for each cloud in
Table~\ref{tab:shear_parameters}. A more general form of
$E_{\rm ext,plane}$ for a homogeneous ellipsoidal cloud is provided in
Appendix~\ref{app:modified_Virial_theorem}
(Eq.~\ref{eq:eext_plane_final}). The term
$ b_{\rm e}(T_0-2\Omega^2_0)R^2_{\rm c}$ indicates the effective
potential energy of galactic gravity and the centrifugal force (see
Appendix~\ref{app:modified_Virial_theorem} for more details). We find
that galactic gravity and the centrifugal force act as a binding force
overall, as the corresponding energy
$ b_{\rm e}(T_0-2\Omega^2_0)MR^2_{\rm c}$ is negative in all cases
(the function $T-2\Omega^2$ is generally negative except in the very
centre, $R_{\rm gal}\ls40$~pc, where there is no cloud; see
Fig.~\ref{fig:a_vs_r}). On the other hand, clouds are supported
against collapse by the gravitationally-induced gas motions in the
plane, whose kinetic energy is
$\frac{1}{2}M(\sigma^2_{\rm gal,r}+\sigma^2_{\rm gal,t})$. The
question then is which of the binding energy of galactic gravity plus
the centrifugal force or the kinetic energy of gravitational motions
is more important, i.e.\ whether $E_{\rm ext,plane}<0$ or
$E_{\rm ext,plane}>0$ (or $E_{\rm ext,plane}=0$).


\begin{table}
  \centering
  \caption{Derived properties of the clouds in NGC4429.}
  \label{tab:shear_parameters}
  \resizebox{\columnwidth}{!}{%
    \begin{tabular}{rccccc}
      \hline\hline
      ID & $\Omega_0$ & $T_0$ & $\sigma_{\rm eff,los}$ & $\log\left(\frac{\rho_{\ast,0}}{\rm M_\odot~pc^{-3}}\right)$  \\
         & (${\rm km~s^{-1}~pc^{-1}}$) & $({\rm km~s^{-1}~pc^{-1}})^2$ & (${\rm km~s^{-1}}$) & \\
       \hline\hline
   1&  $0.68\pm0.01$ &  $0.75\pm0.04$ & $\dots$ &   1.26 \\
   2&  $0.71\pm0.01$ &  $0.57\pm0.05$ & $\phantom{1}3.38\pm1.79$ &   1.33 \\
   3&  $0.72\pm0.01$ &  $0.67\pm0.05$ & $\phantom{1}4.98\pm1.13$ &   1.34 \\
   4&  $0.76\pm0.02$ &  $0.67\pm0.06$ & $\phantom{1}2.15\pm1.93$ &   1.41 \\
   5&  $0.80\pm0.02$ &  $0.67\pm0.08$ & $\phantom{1}2.35\pm2.63$ &   1.47 \\
   6&  $0.69\pm0.01$ &  $0.60\pm0.04$ & $\phantom{1}3.22\pm1.16$ &   1.29 \\
   7&  $0.67\pm0.01$ &  $0.74\pm0.04$ & $\phantom{1}4.31\pm1.00$ &   1.26 \\
   8&  $0.81\pm0.02$ &  $0.69\pm0.09$ & $\phantom{1}1.83\pm2.79$ &   1.50 \\
   9&  $0.75\pm0.02$ &  $0.65\pm0.06$ & $\dots$ &   1.39 \\
  10&  $0.89\pm0.03$ &  $0.81\pm0.14$ & $\phantom{1}2.97\pm1.89$ &   1.59 \\
  11&  $0.68\pm0.01$ &  $0.71\pm0.04$ & $\dots$ &   1.27 \\
  12&  $0.72\pm0.01$ &  $0.59\pm0.05$ & $\dots$ &   1.34 \\
  13&  $0.83\pm0.02$ &  $0.76\pm0.10$ & $\phantom{1}2.24\pm2.29$ &   1.52 \\
  14&  $0.95\pm0.04$ &  $0.95\pm0.20$ & $\phantom{1}2.52\pm1.93$ &   1.64 \\
  15&  $0.77\pm0.02$ &  $0.67\pm0.06$ & $\phantom{1}1.62\pm1.97$ &   1.43 \\
  16&  $0.75\pm0.02$ &  $0.66\pm0.06$ & $\dots$ &   1.40 \\
  17&  $0.81\pm0.02$ &  $0.69\pm0.09$ & $\dots$ &   1.50 \\
  18&  $0.84\pm0.02$ &  $0.68\pm0.10$ & $\phantom{1}3.63\pm1.71$ &   1.53 \\
  19&  $0.63\pm0.01$ &  $0.72\pm0.03$ & $\phantom{1}5.58\pm2.18$ &   1.18 \\
  20&  $0.90\pm0.03$ &  $0.83\pm0.15$ & $\phantom{1}2.51\pm2.10$ &   1.60 \\

      \dots & \dots & \dots & \dots& \dots  \\
217&  $0.65\pm0.01$ &  $0.67\pm0.03$ & $\phantom{1}1.58\pm0.94$ &   1.21 \\
      \hline\hline   
       \end{tabular}
     } \raggedright Notes.\ -- Clouds with no $\sigma_{\rm eff,los}$
     entry are unresolved spatially. Calculations of
     $\sigma_{\rm eff,los}$ assume $b_{\rm e}=\frac{1}{5}$ (spherical
     homogeneous clouds). All uncertainties are quoted at the
     $1\,\sigma$ level, and those of $\sigma_{\rm eff,los}$ have been
     propagated from the uncertainties of both observed and modelled
     quantities (see Eq.~\ref{eq:all_obs_plane_sphere_2}). As noted
     in the text, the uncertainty of the adopted distance $D$ to
     NGC4429 was not propagated through the tabulated uncertainties of
     the quantity $\sigma_{\rm eff,los}$. This is because an error on
     the distance to NGC4429 translates to a systematic (rather than
     random) scaling of some of the measured quantities (no effect on
     the others), here $R_{\rm c}\propto D$, $\Omega_0 \propto D^{-1}$
     and $T_0 \propto D^{-2}$ in
     Eq.~\ref{eq:all_obs_plane_sphere_2}. Oort's constants $A$ and $B$
     can be derived using respectively $A=\frac{T}{4\Omega}$ and
     $B=\frac{T}{4\Omega}-\Omega$. Table~\ref{tab:shear_parameters} is
     available in its entirety in machine-readable form in the
     electronic edition.
\end{table} 

As suggested by Eq.~\ref{eq:eext_plane}, a full derivation of
$E_{\rm ext,plane}$ requires knowledge of $\sigma_{\rm gal,r}$ and
$\sigma_{\rm gal,t}$, the RMS velocities of gravitationally-induced
motions in the plane. Although $\sigma_{\rm gal,r}$ and
$\sigma_{\rm gal,t}$ can not be obtained directly from observations,
we can nevertheless glean some information about them from the
observed quantities $\sigma_{\rm obs,los}$ and $\sigma_{\rm
  gs,los}$. Indeed, if we assume the gas motions induced by the
galactic potential to be isotropic in the plane (i.e.\
$\sigma_{\rm gal,r}=\sigma_{\rm gal,t}$), the RMS velocities of the
in-plane gas motions due to external gravity can easily be derived by
combining Eqs.~\ref{eq:sigma_obslos} and \ref{eq:vrms_gs_los}:
\begin{equation}
  \label{eq:in_plane_rms_velocity}
  \sigma_{\rm gal,r}^2=\sigma_{\rm gal,t}^2\approx\frac{\sigma_{\rm obs,los}^2-\sigma_{\rm gs,los}^2}{\sin^2i}~.
\end{equation}
Substituting Eq.~\ref{eq:in_plane_rms_velocity} into
Eq.~\ref{eq:eext_plane}, we find the net contribution of external
gravity to the gravitational budget of the clouds in NGC4429 to be
positive in most cases (i.e.\ $E_{\rm ext,plane}>0$). Therefore, the
main effect of the external gravity on the clouds of NGC4429 is to
make them less bound (in the plane).

\textbf{\textit{Effective parameters.}} Our MVT
(Eq.~\ref{eq:modified_Virial_theorem_1}) can be written simply
as \begin{equation}
  \begin{split}
    \frac{\ddot{I}}{2} & \approx\left(3M\sigma_{\rm
        sg,los}^2-3b_{\rm s}GM^2/R_{\rm c}\right)+E_{\rm ext}\\
    & \approx\frac{3b_{\rm s}GM^2}{R_{\rm c}}\left(\frac{\sigma_{\rm
          sg,los}^2R_{\rm c}}{b_{\rm s}GM}+\frac{E_{\rm ext}}{3b_{\rm
          s}GM^2/R_{\rm c}}-1\right)
  \end{split}
\end{equation}
(see Eq.~\ref{eq:contribution_external_gravity}). However,
\begin{equation}
  \label{eq:alpha_sg,vir}
  \alpha_{\rm sg,vir}\equiv\frac{\sigma_{\rm sg,los}^2R_{\rm c}}{b_{\rm s}GM}
\end{equation}
(see Eq.~\ref{eq:alpha_vir}), the traditional Virial
parameter regulated by self-gravity only, and we define
\begin{equation}
  \label{eq:beta}
  \beta\equiv\frac{E_{\rm ext}}{3b_{\rm s}GM^2/R_{\rm c}}~,
\end{equation}
the ratio between the contribution of external gravity and the
(absolute value of the) cloud's self-gravitational energy
($\vert U_{\rm sg}\vert=3b_{\rm s}GM^2/R_{\rm c}$), so that
\begin{equation}
  \frac{\ddot{I}}{2}\approx\frac{3b_{\rm s}GM^2}{R_{\rm c}}\left(\alpha_{\rm sg,vir}+\beta-1\right)~.
\end{equation}
This naturally leads us to define an effective Virial parameter
\begin{equation}
  \label{eq:alpha_eff,vir}
  \alpha_{\rm eff,vir}\equiv\alpha_{\rm sg,vir}+\beta
\end{equation}
such that
\begin{equation}
  \label{eq:modified_Virial_theorem_alpha_eff}
  \frac{\ddot{I}}{2}\approx\frac{3b_{\rm s}GM^2}{R_{\rm
      c}}\left(\alpha_{\rm eff,vir}-1\right)~.
\end{equation}
Thus, just like the standard Virial parameter, this effective Virial
parameter informs on the dynamical stability of a cloud. If
$\alpha_{\rm eff,vir}\approx1$, the cloud is gravitationally bound and
in Virial equilibrium even in the presence of the external (i.e.\
galactic) gravitational potential. If $\alpha_{\rm eff,vir}\gg1$, the
cloud is unlikely to be bound (i.e.\ it is transient unless confined
by other forces). If $\alpha_{\rm eff,vir}\lesssim1$, the molecular
cloud is likely to collapse. For clouds that are (marginally)
gravitationally bound, we again require
$\alpha_{\rm eff,vir}\le\alpha_{\rm vir,crit}=2$ \citep{kauffmann2013,
  kauffmann2017}, or equivalently $\beta\le1$ if an internal Virial
equilibrium is established by self-gravity (i.e.\ if
$\alpha_{\rm sg,vir}\approx1$; see Eq.~\ref{eq:alpha_eff,vir}).

Equivalently, from Eq.~\ref{eq:alpha_vir}, we can define an
effective velocity dispersion
\begin{equation}
  \label{eq:effective_velocity_dispersion_2}
  \sigma_{\rm eff,los}^2=\alpha_{\rm eff,vir}\,b_{\rm s}GM/R_{\rm c}~,
\end{equation}
and thus our modified Virial equation
(Eq.~\ref{eq:modified_Virial_theorem_1}) can be simplified to
\begin{equation}
  \label{eq:modified_Virial_theorem_2}
  \frac{\ddot{I}}{2}\approx\left(3M\sigma^2_{\rm eff,los}-3b_{\rm s}GM^2/R_{\rm c}\right)~.
\end{equation}
The parameters $\alpha_{\rm eff,vir}$ (via
Eq.~\ref{eq:modified_Virial_theorem_alpha_eff}) or equivalently
$\sigma_{\rm eff,los}$ (via Eq.~\ref{eq:modified_Virial_theorem_2})
thus embody our MVT and offer a straightforward method to test the
gravitational boundedness of a cloud in the presence of an external
(i.e.\ galactic) gravitational field.

Of course, our expressions are of no use in practice if the external
contribution $E_{\rm ext}$ can not be evaluated (see
Eq.~\ref{eq:contribution_external_gravity}). Indeed, without knowledge
of $\sigma_{\rm gal,z}$, $\sigma_{\rm gal,r}$ and
$\sigma_{\rm gal,t}$, none of $\beta$, $\alpha_{\rm eff,vir}$ or
$\sigma_{\rm eff,los}$ can be evaluated. However, by making
increasingly stringent assumptions, we show in
Appendix~\ref{app:effective_virial_parameter} that it is possible to
evaluate all these quantities from observable quantities alone. We
thus briefly summarise those assumptions and their consequences here,
but refer to Appendix~\ref{app:effective_virial_parameter} for
detailed calculations.

First, we assume clouds are in vertical hydrostatic equilibria, i.e.\
$\sigma^2_{\rm gal,z}\approx b_{\rm e}\nu_0^2Z^2_{\rm c}$
(Eq.~\ref{eq:vertical_rms_velocity}), so that the contribution of the
external potential to the gravitational energy budget of each cloud in
the vertical direction vanishes, i.e.\ $E_{\rm ext,z}\approx 0$ (see
Eq.~\ref{eq:eext_vertical}). Second, we assume the motions associated
with external gravity to be isotropic in the plane, i.e.\
$\sigma^2_{\rm gal,r}=\sigma^2_{\rm gal,t}\approx\frac{\sigma^2_{\rm
    obs,los}-\sigma^2_{\rm gs,los}}{\sin^2i}$
(Eq.~\ref{eq:in_plane_rms_velocity}). Third, we assume
$\sigma^2_{\rm sg,los}\approx\sigma^2_{\rm gs,los}$ (or equivalenty
$\sigma^2_{\rm sg,los}\gg\sigma^2_{\rm gal,z}\cos^2i$; see
Eq.~\ref{eq:vrms_gs_los}) and thus
$\alpha_{\rm sg,vir}\approx\alpha_{\rm gs,vir}$. As shown in
Appendix~\ref{app:effective_virial_parameter}, all three assumptions
taken together lead to
\begin{equation}
  \begin{split}
    \label{eq:all_obs_plane_sphere_2}
    E_{\rm ext} & \approx M\left[\frac{2\left(\sigma_{\rm obs,los}^2-\sigma_{\rm gs,los}^2\right)}{\sin^2i}+b_{\rm e}(T_0-2\Omega_0^2)R_{\rm c}^2\right]~,\\
    \sigma_{\rm eff,los}^2 & \approx\sigma_{\rm gs,los}^2+\frac{1}{3}\left[\frac{2\left(\sigma_{\rm obs,los}^2-\sigma_{\rm gs,los}^2\right)}{\sin^2i}+b_{\rm e}(T_0-2\Omega_0^2)R_{\rm c}^2\right]~{\rm and}\\
    \alpha_{\rm eff,vir} & \approx\frac{\sigma_{\rm gs,los}^2R_{\rm c}}{b_{\rm s}GM}+\frac{R_{\rm c}}{3b_{\rm s}GM}\left[\frac{2\left(\sigma_{\rm obs,los}^2-\sigma_{\rm gs,los}^2\right)}{\sin^2i}+b_{\rm e}(T_0-2\Omega_0^2)R_{\rm c}^2\right]
  \end{split}
\end{equation}
(see Eqs.~\ref{eq:contribution_external_gravity},
\ref{eq:effective_velocity_dispersion_2}, \ref{eq:alpha_eff,vir},
\ref{eq:alpha_gs,vir} \ref{eq:M_gs,vir} and \ref{eq:beta}).  More
general forms for a homogeneous ellipsoidal cloud are provided in
Appendix~\ref{app:effective_virial_parameter}. These
Eqs.~\ref{eq:all_obs_plane_sphere_2} represent our final MVT, whose
power lies in the fact that all of $E_{\rm ext}$,
$\sigma_{\rm eff,los}$ and $\alpha_{\rm eff,vir}$ can be evaluated
directly from observations. Indeed, as mentioned previously, the
measured $M$ ($M_{\rm gas}$), $R_{\rm c}$, $\sigma_{\rm obs,los}$ and
$\sigma_{\rm gs,los}$ are listed for each spatially-resolved cloud in
Table~\ref{tab:gmcs}, while $\Omega_0$, $T_0$ and the resulting
$\sigma_{\rm eff,los}$ (and thus $\alpha_{\rm eff,vir}$; see
Eq.~\ref{eq:effective_velocity_dispersion_2}) are listed in
Table~\ref{tab:shear_parameters}.

The first term on the RHS of Eqs.~\ref{eq:all_obs_plane_sphere_2}
(except for $E_{\rm ext}$) comprises the gas turbulent motions
associated with a cloud's self-gravity, the second term denotes the
gravitational motions associated with external gravity in the plane,
and the last term encompasses the external/galactic forces on the
cloud.

An extended Virial theorem including the background tidal field was
formulated by \citet{chen2016}, but they only evaluated two
representative cases, namely a non-rotating cloud
($\sigma_{\rm eff,los}^2=\sigma^2+b_{\rm e}(T_0-2\Omega_0^2)R_{\rm
  c}^2/3$; Eq.~17 in their paper) and a tidally-locked cloud
($\sigma_{\rm eff,los}^2=\sigma^2+b_{\rm e}T_0R_{\rm c}^2/3$; Eq.~18
in their paper). Our derived MVT is valid for more general cases. In
fact, we obtain the same results as \citet{chen2016} for their two
particular cases. For a non-rotating cloud,
$\sigma_{\rm gal,r}=\sigma_{\rm gal,t}=0$, and we derive
$\sigma_{\rm eff,los}^2=\sigma_{\rm sg,los}^2+b_{\rm
  e}(T_0-2\Omega_0^2)R_{\rm c}^2/3$. For a tidally-locked cloud (i.e.\
Oort's constants $A=0$ and $B=\Omega$),
$(\sigma_{\rm gal,r}^2+\sigma_{\rm gal,t}^2)\approx2b_{\rm
  e}\Omega_0^2R_{\rm c}^2$ (as
$\sigma_{\rm gal,r}^2\approx\sigma_{\rm gal,r}^2\approx b_{\rm
  e}\Omega_0^2R_{\rm c}^2$; see
Eq.~\ref{eq:vrms_gravitational_velocity_appendixB} in
Appendix~\ref{app:effective_virial_parameter}), and we derive
$\sigma_{\rm eff,los}^2=\sigma_{\rm sg,los}^2+b_{\rm e}T_0R_{\rm
  c}^2/3$. We note that the velocity dispersion $\sigma^2$ used by the
extended Virial theorem of \citet{chen2016} should be the internal
turbulent velocity dispersion rather than the observed (i.e.\ total)
velocity dispersion.
  
Overall, to take into account the influence of external gravity on the
dynamical state of a cloud, one should use the effective virial
parameter $\alpha_{\rm vir, eff}$ and effective velocity dispersion
$\sigma_{\rm eff,los}$. The latter quantifies the net kinetic energy
that balances the cloud's (self-)gravitational potential energy. The
kinetic energy obtained using $\sigma_{\rm eff,los}$ includes the
cloud's internal turbulent kinetic energy due to self-gravity as well
as the contributions from the external gravity. If
$\alpha_{\rm eff,vir}>\alpha_{\rm sg,vir}$ or
$\sigma_{\rm eff,los}^2>\sigma_{\rm sg,los}^2$, external gravity acts
against self-gravity and makes the cloud less bound (i.e.\
$E_{\rm ext}>0 $). If $\alpha_{\rm eff,vir}<\alpha_{\rm sg,vir}$ or
$\sigma_{\rm eff,los}^2<\sigma_{\rm sg,los}^2$, external gravity acts
with self-gravity and contributes to the cloud's confinement and/or
collapse (i.e.\ $E_{\rm ext}<0 $). If
$\alpha_{\rm eff,vir}=\alpha_{\rm sg,vir}$ or
$\sigma_{\rm eff,los}^2=\sigma_{\rm sg,los}^2$, then external gravity
has no effects on the cloud's dynamical state (i.e.\ $E_{\rm ext}=0$).
Therefore, the results presented in Figs.~\ref{fig:dynamics_obs} and
\ref{fig:dynamics_gs}, that respectively adopt $\sigma_{\rm obs,los}$
and $\sigma_{\rm gs,los}$, do not reflect the real dynamical states of
the NGC4429 clouds. Specifically, $\sigma_{\rm obs,los}$ embodies gas
motions associated with self-gravity and external gravity, but it
ignores the extra binding energy provided by galactic forces and the
centrifugal force (i.e.\ the term
$b_{\rm e}(T_0-2\Omega_0^2)MR_{\rm c}^2$ in
Eqs.~\ref{eq:modified_Virial_theorem_1},
\ref{eq:contribution_external_gravity} and \ref{eq:eext_plane}, that
is negative in almost all cases), so it overestimates the effect of
external gravity on the clouds. Conversely, $\sigma_{\rm gs,los}$ only
reflects gas motions associated with self-gravity, so it does not
include the contribution of external gravity to a cloud's
gravitational energy budget.


\subsection{Cloud scaling relations using the effective velocity
  dispersion}
\label{sec:larson_sigma_eff}

In consequence, we revisit yet again the three scaling relations that
describe the dynamical states of the clouds in NGC4429, this time
using the effective velocity dispersion $\sigma_{\rm eff,los}$ defined
in Eqs.~\ref{eq:all_obs_plane_sphere_2}. In most cases our derived
$\sigma_{\rm eff,los}$ is larger than the gradient-subtracted velocity
dispersion $\sigma_{\rm gs,los}$, and in all cases it is smaller than
the observed velocity dispersion $\sigma_{\rm obs,los}$ (see
Fig.~\ref{fig:sigma_comp}). This implies that external gravity
generally makes the clouds less bound. We nevertheless note that we
find a few clouds where $\sigma_{\rm eff,los}$ is smaller than
$\sigma_{\rm gs,los}$, suggesting external gravity contributes to the
cloud's confinement and/or collapse in these few cases.
As expected in Eq.~\ref{eq:all_obs_plane_sphere_2},
these clouds all have  $\sigma_{\rm gs,los}$  nearly 
equal to $\sigma_{\rm obs,los}$, and thus low velocity gradients of $0.1- 0.2~
{\rm km~s^{-1}~pc^{-1}}$.

\begin{figure*}
  \includegraphics[width=0.95\textwidth]{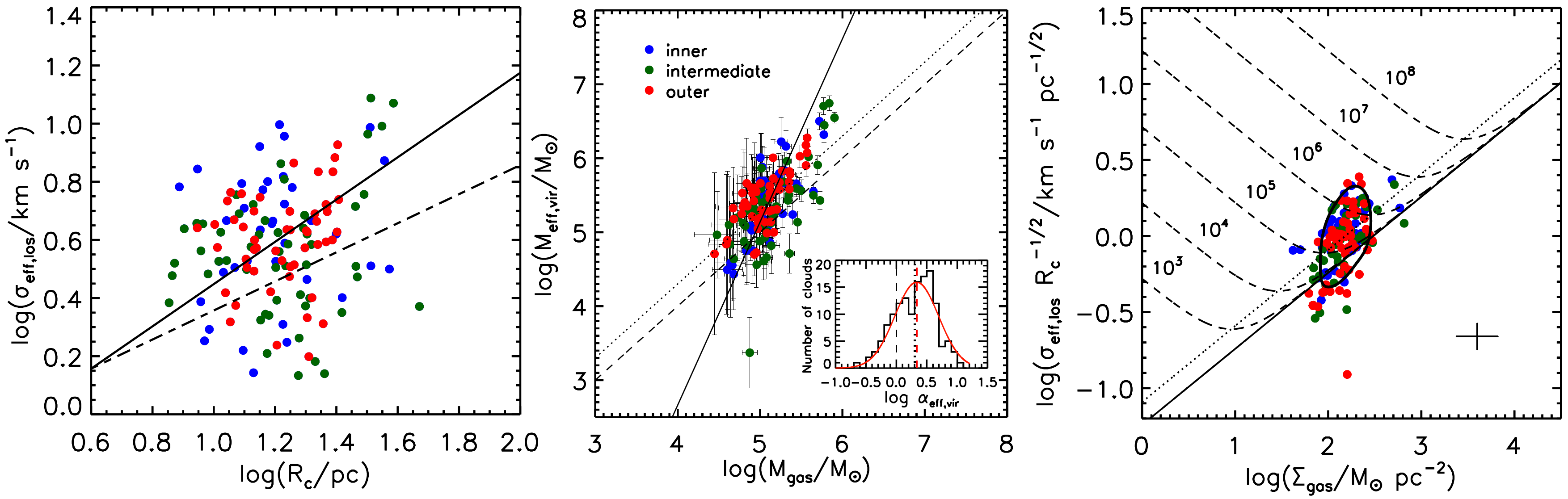}
  \caption{Same as Figs.~\ref{fig:dynamics_obs} and
    \ref{fig:dynamics_gs}, but using our effective measure of velocity
    dispersion $\sigma_{\rm eff,los}$ for axisymmetric clouds.}
  \label{fig:dynamics_eff}
\end{figure*}

The left panel of Fig.~\ref{fig:dynamics_eff} (data points and black
solid line) presents the $\sigma_{\rm eff,los}$ -- $R_{\rm c}$
relation for the $141$ spatially-resolved clouds of NGC4429, where we
have assumed $b_{\rm e}=\frac{1}{5}$ (homogeneous clouds). The
relation appears to have a slightly steeper slope ($0.72\pm0.18$) than
that of MW clouds ($0.5\pm0.05$; \citealt{solomon1987}), but the
correlation is very weak (with a Spearman rank correlation coefficient
of $0.13$).

\begin{figure*}
  \includegraphics[width=0.95\textwidth]{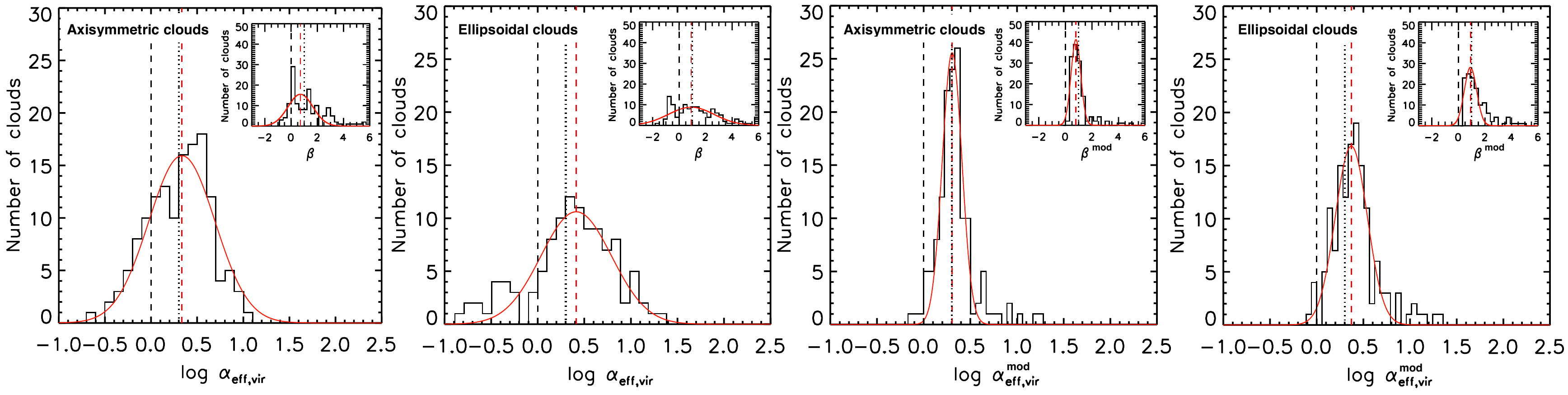}
  \caption{Distributions of the effective virial parameter
    $\alpha_{\rm vir,eff}$ and the external energy parameter $\beta$
    (insets) of the $141$ spatially-resolved clouds of NGC4429,
    calculated from both observations (first and second panels; see
    Section~\ref{sec:role_external_gravity}) and our shear model
    (third and fourth panels; see
    Section~\ref{sec:linear_shear_flow_non-zero_e_ext}), assuming both
    axisymmetric (first and third panels) and ellipsoidal (second and
    fourth panels) clouds.  Log-normal ($\alpha_{\rm vir,eff}$) and
    normal ($\beta$) fits are overlaid (red solid lines). The vertical
    red dashed lines indicate the means of the fits, while the
    vertical black dashed and dotted lines indicate
    $\alpha_{\rm vir}=1$ and $2$ ($\beta=0$ and $1$),
    respectively. }
  \label{fig:alpha_vir}
\end{figure*}

The Virial masses of the spatially-resolved clouds derived using
$\sigma_{\rm eff,los}$, referred to as effective Virial
masses
\begin{equation}
  \label{eq:M_eff,vir}
  M_{\rm eff,vir}\equiv\frac{\sigma_{\rm eff,los}^2R_{\rm c}}{b_{\rm s}G}
\end{equation}
(see Eq.~\ref{eq:Virial_mass}), are compared to the CO-derived gaseous
masses $M_{\rm gas}$ in the middle panel of
Fig.~\ref{fig:dynamics_eff}, where we have again assumed
$b_{\rm s}=\frac{1}{5}$ (spherical homogeneous clouds). A linear fit
between effective Virial and gaseous mass yields (black solid line in
the middle panel of Fig.~\ref{fig:dynamics_eff})
\begin{equation}
  \log\left(\frac{M_{\rm eff,vir}}{\rm M_\odot}\right)=(-7.25\pm1.54)+(2.47\pm0.30)\log\left(\frac{M_{\rm gas}}{\rm M_\odot}\right)~.
\end{equation}
A log-normal fit to the distribution of the resulting effective Virial
parameters,
\begin{equation}
  \label{eq:alpha_eff,vir_2}
  \alpha_{\rm eff,vir}\equiv\frac{M_{\rm eff,vir}}{M}=\frac{M_{\rm eff,vir}}{M_{\rm gas}}
\end{equation}
(see also Eq.~\ref{eq:effective_velocity_dispersion_2}), shown in the
first panel of Fig.~\ref{fig:alpha_vir}, yields a mean
$\langle\alpha_{\rm eff,vir}\rangle=2.15\pm0.12$ and a standard
deviation of $0.35$~dex. This mean is higher than
$\langle\alpha_{\rm sg,vir}\rangle\approx\langle\alpha_{\rm
  gs,vir}\rangle=1.28\pm0.04$ (see Section~\ref{sec:larson_sigma_gs}),
suggesting that the main effect of external gravity on the clouds is
to make them less bound. However, since many NGC4429 clouds have a
mean effective virial parameter close to the critical value regarded
as the boundary between gravitationally-bound and unbound clouds
\citep{kauffmann2013, kauffmann2017}, i.e.\
$\langle\alpha_{\rm eff,vir}\rangle\approx\alpha_{\rm vir,crit}=2$,
the clouds should still be marginally gravitationally bound.

The inset in the first panel of Fig.~\ref{fig:alpha_vir} shows the
distribution of the measured $\beta$ (Eq.~\ref{eq:beta}). A Gaussian
fit yields a mean $\langle\beta\rangle=0.71 \pm0.33$ and a standard
deviation of $1.05$. Given these $\beta\approx1$, the contribution of
external gravity to each clouds' energy budget is generally
significant, on average of the order of (and frequently exceeding) the
self-gravitational energy (see also Eq.~\ref{eq:alpha_eff,vir}). We
will discuss this aspect further in
Section~\ref{sec:equilibrium_between_self-gravity_external-gravity}. However,
we note immediately that a noticeable fraction of spatially-resolved
clouds ($25/141$ or $\approx18\%$) have $\beta\le0$. These negative
$\beta$ could be due to observational uncertainties, and thus
inaccuracies when estimating $\alpha_{\rm eff,vir}$ (or
$E_{\rm ext}$), but some clouds may well have their gas motions
decoupled from global galaxy rotation. Indeed, we found that clouds
with $\beta\le0$ have larger discrepancies between their observed and
modelled angular momenta (with a median projected angular velocity
discrepancy factor of $\approx1.9$ and a median position angle
difference of $\approx24^\circ$; see
Section~\ref{sec:origin_velocity_gradients}) than clouds with
$\beta>0$ (with a median projected angular velocity discrepancy factor
of $\approx1.3$ and a median position angle difference of
$\approx13^\circ$). It thus seems that clouds with $\beta\le0$ only
weakly follow the galaxy orbital rotation.  These clouds are therefore
presumably not as strongly affected by galactic shear and tidal
forces, and they can become more virialised (with
$\langle\alpha_{\rm eff,vir}\rangle\approx0.9$).

The right panel of Fig.~\ref{fig:dynamics_eff} shows the
$\sigma_{\rm eff,los}R_{\rm c}^{-1/2}$ -- $\Sigma_{\rm gas}$ relation
for the $141$ spatially-resolved clouds of NGC4429. The data points
are mostly distributed away from the black solid diagonal line (SVE),
but they are clustered around the black dotted diagonal line. This
again suggests that, although the NGC4429 clouds are not virialised,
they could be marginally gravitationally bound.


\subsection{Cloud scaling relations considering ellipsoidal clouds}
\label{sec:larson_ellipsoidal}

By using a single measure of size for each cloud ($R_{\rm c}$; see
Section~\ref{sec:definition_gmc_properties}), our analysis has so far
implicitly assumed that each cloud is axisymmetric in the orbital
plane. However, the effects of external gravity on a cloud (and its
contribution $E_{\rm ext}$ to a cloud's energy budget) also formally
depend on the actual shape and position angle of the cloud (see
Appendices~\ref{app:modified_Virial_theorem} and
\ref{app:effective_virial_parameter}). To assess the impacts of this
assumption, we now assume instead that each cloud has an ellipsoidal
geometry, with semi-axis $Z_{\rm c}$ perpendicular to the orbital
plane and semi-major axis $X_{\rm c}$ (at a position angle
$\phi_{\rm PA}$ with respect to the radial/galactocentric direction)
and semi-minor axis $Y_{\rm c}$ in the orbital plane.

If an ellipsoidal cloud is homogeneous,
Appendices~\ref{app:modified_Virial_theorem} and
\ref{app:effective_virial_parameter} show that
Eqs.~\ref{eq:all_obs_plane_sphere_2} (that assume vertical
equilibrium, isotropy in the equatorial plane and
$\sigma_{\rm sg,los}\approx\sigma_{\rm gs,los}$) become
\begin{equation}
  \label{eq:all_obs_plane_2}
  \begin{split}
    E_{\rm ext} & \approx M\left[\frac{2\left(\sigma_{\rm obs,los}^2-\sigma_{\rm gs,los}^2\right)}{\sin^2i}\right.\\
   & +\left.b_{\rm e}T_0\left(X_{\rm c}^2\cos^2\phi_{\rm PA}+Y_{\rm c}^2\sin^2\phi_{\rm PA}\right)-b_{\rm e}\Omega_0^2\left(X_{\rm c}^2+Y_{\rm c}^2\right)\vphantom{\frac{\left(\sigma_{\rm obs,los}^2\right)}{\sin^2i}}\right]~,\\
   \sigma_{\rm eff,los}^2 & \approx\sigma_{\rm gs,los}^2+\frac{1}{3}\left[\frac{2\left(\sigma_{\rm obs,los}^2-\sigma_{\rm gs,los}^2\right)}{\sin^2i}\right.\\
    & +\left.b_{\rm e}T_0\left(X_{\rm c}^2\cos^2\phi_{\rm PA}+Y_{\rm c}^2\sin^2\phi_{\rm PA}\right)-b_{\rm e}\Omega_0^2\left(X_{\rm c}^2+Y_{\rm c}^2\right)\vphantom{\frac{\left(\sigma_{\rm obs,los}^2\right)}{\sin^2i}}\right]~{\rm and}\\
    \alpha_{\rm eff,vir} & \approx\frac{\sigma_{\rm gs,los}^2R_{\rm c}}{b_{\rm s}GM}+\frac{R_{\rm c}}{3b_{\rm s}GM}\left[\frac{2\left(\sigma_{\rm obs,los}^2-\sigma_{\rm gs,los}^2\right)}{\sin^2i}\right.\\
    & +\left.b_{\rm e}T_0\left(X_{\rm c}^2\cos^2\phi_{\rm PA}+Y_{\rm c}^2\sin^2\phi_{\rm PA}\right)-b_{\rm e}\Omega_0^2\left(X_{\rm c}^2+Y_{\rm c}^2\right)\vphantom{\frac{\left(\sigma_{\rm obs,los}^2\right)}{\sin^2i}}\right]~.
  \end{split}
\end{equation}
These equations thus represent our final MVT
(Eqs.~\ref{eq:all_obs_plane_sphere_2}) for the case of a homogenous
ellipsoidal cloud.

We note that $X_{\rm c}$, $Y_{\rm c}$ and $\phi_{\rm PA}$ in
Eqs.~\ref{eq:all_obs_plane_2} should be measured in the cloud's
orbital plane (i.e.\ the galaxy's equatorial plane) rather than the
sky plane. To correct for the effects of inclination, we thus create
an image of each cloud deprojected to a face-on view, from which we measure
the semi-major and semi-minor axes analogously to $R_{\rm c}$ in
Section~\ref{sec:definition_gmc_properties} and the position angle
with respect to the radial/galactocentric direction.

\begin{figure*}
  \includegraphics[width=0.95\textwidth]{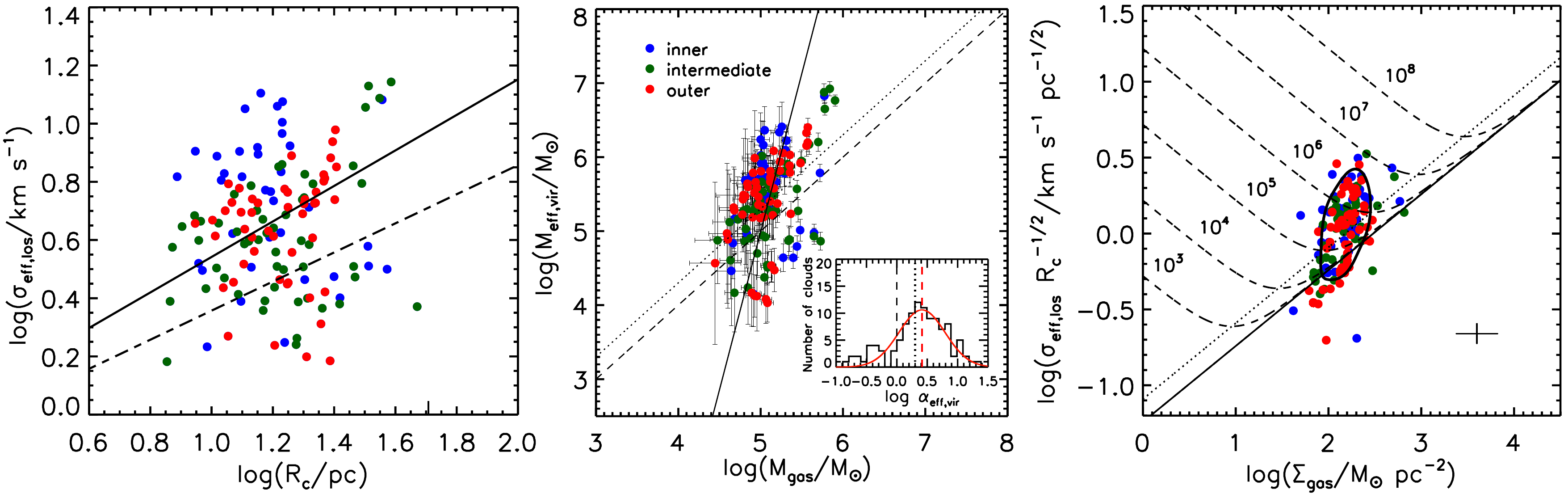}
  \caption{Same as Figs.~\ref{fig:dynamics_obs}, \ref{fig:dynamics_gs}
    and \ref{fig:dynamics_eff}, but using our effective measure of
    velocity dispersion $\sigma_{\rm eff,los}$ for ellipsoidal
    clouds. }
  \label{fig:dynamics_eff_ell}
\end{figure*}

The left panel of Fig.~\ref{fig:dynamics_eff_ell} presents the
$\sigma_{\rm eff,los}$ -- $R_{\rm c}$ relation for the $141$
spatially-resolved clouds of NGC4429 assuming they are ellipsoidal and
$b_{\rm e}=\frac{1}{5}$ (homogeneous clouds). The relation has a slope
of $0.62\pm0.21$, consistent with that of axisymmetric clouds
($0.72\pm0.18$; see Section~\ref{sec:larson_sigma_eff}), and is thus
again slightly steeper than that of MW clouds ($0.5\pm0.05$;
\citealt{solomon1987}), although the correlation is again very weak
(with a Spearman rank correlation coefficient of 0.09. 

The effective Virial masses of the $141$ spatially-resolved clouds
($M_{\rm eff,vir} \equiv \sigma_{\rm eff,los}^2 R_{\rm c} / b_{\rm s}G$) 
derived assuming ellipsoidal shapes 
(see Eq.~\ref{eq:Virial_mass};where $R_{\rm c}$ 
  is defined as $\sqrt{X_{\rm c}Y_{\rm c}}$) are compared to the
CO-derived gaseous masses $M_{\rm gas}$ in the middle panel of
Fig.~\ref{fig:dynamics_eff_ell}. We have again assumed
$b_{\rm s_1}=\frac{1}{5}$ (homogeneous clouds), but calculated
$b_{\rm s_2}$, that quantifies the effects of the ellipticity, separately 
for each cloud using the method provided by
\citet{bertoldi1992} (see Appendix~\ref{app:modified_Virial_theorem}
for more details). We find the exact cloud morphology has negligible
effects on the quantities regarding to the cloud's self-gravity
(e.g.\ $U_{\rm sg}$ and
$\alpha_{\rm sg}$), as $b_{\rm s_2}$ is 
approximately unity ($\langle b_{\rm s_2}\rangle\approx0.95$).
 A linear fit between the effective Virial and
gaseous masses (black solid line in the middle panel of
Fig.~\ref{fig:dynamics_eff_ell}) yields a slope of $4.27\pm0.70$. 
A log-normal fit to the distribution of the effective
Virial parameters ($\alpha_{\rm eff,vir}$) derived assuming ellpsoidal
clouds (see Eqs.~\ref{eq:alpha_eff,vir_2} and
\ref{eq:all_obs_plane_2}), shown in the second panel of
Fig.~\ref{fig:alpha_vir}, yields a mean
$\langle\alpha_{\rm eff,vir}\rangle=2.59\pm0.19$ and a standard
deviation of $0.38$~dex, only slightly larger than that estimated
assuming axisymmetric clouds
($\langle\alpha_{\rm eff,vir}\rangle=2.15\pm0.12$; see
Section~\ref{sec:larson_sigma_eff}), and again higher than
$\langle\alpha_{\rm sg,vir}\rangle\approx\langle\alpha_{\rm
  gs,vir}\rangle=1.28\pm0.04$ (see Section~\ref{sec:larson_sigma_gs}),
suggesting that the main effect of external gravity on the clouds is
to make them less bound irrespective of their exact shapes.

The inset in the second panel of Fig.~\ref{fig:alpha_vir} shows the
distribution of the resulting $\beta$ for the $141$ spatially-resolved
clouds of NGC4429, calculated assuming ellipsoidal clouds.
A Gaussian fit to the distribution yields a mean
$\langle\beta\rangle=0.91\pm0.35$ and a standard deviation of $1.73$,
again slightly larger than that derived assuming axisymmetric clouds
($\langle\beta\rangle=0.71\pm0.33$; see the first panel of
Fig.~\ref{fig:alpha_vir} and Section~\ref{sec:larson_sigma_eff})). As
we shall discuss in Section~\ref{sec:cloud_morphology}, this is
primarily due to the radially-elongated shapes of the NGC4429
clouds. The differences are however minor, and it is still true that
$\langle\alpha_{\rm eff,vir}\rangle\approx\alpha_{\rm vir,crit}=2$ and
$\langle\beta\rangle\approx1$ for ellipsoidal clouds. Therefore, the
evidence remains that the NGC4429 clouds appear to be marginally
gravitationally bound.

The right panel of Fig.~\ref{fig:dynamics_eff_ell} shows the
$\sigma_{\rm eff,los}R_{\rm c}^{-1/2}$ -- $\Sigma_{\rm gas}$ relation
for the $141$ spatially-resolved clouds, derived assuming ellipsoidal
clouds and $b_{\rm e}=\frac{1}{5}$. Just as for axisymmetric clouds,
the data points are generally above the black solid diagonal line
(SVE) but are centered on the black dotted diagonal line. This thus
suggests again that, irrespective of their exact shapes, the NGC4429
clouds are probably not virialised but are likely to be marginally
gravitationally bound.

In summary, the dynamical states of the NGC4429 clouds are regulated
by both self-gravity and external (i.e.\ galactic) gravity. Internal
Virial equilibria between the clouds' turbulent kinetic energies and
their own gravitational energies have been attained, regardless of the
presence of external gravity. The additional contribution of external
gravity to the clouds' gravitational energy budgets includes two
parts: the supporting kinetic energy from gravitational motions
($\frac{1}{2}M\sigma_{\rm gal,z}^2$ in the vertical direction and
$\frac{1}{2}M(\sigma_{\rm gal,r}^2+\sigma_{\rm gal,t}^2)$ in the
plane) and the effective potential energy of the galactic and
centrifugal forces ($-b_{\rm e}M\nu_0^2Z_{\rm c}^2$ in the vertical
direction and $ b_{\rm e}(T_0-2\Omega_0^2)MR_{\rm c}^2$ in the plane).
If we assume the NGC4429 clouds are in vertical hydrostatic equilibria
(i.e.\
$E_{\rm ext,z}\approx M\sigma_{\rm gal,z}^2-b_{\rm e}M\nu_0^2Z_{\rm
  c}^2=0$), gravitational motions are isotropic in the orbital plane
(i.e.\ $\sigma_{\rm gal,r}=\sigma_{\rm gal,t}$) and
$\sigma_{\rm sg,los} \approx \sigma_{\rm gs,los}$, we can calculate
the contributions of external gravity to the clouds' energy budgets
($E_{\rm ext}$) directly from the observations. These are positive in
most cases and on average of the order of the clouds'
self-gravitational energies (i.e.\
$\beta\equiv\frac{E_{\rm ext}}{\vert U_{\rm sg}\vert}\approx1$). The
derived effective virial parameters have a mean of $\approx2$, i.e.\
$\langle\alpha_{\rm eff,vir}\rangle\approx\alpha_{\rm
  vir,crit}=2$. Both results are essentially independent of the exact
cloud shapes (i.e.\ whether we assume axisymmetric or ellpsoidal
clouds), suggesting that the NGC4429 clouds are marginally
gravitationally bound due to the combined effects of self-gravity and
external gravity.

\section{Discussion}
\label{sec:discussion}


\subsection{Shear motions and non-zero $E_{\rm ext}$}
\label{sec:linear_shear_flow_non-zero_e_ext}

As gravitational motions appear to play an important role regulating
the dynamics and boundedness of the clouds in NGC4429, we discuss in
more depth in this section the clouds' motions driven by the external
(i.e.\ galactic) gravitational forces. As in
Appendix~\ref{app:eext_rotating_frame}, we adopt a local Cartesian
coordinate system centred on the centre of mass (COM) of each cloud,
that both orbits around the galaxy centre with the COM (with azimuthal
velocity $\Omega_0R_0$) and rotates on itself (with angular velocity
$\Omega_0$), such that the $x^{\prime}$ axis always points in the
direction of increasing galactocentric radius and the $y^{\prime}$
axis always points in the direction of orbital rotation (see
Fig.~\ref{fig:rotating_frame}). As shown in
Appendix~\ref{app:eext_rotating_frame}, in this rotating frame the
equations of motions driven by external gravity can be written as
\begin{equation}
  \label{eq:aext_xy_rf_2}
  \left\{
    \begin{split}
      a_{\rm ext,x^\prime}^\prime & \approx T_0x^\prime+2\Omega_0v_{\rm gal,y^\prime}^\prime~,\\
      a_{\rm ext,y^\prime}^\prime & \approx -2\Omega_0v_{\rm gal, x^\prime}^\prime~, 
    \end{split}
  \right.
\end{equation}
where $x^\prime$, $v_{\rm gal,x^\prime}^\prime$ and $a_{\rm
  ext,x^\prime}^\prime$ are the components of the position vector 
$\vec{d}_{\rm plane}^\prime$, velocity vector $\vec{v}_{\rm
  gal}^\prime$ and acceleration vector $\vec{a}_{\rm ext}^\prime$
along the $\hat{x}^\prime$ direction, respectively, similarly for
$y^\prime$, $v_{\rm gal,y^\prime}^\prime$ and
$a_{\rm ext,y^\prime}^\prime$. The $T_0x^\prime$ term represents the
tidal force while the terms $2\Omega_0v_{\rm gal,y^\prime}^\prime$ and
$-2\Omega_0v_{\rm gal,x^\prime}^\prime$ represent the Coriolis force.

As discussed in Appendix~\ref{app:eext_rotating_frame}, this set of
coupled differential equations has solution
\begin{equation}
  \label{eq:motion_solutions_2}
  \left\{
    \begin{split}
      x^\prime & ={\rm S}_1\sin(\kappa_0t+\varphi)+{\rm S}_2~,\\
      y^\prime & =\frac{2\Omega_0}{\kappa_0}\,{\rm S}_1\cos(\kappa_0t+\varphi)-2A_0\,{\rm S}_2\,t~+{\rm S}_3~,
    \end{split}
  \right.
\end{equation}
where $\kappa_0$ is the epicyclic frequency evaluated at the cloud's
COM
($\kappa_0^2\equiv\left(R\frac{d\Omega^2(R)}{dR}+4\Omega^2(R)\right)\vert_{R=R_0}$),
$A_0$ is Oort's constant $A$ quantifying shear evaluated at the
cloud's CoM
($A_0\equiv-\frac{R}{2}\frac{d\Omega(R)}{dR}\vert_{R=R_0}$), and
${\rm S}_1$, ${\rm S}_2$ and ${\rm S}_3$ (as well as the arbitrary
phase $\varphi$) are constants that depend on the given boundary
(e.g.\ initial) conditions. Equations~\ref{eq:motion_solutions_2} show
that the gravitational motions associated with external gravity have
two contributions: epicyclic motions around the cloud's COM (i.e.\ the
``guiding centre''; see e.g.\ \citealt{meidt2018}), indicated by the
trigonometric terms ${\rm S}_1\sin(\kappa_0t+\varphi)$ and
$\frac{2\Omega_0}{\kappa_0}\,{\rm S}_1\cos(\kappa_0t+\varphi)$, and
linear shear motion, indicated by the $-2A_0\,{\rm S}_2\,t$ term
\citep[e.g.][]{gammie1991,tan2000,binney2020}.

It is worth noting that, in a model where all fluid elements of a
cloud move on perfectly circular orbits (around the galaxy centre)
determined by the galactic potential, the epicyclic amplitudes vanish
and the gravitational motions are completely dominated by the shear
motions, i.e.\ \begin{equation}
  \label{eq:motion_solutions_shear_2}
  \left\{
    \begin{split}
      x^\prime & = {\rm S}_2~,\\
      y^\prime & = -2A_0\,{\rm S}_2\,t+{\rm S}_3~.
    \end{split}
  \right.
\end{equation}
Hereafter we name this model, where all fluid elements of a cloud are
assumed to populate perfectly circular orbits determined by the
galactic potential, the ``shear model''. We thus define a shear
velocity \begin{equation}
  \label{eq:v_shear}
  v_{\rm shear}\equiv-2A_0\,{\rm S}_2~,
\end{equation}
where ${\rm S}_2$ is the distance of the fluid element from the
cloud's centre along $\hat{x}^\prime$ (see, again,
Fig.~\ref{fig:rotating_frame}). Interestingly, as we shall demonstrate
below, the bulk motions observed in the NGC4429 clouds appear to be
strongly dominated by gravitational shear motions, with little or no
evidence of gravitational epicyclic motions, i.e.\ the fluid elements
of the clouds seem to populate nearly circular orbits (around the
galaxy centre) determined by the galactic potential.

First, the measured velocity gradients across the spatially-resolved
clouds of NGC4429, and the position angles of the rotation axes of
these clouds, are both consistent with those predicted by assuming
purely circular orbital motions (see Fig.~\ref{fig:angmom_comparison}
and Section~\ref{sec:origin_velocity_gradients}, where both the
measured and modelled quantities are calculated in the sky
plane). This provides strong evidence that the bulk motions of the
NGC4429 clouds are dominated by gravitational shear motions.

\begin{figure}
  \includegraphics[width=0.95\columnwidth]{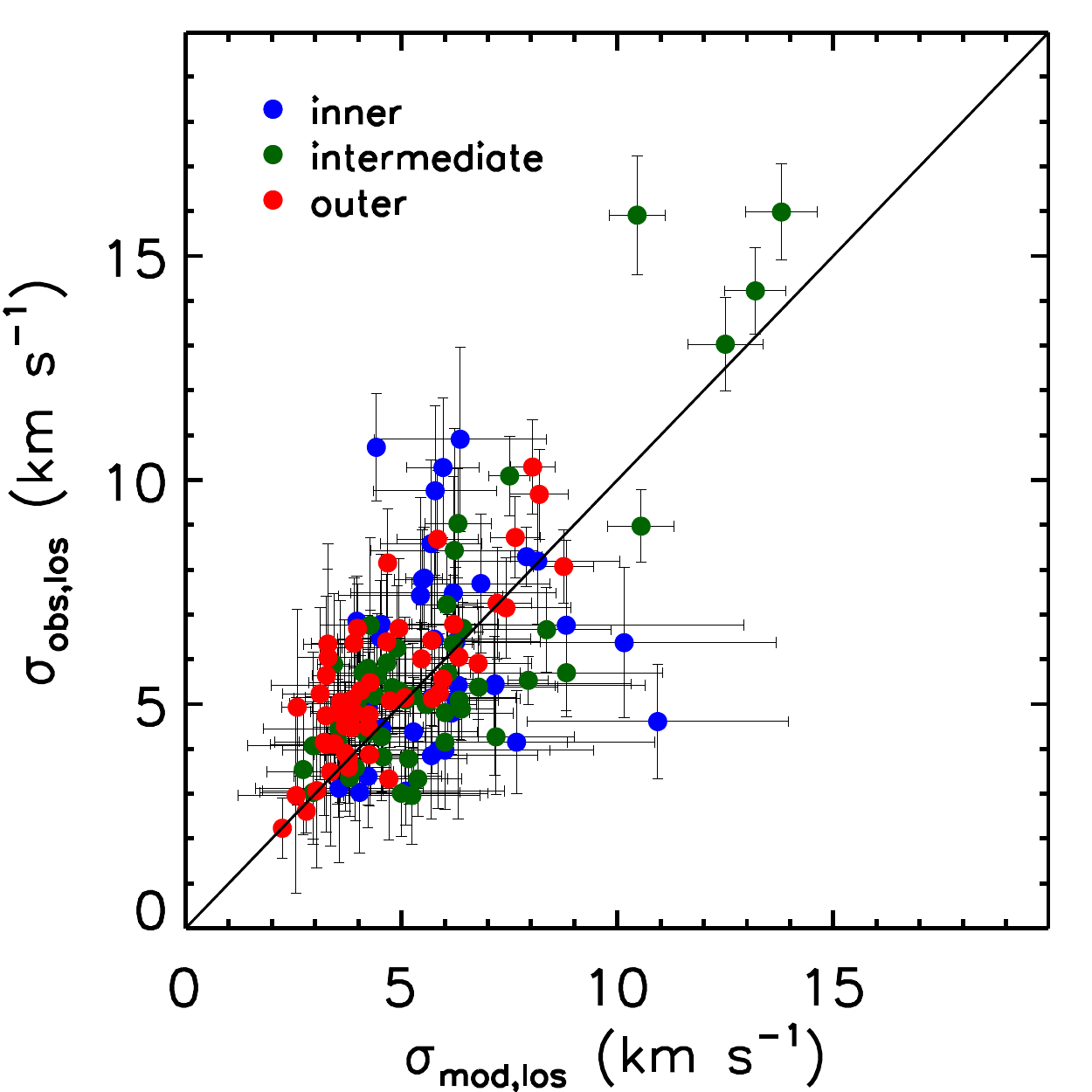}
  \caption{Comparison of the observed and modelled line-of-sight
    velocity dispersion of the $141$ spatially-resolved clouds of
    NGC4429. Data points are colour-coded by region. The black solid
    diagonal line shows the $1:1$ relation.}
  \label{fig:vrms_vs_vpred}
\end{figure}

Second, if all fluid elements of a cloud indeed follow circular orbits
determined by the galactic potential, then we can predict the RMS
velocities of the clouds' gravitational motions in both the radial and
azimuthal directions:
$(\sigma_{\rm gal,r}^{\rm mod})^2=b_{\rm e}\Omega_0^2R_{\rm c}^2$ and
$(\sigma_{\rm gal,t}^{\rm mod})^2=b_{\rm e}(\Omega_0-2A_0)^2R_{\rm
  c}^2$ (see Eqs.~\ref{eq:vrms_gravitational_velocity_appendixB} in
Appendix~\ref{app:effective_virial_parameter}). We can thus also
predict their line-of-sight velocity dispersions using
Eq.~\ref{eq:sigma_obslos}:
\begin{gather}
  \begin{split}
    \sigma^2_{\rm mod,los}\approx & \,\sigma^2_{\rm sg,los}\,+\,\left((\sigma_{\rm gal,r}^{\rm mod})^2\sin^2\theta+(\sigma_{\rm gal,t}^{\rm mod})^2\cos^2\theta\right)\sin^2i\\
    & ~~~~~~~~~~~~ +\,\sigma_{\rm gal,z}^2\cos^2i\\
    \approx & \,\pi b_{\rm s}R_{\rm c}G\Sigma_{\rm gas}+b_{\rm e}R_{\rm c}^2\left(\Omega_0^2\sin^2\theta+(\Omega_0-2A_0)^2\cos^2\theta\right)\sin^2i~,
  \end{split}
\end{gather}
where we have used $\alpha_{\rm sg,vir}\approx1$ (and thus
$\sigma_{\rm sg,los}^2\approx\pi b_{\rm s}R_{\rm c}G\Sigma_{\rm gas}$;
see Eqs.~\ref{eq:internal_turbulent_velocity_dispersion} and
\ref{eq:alpha_sg,vir}) and $\sigma_{\rm gal,z}^2\cos^2i\approx0$. We
compare in Fig.~\ref{fig:vrms_vs_vpred} the observed line-of-sight
velocity dispersions $\sigma_{\rm obs,los}$ of the $141$
spatially-resolved clouds of NGC4429 with those predicted from our
shear model $\sigma_{\rm mod,los}$. We generally find a good agreement
between the two, albeit with a few exceptions. This thus reinforces
our inference that the bulk motions of the NGC4429 clouds are
dominated by gravitational shear motions.

 Lastly,  if all fluid elements of a cloud follow pure epicyclic motions 
described by the trigonometric terms in Eqs.~\ref{eq:motion_solutions_2}, 
the cloud is necessarily in Virial equilibrium \citep{meidt2018}
and thus the contribution of external gravity should vanish, i.e.\
$E_{\rm ext} = 0$ and $\alpha_{\rm eff,vir} \approx \alpha_{\rm sg,vir}
\approx 1$. However, $E_{\rm ext}$ (or equivalently
$\beta$) measured from our observations of spatially-resolved clouds
are clearly not zero (see Section~\ref{sec:role_external_gravity}),
suggesting that the bulk motions within the NGC4429 clouds can not be
dominated by gravitational epicyclic motions. In turn, we expect the
measured $E_{\rm ext}$ (and $\beta$) to more closely match those
predicted from gravitational shear motions only. As our shear model
assumes that all fluid elements of a cloud move on perfectly circular
orbits determined by the galactic potential, this model yields (cf.\
Eqs.~\ref{eq:all_obs_plane_sphere_2})
\begin{equation}
  \label{eq:model_sphere_2}
  \begin{split}
    E_{\rm ext}^{\rm mod} & =4b_{\rm e}A_0^2MR_{\rm c}^2~,\\
    (\sigma_{\rm eff,los}^{\rm mod})^2 & =\sigma_{\rm sg,los}^2+\frac{4b_{\rm e}A_0^2R_{\rm c}^2}{3}\\
    & \approx\sigma_{\rm gs,los}^2+\frac{4b_{\rm e}A_0^2R_{\rm c}^2}{3}~~~{\rm and}\\
    \alpha_{\rm eff,vir}^{\rm mod} & =\frac{\sigma_{\rm sg,los}^2R_{\rm c}}{b_{\rm s}GM}+\frac{4b_{\rm e}A_0^2R_{\rm c}^3}{3b_{\rm s}GM}\\
    & \approx\frac{\sigma_{\rm gs,los}^2R_{\rm c}}{b_{\rm s}GM}+\frac{4b_{\rm e}A_0^2R_{\rm c}^3}{3b_{\rm s}GM}~,
  \end{split}
\end{equation}
where we have again assumed vertical equilibrium, isotropy in the
equatorial plane and $\sigma_{\rm sg,los}\approx\sigma_{\rm gs,los}$.
The detailed derivations of these equations and their more general
forms for a homogeneous ellipsoidal cloud are provided in
Appendix~\ref{app:effective_virial_parameter}.

Unsurprisingly, in the shear model the overall effect of external
gravity primarily depends on the shear arising from the differential
rotation of the galaxy disc (i.e.\ Oort's constant $A$). We note that
$E_{\rm ext}^{\rm mod}$ can be understood as the rotational kinetic
energy of a cloud with angular velocity $\omega_{\rm shear}=-2A_0$, as
generally the rotational kinetic energy
$E_{\rm rot}=\frac{1}{2}I\omega^2$ and $I=2b_{\rm e}MR_{\rm c}^2$ for
a spherical cloud. Our derived $\omega_{\rm shear}$ is the same as
that derived by \citet{goldreich1965} and \citet{fleck1981}, and it
arises naturally when considering fluid element motions near the tidal
radius (see Section~\ref{sec:cloud_morphology}). For a galaxy with a
solid-body circular velocity curve, the external gravity has no effect
on the cloud, i.e.\ $A=0$ and thus $E_{\rm ext}^{\rm mod}=0$.  The
distributions of $\alpha_{\rm vir,eff}^{\rm mod}$ and
$\beta^{\rm mod}\equiv\frac{E_{\rm ext}^{\rm mod}}{\vert U_{\rm
    sg}\vert}$ for axisymmetric and ellipsoidal clouds are shown in
the third and fourth panels of Fig~\ref{fig:alpha_vir}, respectively,
for the $141$ spatially-resolved clouds of NGC4429. For clouds assumed
to be axisymmetric, a log-normal fit to the distribution of
$\alpha_{\rm vir,eff}^{\rm mod}$ yields a mean
$\langle\alpha_{\rm eff,vir}^{\rm mod}\rangle=2.02\pm0.03$ and a
standard deviation of $0.10$~dex, while a Gaussian fit to the
distribution of $\beta^{\rm mod}$ yields a mean
$\langle\beta^{\rm mod}\rangle=0.79\pm0.37$ and a standard deviation
of $0.36$. For clouds assumed to be ellipsoidal, analogous fits yield
$\langle\alpha_{\rm eff,vir}^{\rm mod}\rangle_{\rm
  ellipsoid}=2.35\pm0.07$ and a standard deviation of $0.17$~dex, and
$\langle\beta^{\rm mod}\rangle_{\rm ellipsoid}=0.90 \pm0.28$ and a
standard deviation of $0.46$. Both sets of predictions therefore
compare very well with our measurements
($\langle\alpha_{\rm eff,vir}\rangle=2.15\pm0.12$ and
$\langle\beta\rangle=0.71\pm0.33$ for axisymmetric clouds and
$\langle\alpha_{\rm eff,vir}\rangle_{\rm ellipsoid}=2.59\pm0.19$ and
$\langle\beta\rangle_{\rm ellipsoid}=0.91\pm0.35$ for ellipsoidal
clouds; see Sections~\ref{sec:larson_sigma_eff} and
\ref{sec:larson_ellipsoidal}, respectively), although with less
scatter as expected (our model predictions do not take into account
measurement errors). This thus supports yet again our conclusion that
the bulk motions of the clouds in NGC4429 are primally driven by shear
motions. Indeed, our shear model provides good estimates of
$E_{\rm ext}$, $\alpha_{\rm eff,vir}$ and $\beta$ for the
spatially-resolved clouds of NGC4429.

It is nevertheless worth noting that, while our shear model accounts
for the observed bulk motions of the clouds well, there are also some
discrepancies. Our shear model overestimate the angular velocities of
the spatially-resolved clouds of NGC4429 by a median factor of
$\approx1.5-2.0$, and the modelled and observed position angles have a
median angle difference of $\approx 16^\circ-19^\circ$ (see
Section~\ref{sec:origin_velocity_gradients}). Moreover, there is
considerable scatter about the $1:1$ correlation between the observed
velocity dispersions $\sigma_{\rm obs,los}$ and the modelled velocity
dispersions $\sigma_{\rm mod,los}$ (Fig.~\ref{fig:vrms_vs_vpred}). It
therefore appears that, although the effects of external gravity are
dominant, other factors also noticeably affect the dynamics of clouds,
so that the clouds's fluid elements do not follow pure shear
motions. We discuss one such factor, self gravity, below.


\subsection{Equilibrium between self-gravity and external-gravity}
\label{sec:equilibrium_between_self-gravity_external-gravity}

In previous sections, we established that the contributions of
external gravity to the gravitational energy budgets of the NGC4429
clouds (i.e.\ $E_{\rm ext}$) are clearly non-zero, this whether these
contributions are calculated from observations
(Section~\ref{sec:role_external_gravity}) or our shear model
(Section~\ref{sec:linear_shear_flow_non-zero_e_ext}).  However,
$E_{\rm ext}$ on its own does not determine whether a cloud is
gravitationally bound or not. As a robust threshold between
gravitationally bound and unbound objects, we have adopted a critical
virial parameter $\alpha_{\rm vir,crit}=2$ \citep{kauffmann2013,
  kauffmann2017}. When the effective virial parameter
$\alpha_{\rm eff,vir}$ is equal to this critical value,
$2\vert U_{\rm sg}\vert=2E_{\rm turb}+E_{\rm ext}$
($U_{\rm sg}\equiv-\frac{3b_{\rm s}GM^2}{R_{\rm c}}$), where
\begin{equation}
  E_{\rm turb}\equiv\frac{3}{2}M\sigma_{\rm sg,los}^2
\end{equation}
is the kinetic energy of the turbulent motions associated with
self-gravity (see Eqs.~\ref{eq:alpha_eff,vir}, \ref{eq:beta} and
\ref{eq:alpha_sg,vir}). If a cloud is thus marginally gravitationally
bound (i.e.\ $\alpha_{\rm eff,vir}=\alpha_{\rm vir,crit}=2$) and an
internal Virial equilibrium is established by self-gravity (i.e.\
$\alpha_{\rm sg,vir}\approx1$), as is the case for the NGC4429 clouds,
we further obtain
\begin{equation}
  \label{eq:equilibrium_between_self_external_gravity}
  \left\{
    \begin{split}
      2E_{\rm turb} & +U_{\rm sg}=0~,\\
      E_{\rm ext} & +U_{\rm sg}=0~.
    \end{split}
    \right.
\end{equation}
The top equation indicates an equilibrium between a cloud's
self-gravitational energy and its turbulent kinetic energy, while the
bottom equation indicates an equilibrium between a cloud's
self-gravitational energy and its energy contributed by external
gravity.

In general, one needs to compare $E_{\rm ext}$ with the
self-gravitational energy of a cloud to assess its boundedness. If
$E_{\rm ext}\gg\vert U_{\rm sg}\vert$ (i.e.\ $\beta\gg1$), then external
gravity is much more important and the cloud is not gravitationally
bound (unless other forces are present). If
$E_{\rm ext}\ll\vert U_{\rm sg}\vert$ (i.e.\ $\beta\ll1$), then
self-gravity is much more important and the effects of external
gravity are negligible. If $E_{\rm ext}\approx\vert U_{\rm sg}\vert$
(i.e.\ $\beta\approx1$), then external gravity and self-gravity are
equally important and the cloud reaches a state of equilibrium between
self-gravity and external gravity. Here, we have found that the NGC4429
clouds have $E_{\rm ext}$ comparable to (the absolute values of) their
self-gravitational energies, with both $\langle\beta\rangle\approx1$
(see Section~\ref{sec:role_external_gravity}) and
$\beta^{\rm mod}\approx1$ (see
Section~\ref{sec:linear_shear_flow_non-zero_e_ext}). The energy of
each cloud contributed by external gravity $E_{\rm ext}$ thus roughly
equals its self-gravitational energy and the cloud remains marginally
gravitationally bound.

\textbf{\textit{Tidal radius.}} In the case where the gravitational
motions of the clouds are completely dominated by shear motions, as is
the case for the NGC4429 clouds (see
Section~\ref{sec:linear_shear_flow_non-zero_e_ext}), the bottom
equation of Eqs.~\ref{eq:equilibrium_between_self_external_gravity}
then indicates an equilibrium between a cloud's self-gravitational
energy and its kinetic energy associated with those shear
motions. Another way to assess wether self-gravity or external gravity
is more important is thus to consider the tidal radius of each cloud,
that defines the volume over which self-gravity dominates over
external gravity. Here, we adopt the tidal radius $R_{\rm t}$ defined
by \citet{gammie1991} and \citet{tan2000}, that is the radial distance
from the cloud's center at which the shear velocity due to
differential galactic rotation (i.e.\ our previously-defined
$v_{\rm shear}$; see Eq.~\ref{eq:v_shear}) is equal to the escape
velocity from the cloud:
\begin{equation}
  \label{eq:tidal_radius}
  R_{\rm t}\equiv(1-\beta_{{\rm circ},0})^{-2/3}\left(\frac{2M}{M_{{\rm
        gal},0}}\right)^{1/3}R_0~,
\end{equation}
where as before $M$ is the cloud's mass and $R_0$ the galactocentric
distance of the cloud's CoM in the plane of the disc,
$M_{{\rm gal},0}$ is the total galactic mass interior to $R_0$,
$\beta_{{\rm circ},0}\equiv\frac{d\ln\,V_{\rm
    circ}(R)}{d\ln\,R}\vert_{R=R_0}$, and as before $V_{\rm circ}(R)$
is the galaxy circular velocity curve. Equation~\ref{eq:tidal_radius}
assumes a spherical galaxy mass distribution, i.e.\
$M_{\rm gal}(R)=V_{\rm circ}(R)^2R/G$, and can therefore be simplified to
\begin{equation}
  \label{eq:tidal_radius2}
  R_{\rm t}=\left(\frac{G}{2A_0^2}\right)^{1/3}\,M^{1/3}~.
\end{equation}
The tidal radius defined in this manner is the maximum size of a cloud
(of a given mass $M$) allowed by galactic rotational shear.

\begin{figure}
  \includegraphics[width=0.95\columnwidth]{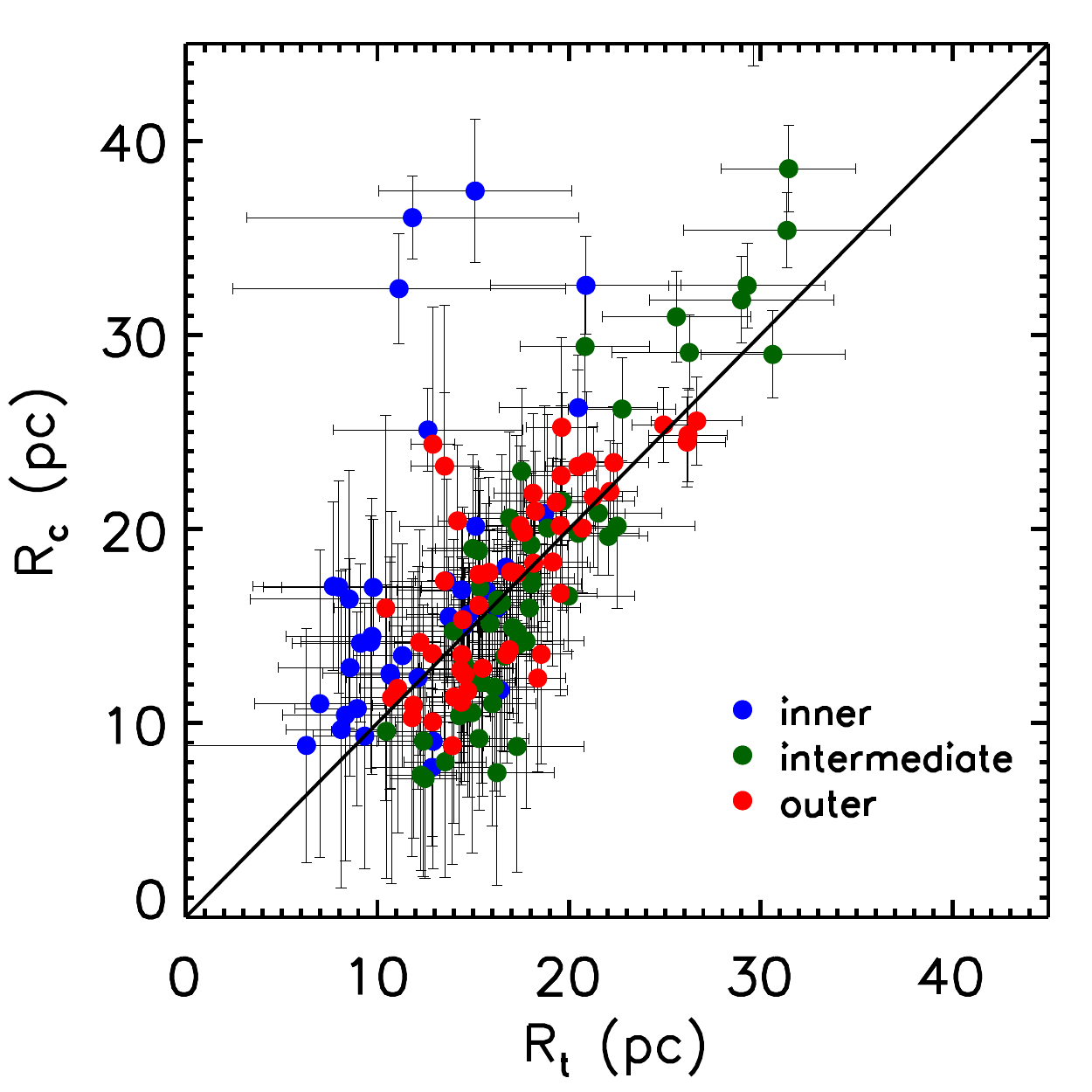}
  \caption{Comparison of the observed cloud size and expected tidal
    radius of the $141$ spatially-resolved clouds of NGC4429. Data
    points are colour-coded by region. The black solid diagonal line
    shows the $1:1$ relation.}
  \label{fig:rc_vs_rt}
\end{figure}

Interestingly, for a cloud with $R_{\rm c}=R_{\rm t}$, we have
\begin{gather}
  \begin{split}
    \label{eq:beta_mod_shear}
    \beta^{\rm mod}(R_{\rm c}=R_{\rm t}) & \equiv\frac{E_{\rm ext}^{\rm mod}(R_{\rm c}=R_{\rm t})}{\vert U_{\rm sg}(R_{\rm c}=R_{\rm t})\vert}\\
    & =\frac{4b_{\rm e}A_0^2M R^2_{\rm t}}{3b_{\rm s}GM^2/R_{\rm t}} \\
    & =\frac{2}{3}\frac{b_{\rm e}}{b_{\rm s}}\\
    & \approx1
  \end{split} 
\end{gather}
(see Eqs.~\ref{eq:model_sphere_2} and \ref{eq:beta}), that is
essentially identical to the measured $\beta$ of the
spatially-resolved clouds of NGC4429 (assuming axisymmetric clouds;
see Sections~\ref{sec:larson_sigma_eff} and
\ref{sec:larson_ellipsoidal}). In our shear model, the tidal radius
given by Eq.~\ref{eq:tidal_radius2} thus approximately corresponds to
the radial distance at which $\beta^{\rm mod}\approx1$. It is thus
clear the reason a cloud with $\beta^{\rm mod}\gg1$ becomes
gravitationally unbound is because the shear motions are so strong
that the (outer) fluid elements manage to escape from the
self-gravitational influence of the cloud. Figure~\ref{fig:rc_vs_rt}
compares the observed sizes (radii $R_{\rm c}$) of the
spatially-resolved clouds of NGC4429 with their tidal radii expected
from Eq.~\ref{eq:tidal_radius2}. There is generally a very good
agreement, albeit with a few exceptions. The NGC4429 clouds therefore
seem to reach their maximum sizes allowed by galactic shear, further
supporting our conclusion that the NGC4429 clouds have reached a rough
equilibrium between self-gravity and external gravity and thereby
manage to remain marginally gravitationally bound. A few clouds in the
inner region have sizes much larger than their tidal radii, suggesting
that these inner clouds can not be gravitationally bound due to shear,
and indeed all these clouds have high $\beta$ ($\beta\approx4$ -- $6$)
and $\alpha_{\rm eff,vir}$ ($\alpha_{\rm eff,vir}\approx5$ -- $7$).

\textbf{\textit{Size and surface density.}} For a cloud to be
marginally gravitationally bound, the contribution of external gravity
to the cloud's energy budget must not exceed the cloud's
self-gravitational energy, i.e.\
$\beta^{\rm mod}=E_{\rm ext}^{\rm mod}/\vert U_{\rm sg}\vert=4b_{\rm
  e}A_0^2MR_{\rm c}^2/\vert-3b_{\rm s}GM^2/R_{\rm c}\vert=4b_{\rm
  e}A_0^2R_{\rm c}/3\pi b_{\rm s}G\Sigma_{\rm gas}\le1$ (see
Eqs.~\ref{eq:model_sphere_2} and \ref{eq:beta}). This implies that, at
a given surface density, there is a maximum size ($R_{\rm shear}$) for
a cloud to stay marginally bound against tidal/shear disruptions:
\begin{equation}
  \label{eq:maximual_tidally_stable_size}
  R_{\rm c}\le R_{\rm shear}\approx\frac{3\pi b_{\rm s}G\Sigma_{\rm gas}}{4b_{\rm e}A_0^2}~.
\end{equation}
Equivalently, at a given size, there is a minimum surface density
($\Sigma_{\rm shear}$) for a cloud to remain marginally bound:
\begin{equation}
  \label{eq:minimal_tidally_stable_surface_density}
  \Sigma_{\rm gas}\ge\Sigma_{\rm shear}\approx\frac{4b_{\rm e}A_0^2R_{\rm c}}{3\pi b_{\rm s}G}~.
\end{equation}

The spatially-resolved clouds of NGC4429 have a mean surface density
$\langle\Sigma_{\rm gas}\rangle\approx160$~{\Msolar}~pc$^{-2}$ and a
mean Oort's constant $A_0$ (i.e.\ shear)
$\langle A_0\rangle\approx0.3$~km~s$^{-1}$~pc$^{-1}$. A simple
calculation using Eq.~\ref{eq:maximual_tidally_stable_size}
(and assuming $b_{\rm s}=b_{\rm e}=\frac{1}{5}$ for spherical
homogeneous clouds as usual) then suggests that, if limited by shear,
the mean size $\langle R_{\rm shear}\rangle$ of the clouds in NGC4429
should be $\approx18$~pc, that matches extremely well the observed
mean size $\langle R_{\rm c}\rangle\approx17$~pc.
We thus find again that typical clouds in NGC4429 reach their maximal
sizes (or minimum surface densities) allowed by shear, and are thus
not limited by other processes (shear rules!).

Finally, as we have pointed out above, the effects of self-gravity are
generally of the same order as those of external (i.e.\ galactic)
gravity: $\beta\approx1$ (see Section~\ref{sec:larson_sigma_eff}) and
$R_{\rm c} \approx R_{\rm t}$. The motions of the fluid elements
within these marginally gravitationally-bound clouds will therefore
not completely follow those prescribed by external gravity alone
(i.e.\ the shear motions governed by
Eqs.~\ref{eq:motion_solutions_shear_2}). This is again expected, as if
the bulk motions of cloud fluid elements were to exactly follow pure
shear motions, the clouds could not be (marginally) gravitationally
bound.

Therefore, the equations of (bulk) motions must include additional
terms due to self-gravity (cf.\ Eqs.~\ref{eq:aext_xy_rf_2}):
\begin{equation}
  \label{eq:aext_xy_rf_self}
  \left\{
    \begin{split}
      a_{\rm ext,x^\prime}^\prime & \approx T_0x^\prime+2\Omega_0v_{\rm gal,y^\prime}^\prime-\frac{\partial\Phi_{\rm sg}}{\partial x^\prime}~,\\
      a_{\rm ext,y^\prime}^\prime & \approx-2\Omega_0v_{\rm gal,x^\prime}^\prime-\frac{\partial\Phi_{\rm sg}}{\partial y^\prime}~, 
    \end{split}
  \right.
\end{equation}
where $\Phi_{\rm sg}$ is the cloud's own (self) gravitational
potential. Solving these coupled differential equations is very
difficult and beyond the scope of this paper, although we do derive
approximate analytic solutions in Section~\ref{sec:cloud_morphology}
for a particular case.
We refer readers to \citet{julian1966}, \citet{gammie1991} and
\citet{binney2020} for some numerical solutions.

\citet{gammie1991} suggested that
Eqs.~\ref{eq:motion_solutions_shear_2} can provide good-zeroth order
solutions to Eqs.~\ref{eq:aext_xy_rf_self} at large radii (where
$R_{\rm c}\ge R_{\rm t}$). Therefore, the bulk motions of
gravitationally-unbound ($R_{\rm c} \gg R_{\rm t}$)
clouds should roughly approximate the gravitational shear motions
described by Eqs.~\ref{eq:motion_solutions_shear_2}. However, unlike
gravitationally-unbound clouds, where the discrepancies between bulk
motions and shear motions are expected to be negligible, marginally
gravitationally-bound clouds should have bulk motions that deviate
considerably from shear motions. This is because the discrepancies
between bulk motions and shear motions should increase with the
importance of self-gravity. Indeed, while the bulk motions of the
NGC4429 clouds do approximately follow gravitational shear motions,
noticeable deviations are also found (see
Section~\ref{sec:linear_shear_flow_non-zero_e_ext}). We provide
approximate solutions for this case below.

\subsection{Cloud morphology}
\label{sec:cloud_morphology}

Cloud morphology may reflect the origin of the gas motions
\citep[e.g.][]{meidt2018} and the physical mechanisms injecting energy
into the gas on cloud scales \citep[e.g.][]{koda2006}. To quantify the
morphology of the $141$ spatially-resolved clouds of NGC4429, we
considered their major and minor axes (and thus their axis ratios and
position angles with respect to the radial/galactocentric direction)
as measured in the plane of the sky (i.e.\ deprojected) in
Section~\ref{sec:larson_ellipsoidal}.

\begin{figure*}
  \includegraphics[width=0.9\textwidth]{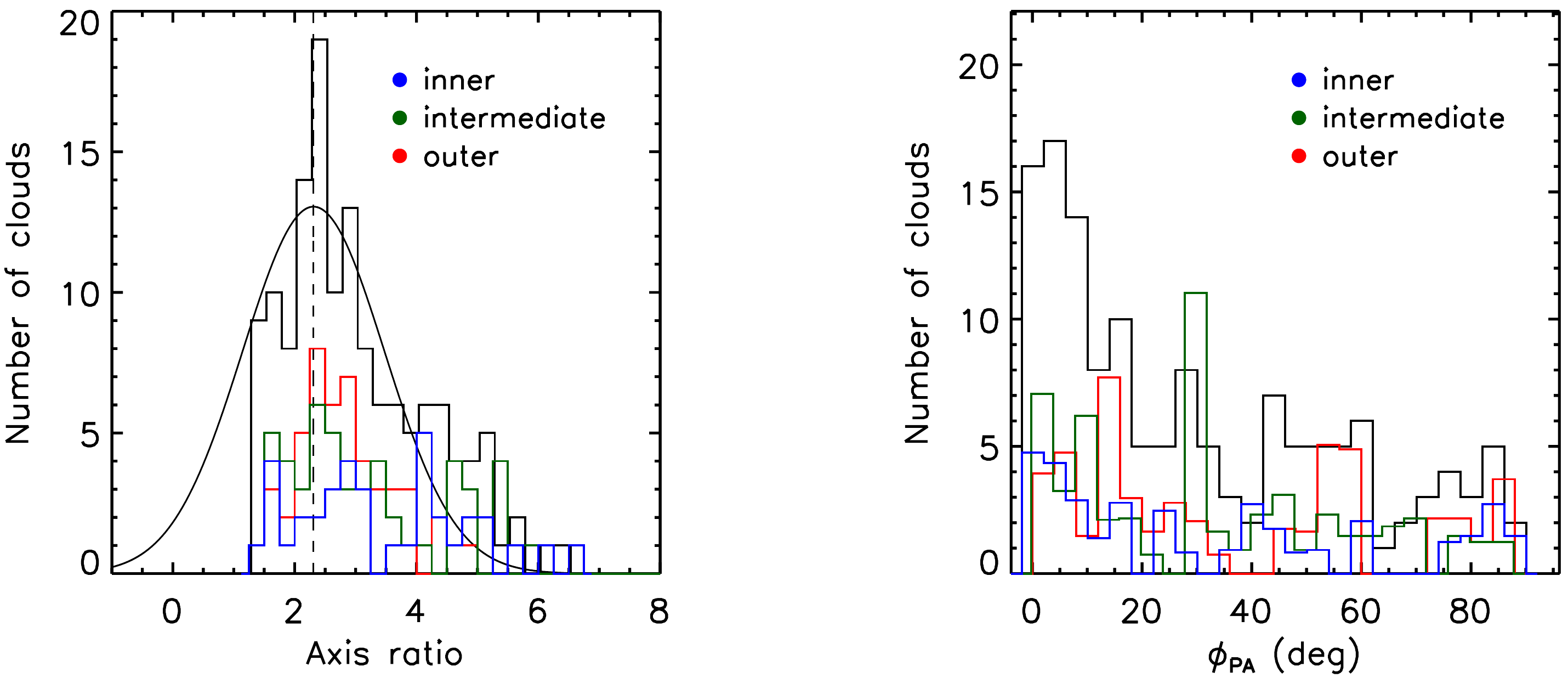}
  \caption{Distribution of deprojected axis ratios ({\it left}) and
    position angles $\phi_{\rm PA}$ between the (morphological) major
    axes and the direction to the galaxy centre ({\it right}) for the
    $141$ spatially-resolved clouds of NGC4429 (black histograms), and
    for only the clouds in the inner (blue histograms), intermediate
    (green histograms) and outer (red histograms) region of the
    galaxy, respectively. $\phi_{\rm PA}=0^\circ$ is the radial
    (i.e.\ galactocentric) direction, while $\phi_{\rm PA}=90^\circ$
    is the azimuthal direction. The black vertical dashed line in the
    left panel indicates the mean axis ratio derived from a Gaussian
    fit (black solid line).}
  \label{fig:cloud_morphology}
\end{figure*}

The distribution of the deprojected clouds' axis ratios
is shown in the left panel of Fig.~\ref{fig:cloud_morphology}. A
Gaussian fit to the distribution yields a mean of $2.3\pm0.2$,
suggesting that the clouds of NGC4429 are significantly elongated.
Moreover, the clouds in the inner and intermediate regions are more
elongated (mean axis ratio of $2.9$ and $2.6$, respectively) than the
clouds in the outer region (mean axis ratio of $2.2$).
  
The distribution of deprojected position angles $\phi_{\rm PA}$ is
shown in the right panel of Fig.~\ref{fig:cloud_morphology}. The
distribution peaks at $5^\circ$, with a mean
$\langle\phi_{\rm PA}\rangle\approx32^\circ$, confirming the
impression from Fig.~\ref{fig:gmcs} that
the clouds of NGC4429 are preferentially elongated in the radial
(i.e.\ galactocentric) direction. In fact, the clouds at small radii
tend to have smaller $\phi_{\rm PA}$, i.e.\ they are even more
preferentially elongated in the direction of the galaxy centre. The
mean angle $\phi_{\rm PA}$ of the clouds in the inner, intermediate
and outer region is $28^\circ$, $32^\circ$ and $34^\circ$,
respectively.

It is worth noting that the tendency for the clouds to align with the
radial direction could at least partially be due to an artefact of
{\cprops}. As {\cprops} tends to assign the pixels with the shortest
``distances'' (through the 3D data cube) to the same cloud, the clouds
identified by {\cprops} could be preferentially elongated along the
isovelocity contours, that are often nearly radial in NGC4429 (see
Fig.~\ref{fig:gmc_vfield}).
Having said that, we note that the clouds identified by {\cprops} in
other galaxies do not seem to exhibit such a tendency (e.g.\ NGC4526,
\citealt{utomo2015}; M33, \citealt{gratier2012}; NGC6946,
\citealt{wu2017}). This thus suggests that the observed trend of the
clouds of NGC4429 to be radially elongated could be real.


If the observed tendency is real, what are the physical mechanisms
that could cause such a strong radial elongation of the clouds of
NGC4429? It is interesting to note that, according to
Eq.~\ref{eq:beta_mod_shear},
\begin{equation}
  \label{eq:eq:beta_mod_shear_radius}
  \beta^{\rm mod}(R_{\rm c})=\beta^{\rm mod}(R_{\rm
    t})\left(\frac{R_{\rm c}}{R_{\rm
        t}}\right)^3\approx\left(\frac{R_{\rm c}}{R_{\rm t}}\right)^3~,
\end{equation}
which suggests that if $R_{\rm c}<R_{\rm t}$ then $\beta^{\rm mod}<1$
(and vice-versa) and if $R_{\rm c}>R_{\rm t}$ then $\beta^{\rm mod}>1$
(and vice-versa). If a cloud is primarily dominated by self-gravity
(i.e.\ $R_{\rm c}\ll R_{\rm t}$ or $\beta^{\rm mod}\ll1$), the effects
of external gravity are negligible and the cloud should be roughly
round.
On the other hand, if a cloud is largely dominated by external gravity
(i.e.\ $R_{\rm c}\gg R_{\rm t}$ or $\beta^{\rm mod}\gg1$), the cloud
should be elongated in the azimuthal direction due to strong shear
motions \citep{meidt2018}. However, the NGC4429 clouds are neither
round nor azimuthally elongated, suggesting their morphologies can not
be regulated by either self-gravity and/or external gravity alone.
  
It is thus interesting to investigate the geometry of a marginally
gravitationally-bound cloud, as is the case for the bulk of the
NGC4429 clouds, where both self-gravity and external gravity are
important (i.e.\ $R_{\rm c}\approx R_{\rm t}$ and
$\beta\approx\beta^{\rm mod}\approx1$; see
Sections~\ref{sec:larson_sigma_eff} and
\ref{sec:equilibrium_between_self-gravity_external-gravity}). For
this, we must solve the equations of motions given by
Eqs.~\ref{eq:aext_xy_rf_self}, that include both self-gravity and
external gravity terms. If we can calculate the motions of the fluid
elements near the external edge of each cloud, these motions will
define the approximate overall shapes of marginally
gravitationally-bound clouds.

Exact analytic solutions to Eqs.~\ref{eq:aext_xy_rf_self} may not be
possible, so we instead turn to a mathematical technique analogous to
perturbation theory to find approximate solutions. We define a new
dimensionless variable $\epsilon\equiv1$ and rewrite
Eqs.~\ref{eq:aext_xy_rf_self} as
\begin{equation}
  \label{eq:aext_xy_rf_self_particular}
  \left\{
    \begin{split}
      a_{\rm ext,x^\prime}^\prime & \approx T_0x^\prime+2\Omega_0v_{\rm gal,y^\prime}^\prime-\frac{\partial\Phi_{\rm sg}}{\partial x^\prime}~,\\
      a_{\rm ext,y^\prime}^\prime & \approx-2\Omega_0\epsilon v_{\rm gal,x^\prime}^\prime-\frac{\partial\Phi_{\rm sg}}{\partial y^\prime}~. 
    \end{split}
  \right.
\end{equation}
Approximate solutions to the above equations
can be written as \begin{equation}
  \label{eq:solution_rf_self_particular}
  \left\{
    \begin{split}
      x^\prime(t) & \approx x^{\prime\,(0)}(t)+\epsilon x^{\prime\,(1)}(t)~,\\
      y^\prime(t) & \approx y^{\prime\,(0)}(t)+\epsilon y^{\prime\,(1)}(t)~, 
    \end{split}
  \right.
\end{equation}
where analogously to perturbation theory we will refer to
$x^{\prime\,(0)}$ and $y^{\prime\,(0)}$ as the zeroth-order solutions
and to $x^{\prime\,(1)}$ and $y^{\prime\,(1)}$ as the first-order
solutions, although the latter are not necessarily smaller than the
former. Substituting Eqs.~\ref{eq:solution_rf_self_particular} into
Eqs.~\ref{eq:aext_xy_rf_self_particular}, we can separate the zeroth-
and first-order equations in $\epsilon$:
\begin{equation}
  \label{eq:aext_xy_rf_self_particular_zero}
  \left\{
    \begin{split}
      \ddot{x}^{\prime\,(0)} & =T_0x^{\prime\,(0)}+2\Omega_0\dot{y}^{\prime\,(0)}-\frac{\partial\Phi_{\rm sg}}{\partial x^\prime}~,\\
      \ddot{y}^{\prime\,(0)} & =-\frac{\partial\Phi_{\rm sg}}{\partial y^\prime}~,
    \end{split}
  \right.
\end{equation}
and 
\begin{equation}
  \label{eq:aext_xy_rf_self_particular_first}
  \left\{
    \begin{split}
      \ddot{x}^{\prime\,(1)} & =T_0x^{\prime\,(1)}+2\Omega_0\dot{y}^{\prime\,(1)}~,\\
      \ddot{y}^{\prime\,(1)} & =-2\Omega_0\dot{x}^{\prime\,(0)}~.
    \end{split}
  \right.
\end{equation}
We note that the solutions to
Eqs.~\ref{eq:aext_xy_rf_self_particular_zero} --
\ref{eq:aext_xy_rf_self_particular_first} provide solutions to
Eqs.~\ref{eq:aext_xy_rf_self_particular} for only a particular case,
and they are only approximate solutions as the second-order term in
$\epsilon$ is assumed to be negligible, i.e.\
$-2\Omega_0\epsilon^2\dot{x}^{\prime\,(1)}\approx0$ (assumptions we
will justify below).

We first solve the zeroth-order equations. While finding a general
analytic solution to the Eqs.~\ref{eq:aext_xy_rf_self_particular_zero}
is beyond the scope of this paper, there must exist a particular
cloudcentric radius $R_{\rm circ}$
where to zeroth order the fluid element has uniform circular motion of
a particular angular frequency $\omega_{\rm circ}$ (and arbitrary
phase $\psi$). We thus postulate
\begin{equation}
  \label{eq:solution_rf_self_zero_postulate}
  \left\{
    \begin{split}
      x^{\prime\,(0)}(t) & =R_{\rm circ}\cos(\omega_{\rm circ}t+\psi)~,\\
      y^{\prime\,(0)}(t) & =R_{\rm circ}\sin(\omega_{\rm circ}t+\psi)~. 
    \end{split}
  \right.
\end{equation}
Substituting Eqs.~\ref{eq:solution_rf_self_zero_postulate} into the
first equation of Eqs.~\ref{eq:aext_xy_rf_self_particular_zero}, we
find that the first and second terms on the RHS (i.e.\ the tidal and
Coriolis force terms) cancel out only for an angular frequency
$\omega_{\rm circ}=-2A_0$ (as $T=4A\Omega$). While shear does not lead
to circular motions, that is of course simply equal to
$v_{\rm shear}/x^{\prime\,(0)}$ (see Eq.~\ref{eq:v_shear}), i.e.\
\begin{equation}
  \label{eq:omega_circ}
  \omega_{\rm circ}=-2A_0=v_{\rm shear}/x^{\prime\,(0)}=\omega_{\rm shear}
\end{equation}
(the same $\omega_{\rm shear}$ defined in
Section~\ref{sec:linear_shear_flow_non-zero_e_ext}). The term on the
left-hand side and the third term on the RHS then lead to a condition
on $R_{\rm circ}$. Assuming the entire cloud mass is contained within
$R_{\rm circ}$, one obtains
\begin{equation}
  \label{eq:R_circ}
  \begin{split}
    R_{\rm circ} & =\left(\frac{GM}{4A_0^2}\right)^{1/3}\\
    & =2^{-\frac{1}{3}}\,R_{\rm t}
  \end{split}
\end{equation}
(see Eq.~\ref{eq:tidal_radius2}). The same condition is obtained by
substituting Eqs.~\ref{eq:solution_rf_self_zero_postulate} into the
second equation of Eqs.~\ref{eq:aext_xy_rf_self_particular_zero}. It
is trivial to show that at this radius, $v_{\rm shear}=-v_{\rm circ}$,
where $v_{\rm circ}$ is the circular velocity due to the cloud alone
(i.e.\
$v_{\rm shear}(R_{\rm circ})=-2A_0R_{\rm circ}=-(GM/R_{\rm
  circ})^{1/2}=-v_{\rm circ}(R_{\rm circ})$ for $R_{\rm circ}$ given
by Eq.~\ref{eq:R_circ}).

In other words, the (zeroth-order in $\epsilon$) solution to
Eqs.~\ref{eq:aext_xy_rf_self_particular_zero} is
\begin{equation}
  \label{eq:solution_rf_self_zero}
  \left\{
    \begin{split}
      x^{\prime\,(0)}(t) & =R_{\rm circ}\cos(-2A_0t+\psi)~,\\
      y^{\prime\,(0)}(t) & =R_{\rm circ}\sin(-2A_0t+\psi)~,
    \end{split}
  \right.
\end{equation}
where $R_{\rm circ}$ is given by Eq.~\ref{eq:R_circ}. This orbit is
thus intuitive to understand. In the cloud rotating frame, to zeroth
order, the fluid element will have uniform circular motion at the
radius $R_{\rm circ}$ where the shear velocity due to the external
(i.e.\ galactic) potential $v_{\rm shear}$ is equal to the cloud's own
circular velocity $v_{\rm circ}$, and thus the shear angular velocity
$\omega_{\rm shear}$ is equal to the cloud's angular velocity
$\omega_{\rm circ}$ (i.e.\
$v_{\rm shear}(R_{\rm circ})/R_{\rm circ}=v_{\rm circ}(R_{\rm
  circ})/R_{\rm circ}$).

Equally important, the radius of this circular orbit is very close to
the tidal radius and thus the external edge of the cloud
($R_{\rm circ}\approx0.8\,R_{\rm t}$ according to
Eq.~\ref{eq:R_circ}). Our solutions can thus indeed help us understand
the outer shapes of marginally-bound clouds.

We note that our zeroth-order solutions in $\epsilon$ above are
different from those found in \citet{goldreich1982} and
\citet{gammie1991} (Eqs.~\ref{eq:motion_solutions_shear_2} in this
paper). This is because we introduced $\epsilon$ in the azimuthal
Coriolis force term while they applied $\epsilon$ to the self-gravity
terms, and because we are considering a particular case where the
fluid element's shear velocity is equal to its circular velocity.
Equations~\ref{eq:solution_rf_self_zero} thus suggest that the fluid
element should have a circular orbit about the cloud's CoM (i.e.\ the
cloud should be round near its tidal radius) if the Coriolis force in
the azimuthal direction is neglected (the
$-2\Omega_0v_{\rm gal,x^\prime}^\prime$ term in the second equation of
Eqs.~\ref{eq:aext_xy_rf_self}). This is not surprising, since as we
have shown the Coriolis force ($2\Omega_0v_{\rm gal,y^\prime}^\prime$)
cancels out the tidal force ($T_0x^\prime$) in the radial direction,
hence only the cloud's self-gravity needs to be considered.

Having solved the zeroth-order equations of motion
(Eqs.~\ref{eq:aext_xy_rf_self_particular_zero}), we can now solve the
first-order equations in $\epsilon$
(Eqs.~\ref{eq:aext_xy_rf_self_particular_first}). Substituting the
first equation of Eqs.~\ref{eq:solution_rf_self_zero} into the second
equation of Eqs.~\ref{eq:aext_xy_rf_self_particular_first} and
imposing that the fluid element follows the zeroth-order solution at
$t=0$ (i.e.\ $\dot{y}^{\prime\,(1)}(t=0)=y^{\prime\,(1)}(t=0)=0$)
yields a solution for $y^{\prime\,(1)}(t)$. Substituting this in turn
into the first equation of
Eqs.~\ref{eq:aext_xy_rf_self_particular_first} and imposing again that
the fluid element follows the zeroth-order solution at $t=0$ (i.e.\
$\dot{x}^{\prime\,(1)}(t=0)=x^{\prime\,(1)}(t=0)=0$) yields a solution
for $x^{\prime\,(1)}(t)$. These first-order solution are
\begin{equation}
  \label{eq:solution_rf_self_one}
  \begin{split}
    x^{\prime\,(1)}(t) & \approx\frac{\Omega_0}{A_0}R_{\rm circ}\,\bigg[\frac{A_0}{A_0+\Omega_0}\big(\cos(\psi)\cosh(\sqrt{T_0}\,t)\\
    & \hspace*{17mm}-\sqrt{\frac{\Omega_0}{A_0}}\sin(\psi)\sinh(\sqrt{T_0}\,t)\big)\\
    & \hspace*{17mm}+\frac{\Omega_0}{A_0+\Omega_0}\cos(-2A_0t+\psi)-\cos(\psi)\bigg]~,\\
  y^{\prime\,(1)}(t) & \approx\frac{\Omega_0}{A_0}R_{\rm circ}\left[\sin(-2A_0t+\psi)+2A_0\cos(\psi)\,t-\sin(\psi)\right]~. 
  \end{split}
\end{equation}

We therefore have complete zeroth- and first-order solutions in
$\epsilon$ of the equations of motion
Eqs.~\ref{eq:aext_xy_rf_self_particular}, for fluid elements
originally in uniform circular rotation around the cloud's CoM at a
cloudcentric radius of $R_{\rm circ}$.

In practice, with our treatment in term of $\epsilon$, we have
neglected the firt-order Coriolis force term in the azimuthal
direction (i.e.\ the $-2\Omega_0v_{\rm gal,x^\prime}^{\prime\,(1)}$
term in the second equation of Eqs.~\ref{eq:aext_xy_rf_self}). Our
solutions will thus only be valid as long as this term remains small
compared to $\ddot{y}^\prime$. A comparison of these two terms shows
that this remains the case for times up to several
$t_{\rm shear}\equiv1/2A_0$ for nearly all phases $\psi$, when the
fluid element remains relatively close to the circle of radius
$R_{\rm circ}$.

\begin{figure}
  \includegraphics[width=0.95\columnwidth]{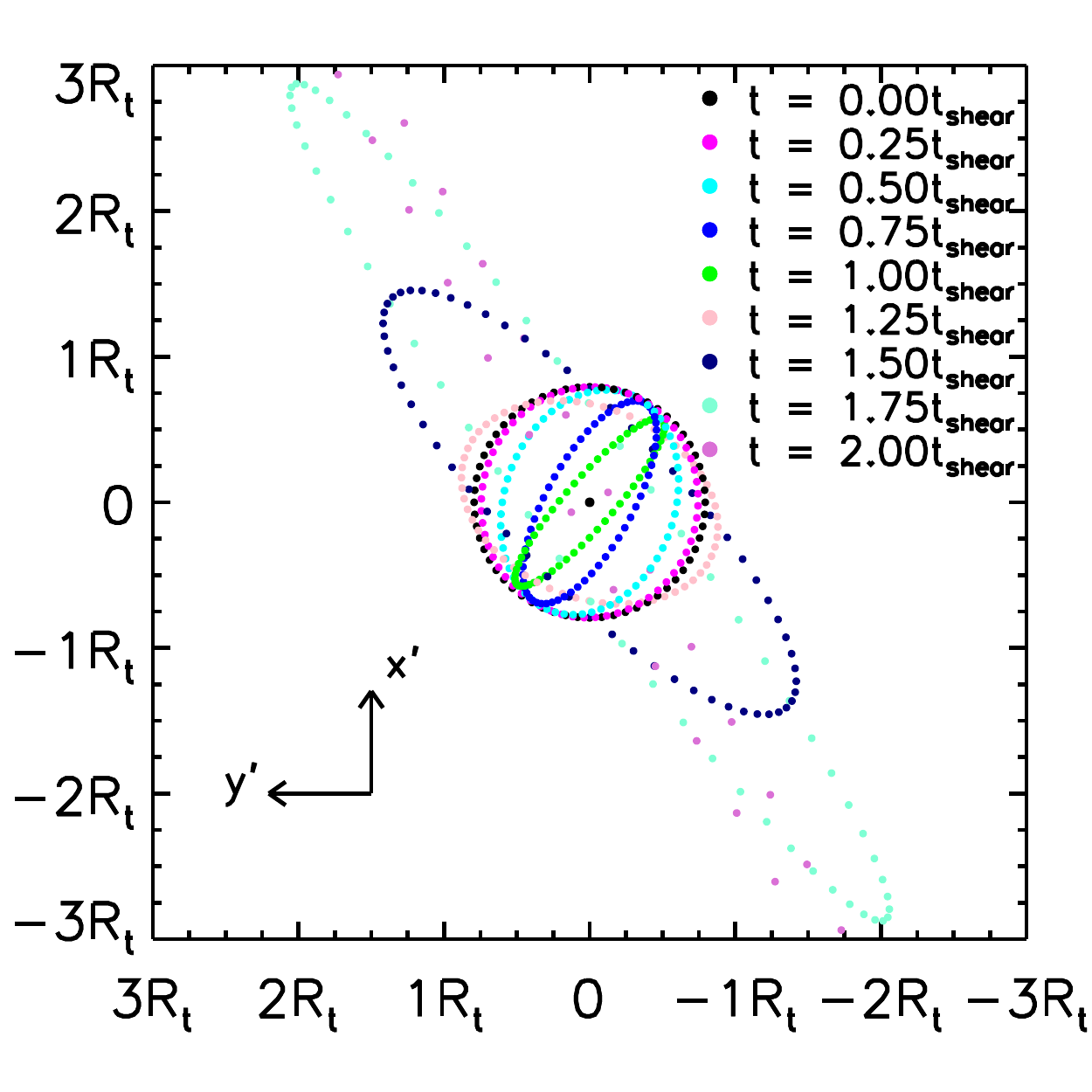}
  \caption{Orbits resulting from the zeroth- and first-order solutions
    to the equations of motion
    Eqs.~\ref{eq:aext_xy_rf_self_particular}, for fluid elements
    originally in uniform circular rotation around the cloud's CoM at
    a cloudcentric radius of $R_{\rm circ}$, in the rotating frame
    adopted in Appendix~\ref{app:eext_rotating_frame} (see
    Fig.~\ref{fig:rotating_frame}). Several orbits are shown, sampling
    all phases uniformly and colour-coded as a function of time. The
    initially circular ring becomes increasingly elongated over time
    and is nearly always elongated radially. The small black solid
    circle marks the galaxy centre while the large black dashed circle
    shows the original configuration of the fluid elements
    (zeroth-order solution).}
  \label{fig:orbit_shear}
\end{figure}

By sampling the phases $\psi$ uniformly, 
Figure~\ref{fig:orbit_shear} therefore shows how a circular ring of
matter initially at a cloudcentric radius $R_{\rm circ}$ evolves over
time (colour-coded), up to a time $t=2\,t_{\rm shear}$. As expected from
our solutions, particularly the diverging term in $x^{\prime\,(1)}(t)$
(see Eqs.~\ref{eq:solution_rf_self_one}), the fluid element orbits and
thus the ring become increasingly elongated over time, this almost
always in the radial direction (i.e.\ along $\hat{x}^\prime$), more so
but not exclusively at late times.


Therefore, contrary to naive expectations, clouds with sizes
$\approx R_{\rm t}$ and thus $\approx R_{\rm circ}$, that are
necessarily marginally-bound, should have shapes that are {\em
  radially} elongated. 
This state thus presumably represents an intermediate state between i)
small strongly-bound clouds that are expected to be spherical (due to
self-gravity) and ii) large unbound gas accumulations that are
expected to be azimuthally elongated (due to shear). In other words,
the general radial elongation of the NGC4429 clouds is fully
consistent with the fact that the clouds extend to typically their
tidal radii and are typically marginally gravitationally bound, with
roughly equal impact from self- and external gravity (i.e.\
$\beta\approx1$).

Interestingly, the numerical solutions of \citet{julian1966} and
\citet{binney2020} suggest a similar result. By solving
Eqs.~\ref{eq:aext_xy_rf_self} numerically, they derived density
patterns for a shear flow under both  self-  and external
gravity forces.
 While the outer contours at lower surface densities (where self-gravity is
much less important than external gravity) are elongated azimuthally
as expected, their results demonstrate that the innermost contours at
higher densities (where self-gravity  may be 
 as important as external gravity) are {\em radially}
elongated (see Figs.~$7$ -- $9$ in \citealt{julian1966} and Fig.~$9$
in \citealt{binney2020}). These works thus reinforce our approximate
analytical solutions above.

\subsection{Cloud scale height}
\label{sec:cloud_scale_height}

Our current analysis is based on the common assumption that the clouds
of NGC4429 are in vertical hydrostatic equilibria, i.e.\
$M(\sigma^2_{\rm gal,z}- b_{\rm e}\nu_0^2Z^2_{\rm c})\approx0$ (see
Eq.~\ref{eq:vertical_rms_velocity}). If this assumption is valid, we
can derive the scale height of each cloud ($Z_{\rm c}$) from estimates
of $\nu_0$ and $\sigma_{\rm gal,z}$. As before, $\nu_0$ is obtained
directly from our stellar mass model (i.e.\
$\nu_0^2=4\pi G\rho_{\ast,0}$, where $\rho_{\ast,0}$ is provided by
our our MGE model; see
Appendix~\ref{app:stellar_density_calculation}). According to
Eq.~\ref{eq:vrms_gs_los},
$\sigma^2_{\rm gal,z}\cos^2i\approx\sigma^2_{\rm gs,los}-\sigma^2_{\rm
  sg,los}$, so if $\sigma_{\rm sg,los}$ can indeed be derived from
Eq.~\ref{eq:internal_turbulent_velocity_dispersion} (i.e.\
$\sigma^2_{\rm sg,los}=\pi b_{\rm s}GR_{\rm c}\Sigma_{\rm gas}$), we
can obtain $\sigma^2_{\rm gal,z}$.

We note that the uncertainties of our measured physical quantities
(i.e.\ $\sigma^2_{\rm gs,los}$, $R_{\rm c}$ and $\Sigma_{\rm gas}$)
can be significant and prevent us from accurately estimating
$\sigma_{\rm gal,z}$ for individual clouds, as we find negative
$\sigma^2_{\rm gal,z}$ in a few cases. Instead, we therefore consider
only the average quantities for $\sigma^2_{\rm gal,z}$ and
$Z_{\rm c}$. We derive a mean
$\langle\sigma^2_{\rm gal,z}\rangle=18\pm2$~km$^2$~s$^{-2}$ for the
$141$ spatially-resolved clouds of NGC4429   (assuming
  $b_{\rm s}=\frac{1}{5}$ for spherical homogeneous clouds), whose
mean line-of-sight projection
($\langle\sigma^2_{\rm gal,z}\cos^2i\rangle\approx2$~km$^2$~s$^{-2}$)
is indeed relatively small compared to
$\langle\sigma^2_{\rm gs,los}\rangle\approx11$~km$^2$~s$^{-2}$ and
$\langle\sigma^2_{\rm sg,los}\rangle\approx8$~km$^2$~s$^{-2}$.

Utilising our derived
$\langle\sigma^2_{\rm gal,z}\rangle=18\pm2$~km$^2$~s$^{-2}$ and
$\langle\rho_{\ast,0}\rangle\approx33$~{\Msolar}~pc$^{-3}$ (and
further assuming $b_{\rm e}=\frac{1}{5}$ for spherical homogeneous
clouds), we derive a mean cloud scale height
$\langle Z_{\rm c}\rangle\approx7$~pc, that is clearly smaller than
the average cloud radius $\langle R_{\rm c}\rangle\approx16$~pc (see
Section~\ref{sec:probability_distribution_functions_gmc_properties}). Consequently,
the clouds of NGC4429 are not strictly spherical, but more likely to
be elongated (i.e.\ flattened) in the plane, if the clouds are indeed
in vertical hydrostatic equilibria.

\begin{figure}
  \includegraphics[width=0.95\columnwidth]{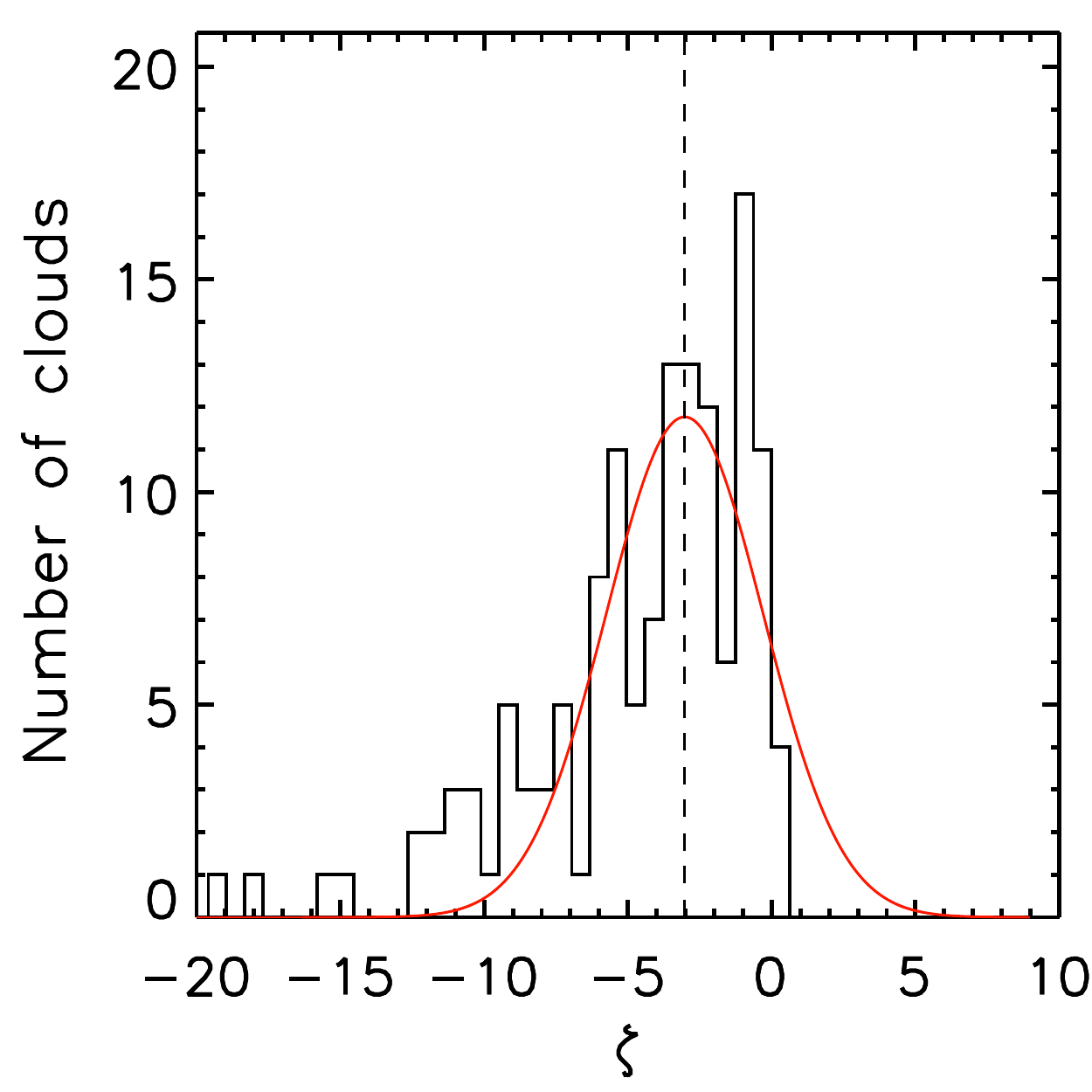}
  \caption{Distribution of $\zeta$, defined as the the ratio between
    the vertical contribution of the external potential to a cloud's
    energy budget and the (absolute value of the) cloud's
    self-gravitational energy, assuming the cloud is roundish (i.e.\
    $Z_{\rm c}=R_{\rm c}$), for the $141$ spatially-resolved clouds of
    NGC4429. The black dashed vertical line shows the mean of a
    Gaussian fit (red solid line) to the distribution.}
  \label{fig:zeta_distribution}
\end{figure}

The above analysis aimed to derive the scale heights of the clouds
$Z_{\rm c}$ by assuming that the clouds are in vertical hydrostatic
equilibria. However, we can also investigate the vertical equilibrium
state of the clouds by assuming $Z_{\rm c}=R_{\rm c}$ instead. For
such a roundish cloud, the contribution of the vertical component of
the external potential to the cloud's gravitational energy budget is
$\approx M(\sigma^2_{\rm gal,z}- b_{\rm e}\nu_0^2R^2_{\rm c})$. We
thus define $\zeta$ as the ratio between
$M(\sigma^2_{\rm gal,z}- b_{\rm e}\nu_0^2R^2_{\rm c})$ and
the (absolute value of the) self-gravitational energy of a roundish
cloud ($U_{\rm sg}=-3 b_{\rm s}GM^2/R_{\rm c}$):
\begin{equation}
  \zeta\equiv\frac{M(\sigma^2_{\rm gal,z}- b_{\rm e}\nu_0^2R^2_{\rm c})}{3 b_{\rm s}GM^2/R_{\rm c}}~.
\end{equation}
The distribution of $\zeta$ for the $141$ spatially-resolved clouds of
NGC4429 is presented in Fig.~\ref{fig:zeta_distribution}, where we
have assumed $b_{\rm s}=b_{\rm e}=\frac{1}{5}$ (spherical homogeneous
clouds) as usual. We find that $\zeta$ is not negligible, but
significantly below zero. A Gaussian fit to the $\zeta$ distribution
yields a mean $\langle\zeta\rangle=-1.97\pm0.55$. This implies that
the effect of external gravity on roundish clouds is to compress them
in the vertical direction. In other words, if the clouds of NGC4429
were roundish, the shear in the plane of the galaxy
would be overwhelmed by compression in the vertical direction, and the
net effect of external gravity would be to contribute to the
(vertical) collapse of the clouds (as
$\vert\langle\beta\rangle\vert<\vert\langle\zeta\rangle\vert$). This
probably explains why the clouds of NGC4429 appear to be flattened in
the plane.

Indeed, a cloud in a thin disc is more likely to exhibit an elongated
structure in the plane rather than be spherical, as the force applied
by the background galactic potential in the vertical direction far
exceeds the forcing experienced in the plane
\citep[e.g.][]{meidt2018}. In fact, such elongations of molecular
clouds in the equatorial plane have been observed in a sample of more
than $500$ MW clouds \citep{koda2006}.


\section{Conclusions}
\label{sec:conclusions}

Using our modified version of the {\cpropstoo} code, more robust and
efficient to identify GMCs in complex and crowded environments, and
$^{12}$CO$(J=3-2)$ ALMA observations at $14\times11$~pc$^2$
resolution, we identified $217$ GMCs ($141$ spatially resolved) in the
central molecular gas disc of the lenticular galaxy NGC4429. To
investigate the dynamical states of the GMCs, we developed and
utilised a modified Virial theorem that fully accounts for the impacts
of the background galactic potential. The main results are as follows:

\begin{enumerate}
\item The GMCs of NGC4429 appear to have smaller sizes ($7$ --
  $50$~pc), lower gaseous masses ($0.3$ -- $8\times10^5$~{\Msolar}),
  higher gaseous mass surface densities ($40$ --
  $650$~{\Msolar}~pc$^{-2}$) and higher observed linewidths ($2$ --
  $16$~km~s$^{-1}$) than the GMCs of the Milky Way disc and other
  Local Group galaxies.

\item Cloud properties exhibit several trends with galactocentric
  distance. Specifically, except for the three innermost resolved
  clouds at $R_{\rm gal}<100$~pc, the GMCs at small radii tend to have
  smaller sizes, lower gaseous masses, higher gaseous mass surface
  densities and higher observed linewidths than clouds farther
  out. However, we also find that all these quantities drop abruptly
  in the outermost region of the molecular gas disc
  ($R_{\rm gal}\gs375$~pc).

\item The GMCs of NGC4429 appear to be elongated (mean axis ratio of
  $\approx2.3 \pm 0.2$)  and
  are preferentially aligned in the radial direction (i.e.\ toward the
  galactic centre). The clouds also appear to be flattened in the
  plane of the galaxy.

\item The cloud mass distribution follows a truncated power law with
  slope $-2.18\pm0.21$ and truncation mass
  $(8.8\pm1.3)\times10^5$~{\Msolar}, suggesting most of the molecular
  mass of NGC4429 is in low-mass clouds. We find a slight variation of
  the mass spectrum with galactocentric distance, suggesting massive
  clouds are more favoured at intermediate radii
  ($220<R_{\rm gal}<330$~pc).

\item Strong velocity gradients are observed within individual GMCs
  ($\omega\approx0.05$ -- $0.91$~km~s$^{-1}$~pc$^{-1}$), significantly
  larger than those of GMCs in the MW and other Local Group
  galaxies. A steep size -- line width relation (with a power-law
  index $0.82\pm0.13$) and large observed Virial parameters
  ($\langle\alpha_{\rm obs,vir}\rangle\approx4.04\pm0.22$) are also
  found for the clouds of NGC4429. However, we argue the large
  velocity gradients, steep size -- line width relation and large
  observed Virial parameters are all a consequence of gas motions
  driven by the background galactic potential (i.e.\ local circular
  orbital rotation), not the clouds' self-gravity. To remove the
  contribution of galaxy rotation from the clouds' linewidths and
  derive linewidths quantifying turbulence only, we measure the
  gradient-subtracted linewidths of the clouds $\sigma_{\rm
    gs,los}$. Using this measure, an internal Virial equilibrium
  appears to have been reached betweem the clouds' turbulent kinetic
  energies ($E_{\rm turb}$) and their self-gravitational energies
  ($U_{\rm sg}$), i.e.\
  $\langle\alpha_{\rm sg,vir}\rangle\approx\langle\alpha_{\rm
    gs,vir}\rangle\approx1.28\pm0.04$.

\item However, we argue that neither $\alpha_{\rm obs,vir}$ nor
  $\alpha_{\rm sg,vir}$ reflects the true dynamical state of a
  cloud. We thus discuss and revisit the conventional Virial theorem,
  deriving a modified theorem that explicitly takes into account both
  the self-gravity of the clouds and the effects of the external
  (galactic) gravitational potential in the vertical direction and the
  plane separately. This allows us to define an effective velocity
  dispersion $\sigma_{\rm eff,los}$ and an effective Virial parameter
  $\alpha_{\rm eff,vir}\equiv\alpha_{\rm sg,vir}+\frac{E_{\rm
      ext}}{\vert U_{\rm sg}\vert}$, that provide straightforward
  measurable diagnostics of cloud boundedness in the presence of a
  non-negligeable external potential.
  
\item Using our new diagnostics, we find the contributions of external
  gravity to the clouds' energy budgets $E_{\rm ext}$ are generally
  much larger than zero. This is because the bulk motions of the
  clouds are dominated by gravitational shear motions rather than
  epicyclic motions.
  The clouds of NGC4429 are in a critical state in which the energy
  contributed by external gravity $E_{\rm ext}$ is approximately equal
  to the self-gravitational energy, i.e.\
  $\frac{E_{\rm ext}}{\vert U_{\rm sg}\vert}\approx 1$. As such, the
  clouds are not virialised but remain marginally gravitationally
  bound, with a mean effective Virial parameter
  ($\langle\alpha_{\rm eff,vir}\rangle\approx2.15\pm0.12$ and
  $\langle\alpha^{\rm mod}_{\rm eff,vir}\rangle\approx2.02\pm0.03$)
  close to the threshold between gravitationally-bound and unbound
  objects ($\alpha_{\rm vir,crit}=2$). This is also true when the
  elongated shapes of the clouds are taken into account
  ($\langle\alpha_{\rm eff,vir}\rangle\approx2.65\pm0.15$ and
  $\langle\alpha^{\rm mod}_{\rm eff,vir}\rangle\approx2.46\pm0.06$ for
  ellipsoidal clouds). As the clouds appear to reach an equilibrium
  between self-gravity and external gravity, they also have sizes
  consistent with their tidal radii (i.e.\
  $R_{\rm c}\approx R_{\rm t} $) and are radially elongated (with an
  average axis ratio of $\approx2$). Overall, external gravity appears
  to be as important as self-gravity to regulate the morphologies,
  dynamics and thus ultimately the fates of the clouds.


\item Galactic rotational shear appears to play a dominant role
    to regulate the properties of the clouds of NGC4429. Our shear 
    model predicts that, as rotational
    shear increases, the contribution of external gravity to a cloud's
    energy budget $E^{\rm mod}_{\rm ext}$ also increases and the cloud becomes less bound,
    leading to a maximum size (or equivalently a minimum gaseous mass
    surface density) for the cloud to remain marginally bound:
    $R_{\rm shear}\approx3\pi b_{\rm s}G\Sigma_{\rm gas}/4 b_{\rm
      e}A_0^2$
    ($\Sigma_{\rm shear}\approx4 b_{\rm e}A_0^2R_{\rm c}/3\pi b_{\rm
      s}G$), that matches very well the observed sizes of the clouds
    of NGC4429.
\end{enumerate}

\section*{Acknowledgements}

We thank  Michele Cappellari, Marc Sarzi,
Christopher McKee and James Binney for valuable
discussions. MB was supported by STFC consolidated grant 'Astrophysics
at Oxford" ST/H002456/1 and ST/K00106X/1. MC acknowledges support from
a Royal Society University Research Fellowship. TAD acknowledges
support from a STFC Ernest Rutherford Fellowship. MDS acknowledges
support from a STFC DPhil studentship ST/N504233/1. KO was supported
by Shimadzu Science and Technology Foundation. This publication arises
from research funded by the John Fell Oxford University Press Research
Fund.

This paper makes use of the following ALMA data:
ADS/JAO.ALMA\#2013.1.00493.S. ALMA is a partnership of ESO
(representing its member states), NSF (USA) and NINS (Japan), together
with NRC (Canada) and NSC and ASIAA (Taiwan) and KASI (Republic of
Korea), in cooperation with the Republic of Chile. The Joint ALMA
Observatory is operated by ESO, AUI/NRAO and NAOJ. This paper also
makes use of observations made with the NASA/ESA Hubble Space
Telescope, and obtained from the Hubble Legacy Archive, which is a
collaboration between the Space Telescope Science Institute
(STScI/NASA), the Space Telescope European Coordinating Facility
(ST-ECF/ESA) and the Canadian Astronomy Data Centre
(CADC/NRC/CSA). This research has made use of the NASA/IPAC
Extragalactic Database (NED) which is operated by the Jet Propulsion
Laboratory, California Institute of Technology, under contract with
the National Aeronautics and Space Administration.

\section*{DATA  AVALABILITY}
The data underlying this article  is available in the ALMA archive (\url{https://almascience.eso.org/asax/})
under project code 2013.1.00493.S.




\bibliographystyle{mnras}
\bibliography{references} 


\clearpage

\appendix


\section{Modified Virial Theorem}
\label{app:modified_Virial_theorem}

\begin{figure}
  \includegraphics[width=0.95\columnwidth]{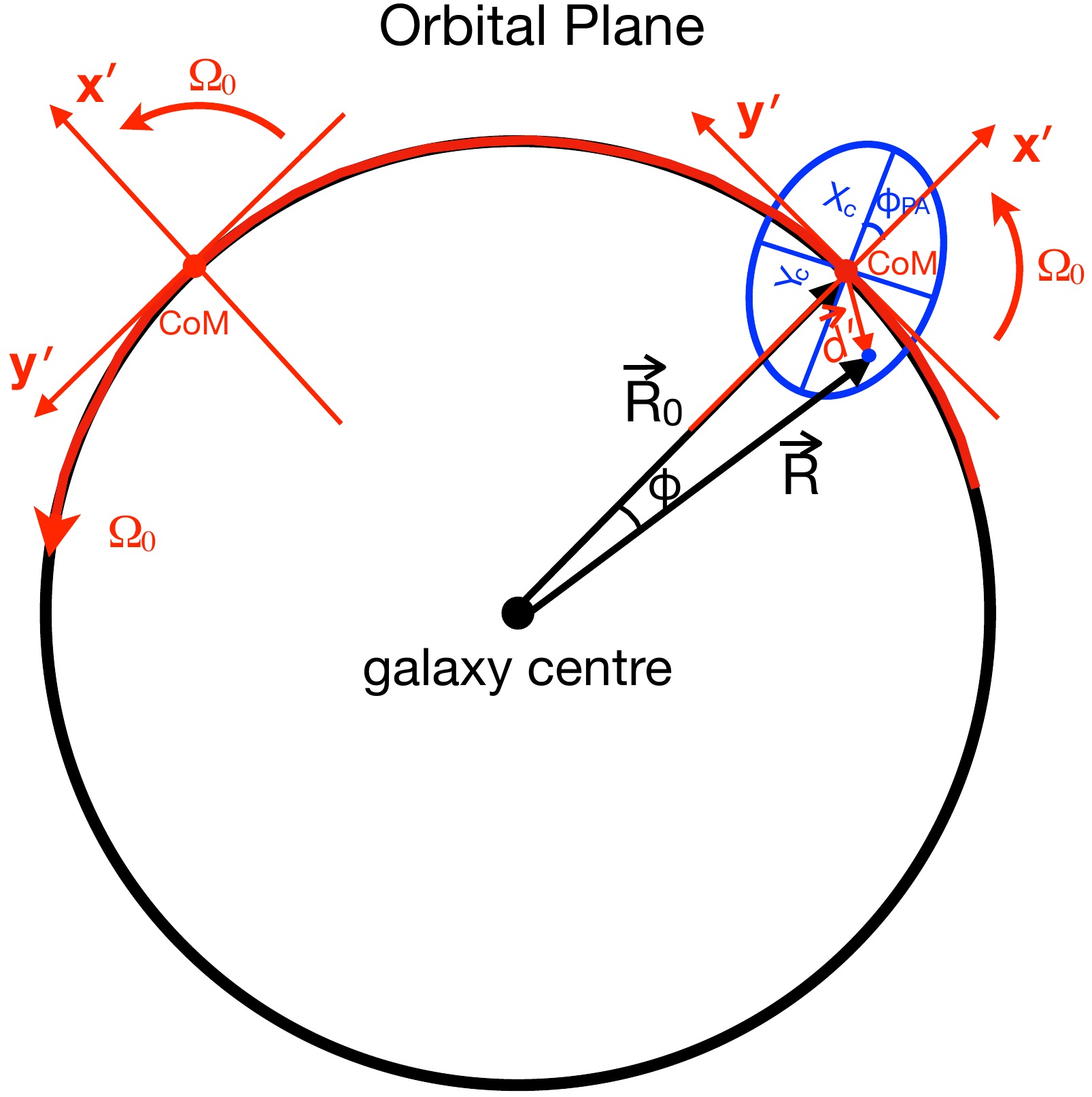}
  \caption{Schematic diagram of our rotating frame of reference in the
    orbital plane (i.e.\ the mid-plane of the galaxy disc). This
    rotating frame is a local Cartesian coordinate system centred at
    the cloud's CoM, that both orbits around the galaxy centre with
    the cloud's CoM (with angular velocity $\Omega_0$, the circular
    orbital angular velocity of the cloud's CoM) and rotates on itself
    (with the same angular velocity $\Omega_0$),
    such that the $x^\prime$ axis always points in the direction of
    increasing galactocentric radius and the $y^\prime$ axis always
    points in the direction of the orbital rotation at the cloud's
    CoM.  We assume a homogenous ellipsoidal cloud, whose semi-major
    axis $X_{\rm c}$ and semi-minor axis $Y_{\rm c}$ are located in
    the orbital plane. The semi-major axis $X_{\rm c}$ makes an angle
    $\phi_{\rm PA}$ with respect to the radial (i.e.\ $\hat{x}^\prime$
    or $\vec{R}_0$) direction.}
  \label{fig:rotating_frame}
\end{figure}

Our goal in this appendix is to derive a modified Virial theorem
(MVT), that encompasses not only a cloud's self-gravity, but also the
effects of the external (i.e.\ galactic) potential. We thus envision
each cloud as a continuous structure with well-defined borders in
position- and velocity-space, located in a rotating gas disc with a
circular velocity determined by an axisymmetric background galactic
gravitational potential ($\Phi_{\rm gal}$). We assume each cloud has a
homogeneous density distribution and an ellipsoidal geometry. The
cloud's centre of mass (CoM) and its two semi-axes (semi-major axis
$X_{\rm c}$ and semi-minor axis $Y_{\rm c}$) are assumed to be located
in the orbital plane (i.e.\ the mid-plane of the galaxy disc; see
Fig.~\ref{fig:rotating_frame}). We assume each fluid element of a
cloud experiences two kinds of motions: (1) random turbulent motions
(velocity dispersion $\sigma_{\rm sg}$) arising from self-gravity
(i.e.\ the cloud's own gravitational potential $\Phi_{\rm sg}$) and
(2) bulk gravitational motions (velocity $\vec{v}_{\rm gal}$ and root
mean square (RMS) velocity $\sigma_{\rm gal} $) arising from the
external gravity (i.e.\ the galactic gravitational potential
$\Phi_{\rm gal}$). We neglect thermal motions, as they are often
small compared to turbulent motions in a cold gas cloud
\citep[e.g.][]{fleck1980}. The turbulent motions due to self-gravity
are expected to be quasi-isotropic in three dimensions
\citep{field2008, ballesterosparedes2011}, while the gas motions
induced by the external gravitational potential are often
non-isotropic \citep{meidt2018}. We assume the cloud's own
gravitational potential $\Phi_{\rm sg}$ to be (statistically)
independent of the local external gravitational potential
$\Phi_{\rm gal}$, and the motions due to self-gravity
($\sigma_{\rm sg}$) to be uncorrelated with the motions due to
external gravitational potential ($\sigma_{\rm gal}$), as suggested by
\citet{meidt2018}. We ignore external pressure and magnetic fields,
and consider only the effects of self-gravity and external gravity.

Assuming the surface terms are negligible \citep{larson1981}, the
general form of the Virial theorem for a cloud is
\begin{equation}
  \label{eq:general_Virial_theorem}
  \frac{\ddot{I}}{2}=2E_{\rm k}+\int_{V}{\left(\vec{a}(\vec{d})\cdot\vec{d}\right)dm}
\end{equation}
(see e.g.\ Eqs.~14.6 and 14.7 of \citealt{lequeux2005}), where $I$ is
the cloud's moment of inertia, $E_{\rm k}$ its total kinetic energy,
$\vec{d}$ the position vector of a fluid element inside the cloud with
respect to the cloud's CoM, $\vec{a}\equiv\ddot{\vec{d}}$ the
acceleration of the fluid element inside the cloud,
$dm \equiv \rho dV$ the mass of the fluid element, and the integral is
taken over all fluid elements within the volume $V$ of the cloud of
total mass $M$ ($\,\int_{V}{dm}=M\,$). We use $\vec{d}$ rather than
the usual variable $\vec{r}$ to avoid confusion with the position
vector (with respect to the galactic centre) in the plane of the disc
$\vec{R}$ and its associated magnitude $R$, where $\vec{R}_0$ and
$R_0$ are evaluated at the cloud's CoM (see
Fig.~\ref{fig:rotating_frame}). The equilibrium condition associated
with Eq.~\ref{eq:general_Virial_theorem} should be that the
time-averaged $\ddot{I}(t)$ is equal to zero, i.e.\
$\langle\ddot{I}(t)\rangle=0$ \citep{mckee1999,binney2008}.  However,
it is unclear how one can evaluate the resulting long-term average if
the system is not in a time-independent state. We therefore adopt
instead the instantaneous equilibrium condition $\ddot{I}=0$, commonly
adopted across several works \citep[e.g.][]{lequeux2005,
  ballesterosparedes2006b}. As such, $\ddot{I}>0$ indicates that the
cloud is expanding, while $\ddot{I}<0$ indicates that the cloud is
contracting \citep{ballesterosparedes2006b}.

The above Virial equation can be split into two independent parts
based on our assumptions that $\Phi_{\rm sg}$ and $\Phi_{\rm gal}$ are
independent (i.e.\ $\vec{a}=-\nabla\Phi_{\rm sg}+\vec{a}_{\rm ext}$,
where $\vec{a}_{\rm ext}$ is the external acceleration due to galactic
forces only) and $\sigma_{\rm sg}$ and $\sigma_{\rm gal}$ are
uncorrelated (i.e.\
$E_{\rm k}=\frac{1}{2}M(\sigma^2_{\rm sg}+\sigma^2_{\rm gal}$)):
\begin{gather}
  \label{eq:separate_Virial_equation}
  \begin{split}
    \frac{\ddot{I}_{\rm}}{2}= & \left[3M\sigma_{\rm sg,los}^2+\int_{V}{\left(-\nabla\Phi_{\rm sg}(\vec{d})\cdot\vec{d}\right)dm}\right] \\
    + & \left[M\sigma_{\rm gal}^2+\int_{V}{\left(\vec{a}_{\rm ext}(\vec{d})\cdot\vec{d}\right)dm}\right]~,
  \end{split}
\end{gather}
where $\sigma_{\rm sg,los}$ is the line-of-sight (i.e.\
one-dimensional) turbulent velocity dispersion due to self-gravity
($\sigma_{\rm sg,los}^2\equiv\frac{1}{3}\sigma_{\rm sg}^2$) and
$\sigma_{\rm gal}$ the RMS velocity of gravitational motions
associated with external gravity
($\sigma_{\rm gal}\equiv\frac{1}{M}\int_{V}{(\vec{v}_{\rm
    gal}-\overline{\vec{v}}_{\rm gal})^2\,dm}$, where
$\overline{\vec{v}}_{\rm gal}$ is the mean velocity of the cloud's
gravitational motions due to external gravity). The first term in
square brackets on the right-hand side (RHS) of
Eq.~\ref{eq:separate_Virial_equation} comprises the energy terms
regulated by self-gravity, while the second term in square brackets
contains the contribution of external gravity to the cloud's energy
budget $E_{\rm ext}$.


\textbf{\textit{Self-gravity.}} The integration of the self-gravity
term on the RHS of Eq.~\ref{eq:separate_Virial_equation} is
straightforward:
\begin{equation}
  \label{eq:self_gravity_integration}
  \int_{V}{\left(-\nabla\Phi_{\rm sg}(\vec{d})\cdot\vec{d}\right)dm}=-\frac{3b_{\rm s}GM^2}{R_{\rm c}}~,
\end{equation}
where $G$ is the gravitational constant, $R_{\rm c}$ the measured
cloud's radius ($R_{\rm c}\equiv\sqrt{X_{\rm c}Y_{\rm c}}\,$) and
$b_{\rm s}$ a geometrical factor that quantifies the effects of
inhomogeneities and/or non-sphericity of the cloud mass distribution
on its self-gravitational energy. For a cloud in which the isodensity
contours are homoeoidal ellipsoids,
$b_{\rm s}=b_{\rm s_1}b_{\rm s_2}$, where $b_{\rm s_1}$ quantifies the
effects of the inhomogeneities and $b_{\rm s_2}$ those of the
ellipticity. \citet{bertoldi1992} derived
$b_{\rm s_1}=\frac{(1-\psi/3)}{(5-2\psi)}$ for a cloud with a radial
mass volume density profile $\rho(r)\propto r^{-\psi}$, while
$b_{\rm s_2}=\frac{R_{\rm m}}{R_{\rm d}}\frac{\arcsin(e)}{e}$ for an
ellipsoidal cloud, where $R_{\rm m}$ is the observed (i.e.\ projected)
cloud radius averaged over all possible cloud orientations (i.e.\
averaged over $4\pi$ steradians), $R_{\rm d}$ is the deprojected cloud
radius and $e$ is the cloud's eccentricity
($e\equiv\sqrt{1-(Y_{\rm c}/X_{\rm c})^2}\,$). The
$R_{\rm m}/R_{\rm d}$ ratio depends on the cloud's aspect ratio
$Z_{\rm c}/\sqrt{X_{\rm c}Y_{\rm c}}\,$, where $Z_{\rm c}$ is the
cloud's scale height, and $R_{\rm m}/R_{\rm d}=1$ when
$Z_{\rm c}/\sqrt{X_{\rm c}Y_{\rm c}}=1$ (see Fig.~2 in
\citealt{bertoldi1992}). For a homogeneous spherical cloud,
$b_{\rm s}=\frac{1}{5}$.

An equation suitable for a cloud regulated by self-gravity only is
thereby obtained:
\begin{equation}
  \label{eq:virial_theorem_self_appendix}
  \frac{\ddot{I}}{2}=3M\sigma_{\rm sg,los}^2-\frac{3b_{\rm s}GM^2}{R_{\rm c}}~.
\end{equation}
For a self-gravitating cloud in equilibrium (i.e.\ $\ddot{I}=0$), this
yields
\begin{equation}
  \label{eq:sigma_sg_appendix}
    \sigma_{\rm sg,los}^2=b_{\rm s}GM/R_{\rm c}~,
\end{equation}
or equivalently
\begin{equation}
  M=\frac{\sigma_{\rm sg,los}^2R_{\rm c}}{b_{\rm s}G}\equiv M_{\rm sg,vir}
\end{equation}
(cf.\ Eq.~\ref{eq:Virial_mass}).

\textbf{\textit{External gravity.}} The contribution of external
gravity to a cloud's energy budget is given by the second term in
square brackets on the RHS of Eq.~\ref{eq:separate_Virial_equation}:
\begin{equation}
  \label{eq:e_ext_appendix}
  E_{\rm ext}=M\sigma_{\rm gal}^2+\int_{V}{\left(\vec{a}_{\rm ext}(\vec{d})\cdot\vec{d}\right)dm}~.
\end{equation}
If we assume the gravitational motions in the plane to be separable
from those in the vertical direction (perpendicular to the orbital
plane), the above equation can easily be separated into two parts, one
in the vertical direction ($E_{\rm ext,z}$) and the other in the
orbital plane ($E_{\rm ext,plane}$):
\begin{gather}
  \label{eq:split_eext_nrf}
  \begin{split}
    E_{\rm ext}= & \underbrace{\left(M\sigma_{\rm gal,z}^2+\int_{V}{\left(a_{\rm ext,z}d_{\rm z}\right)dm}\right)}_{\rm external,~vertical~direction} \\ 
    + & \underbrace{\left(M\sigma_{\rm gal,plane}^2+\int_{V}{\left(\vec{a}_{\rm ext, plane}\cdot\vec{d}_{\rm plane}\right)dm}\right)}_{\rm external,~plane}~,
  \end{split}
\end{gather}
where $\sigma_{\rm gal,z}^2$, $a_{\rm ext,z}$ and $d_{\rm z}$ are the
components of $\sigma_{\rm gal}^2$, $\vec{a}_{\rm ext}$ and $\vec{d}$
along the vertical direction, respectively, and
$\sigma_{\rm gal,plane}^2$, $\vec{a}_{\rm ext,plane}$ and
$\vec{d}_{\rm plane}$ are the components of $\sigma_{\rm gal}^2$,
$\vec{a}_{\rm ext}$ and $\vec{d}$ in the orbital plane, respectively.

We derive $E_{\rm ext,z}$ (the first term on the RHS of
Eq.~\ref{eq:split_eext_nrf}) in Appendix~\ref{app:eext_vertical} (see
Eq.~\ref{eq:eext_z_nrf}) for a homogenous ellipsoidal cloud whose
semi-major axis $X_{\rm c}$ and semi-minor axis $Y_{\rm c}$ are
located in the orbital plane, yielding
\begin{equation}
  \label{eq:eext_z_final}
  E_{\rm ext,z}\approx M\left(\sigma_{\rm gal,z}^2-b_{\rm e}\nu_0^2Z_{\rm c}^2\right)~,
\end{equation}
where as before $Z_{\rm c}$ is the scale height of the cloud
($Z_{\rm c}=\sqrt{X_{\rm c}Y_{\rm c}}\equiv R_{\rm c}$ for a spherical
cloud), $\nu_0^2\equiv4\pi G\rho_{\ast,0}$ (formally the total mass
volume density evaluated at the cloud's CoM, but we use here
$\rho_{\ast,0}$, the stellar mass volume density $\rho_\ast$ evaluated
at the cloud's CoM using our MGE model, as it is accurately
constrained; see Appendix~\ref{app:stellar_density_calculation}), and
$b_{\rm e}$ is a geometrical factor that quantifies the effects of
inhomogeneities of the cloud mass distribution (analogously to
$b_{\rm s_1}$ but for the external gravity term;
$b_{\rm e}=\frac{1}{5}$ again for a homogeneous cloud).

Similarly, we derive $E_{\rm ext,plane}$ (the second term on the RHS
of Eq.~\ref{eq:split_eext_nrf}) for a homogenous ellipsoidal cloud in
Appendix~\ref{app:eext_rotating_frame} (see
Eq.~\ref{eq:eext_plane_rf}), yielding
\begin{gather}  
  \begin{split}
  \label{eq:eext_plane_final}
    E_{\rm ext,plane} & \approx M\Big(\sigma_{\rm gal,r}^2+\sigma_{\rm gal,t}^2+b_{\rm e}T_0\,(X_{\rm c}^2\cos^2\phi_{\rm PA}+Y_{\rm c}^2\sin^2\phi_{\rm PA})\\
    & ~~~~~~~~~~ -b_{\rm e}\Omega_0^2\,(X_{\rm c}^2+Y_{\rm c}^2)\Big)~,
  \end{split}
\end{gather}
where $\sigma_{\rm gal,r}^2$ and $\sigma_{\rm gal,t}^2$ are the RMS
velocities of gas motions due to external gravity in respectively the
radial (i.e.\ $\hat{r}$, the direction pointing from the galaxy centre
to the cloud's CoM, thus parallel to $\vec{R}_0$) and the azimuthal
(i.e.\ $\hat{t}$, the direction along the orbital rotation, thus
perpendicular to $\hat{r}$ and $\vec{R}_0$) direction as measured in
an inertial frame (i.e.\ by a distant observer),
$(\sigma_{\rm gal,r}^2+\sigma_{\rm gal,t}^2)$ is thus the RMS velocity
of in-plane gravitational motions (i.e.\ $\sigma_{\rm gal,plane}^2$)
measured in the inertial frame, $\phi_{\rm PA}$ is the angle the
semi-major axis $X_{\rm c}$ makes with respect to the radial (i.e.\
$\hat{r}$ or $\vec{R}_0$) direction (see
Fig.~\ref{fig:rotating_frame}), $\Omega_0$ is the circular orbital
angular velocity of the cloud's CoM,
$T_0\equiv-R\frac{d\Omega^2(R)}{dR}\vert_{R=R_0}$
\citep[e.g.][]{stark1978} is the tidal acceleration per unit length in
the radial direction $T$ evaluated at the cloud's CoM, and $R$ is the
galactocentric distance in the plane of the disc while $R_0$ is that
at the cloud's CoM. We note that here and throughout, $\Omega(R)$ is
the theoretical quantity
$\Omega(R)\equiv\sqrt{\frac{1}{R}\frac{d\Phi_{\rm gal}}{dR}}$ defined
by the galaxy potential $\Phi_{\rm gal}$, i.e.\ it is the angular
velocity of a fluid element moving in perfect circular motion
($\Omega(R)=V_{\rm circ}(R)/R$, where $V_{\rm circ}(R)$ is the
circular velocity curve) rather than the observed angular velocity of
the fluid element ($V_{\rm rot}(R)/R$, where $V_{\rm rot}$ is the
observed rotation curve). For an axisymmetric cloud (i.e.\
$X_{\rm c}=Y_{\rm c}=R_{\rm c}$), we then have
\begin{equation}  
  \label{eq:eext_plane_final_sphere}
  E_{\rm ext,plane}\approx M\Big(\sigma_{\rm gal,r}^2+\sigma_{\rm gal,t}^2+b_{\rm e}(T_0-2\Omega_0^2)\,R_{\rm c}^2\Big)~.
\end{equation}

Combining Eqs.~\ref{eq:eext_z_final} and 
\ref{eq:eext_plane_final}, we obtain the total contribution
of external gravity to a cloud's energy budget:
\begin{gather}
  \begin{split}
    \label{eq:eext}
    E_{\rm ext} = & \,E_{\rm ext,z}+E_{\rm ext,plane} \\ 
    \approx & \,M\left(\sigma_{\rm gal,z}^2-b_{\rm e}\nu_0^2Z_{\rm c}^2\right)+M\left(\sigma_{\rm gal,r}^2+\sigma_{\rm gal,t}^2\right) \\
    + & \,M\left(b_{\rm e}T_0\,(X_{\rm c}^2\cos^2\phi_{\rm PA}+Y_{\rm c}^2\sin^2\phi_{\rm PA})-b_{\rm e}\Omega_0^2\,(X_{\rm c}^2+Y_{\rm c}^2)\right)~.
  \end{split}
\end{gather}
For an axisymmetric cloud (i.e.\ $X_{\rm c}=Y_{\rm c}=R_{\rm c}$), we
then have
\begin{equation}  
  \label{eq:eext_sphere}
  E_{\rm ext}\approx M\left(\sigma_{\rm gal,z}^2-b_{\rm e}\nu_0^2Z_{\rm c}^2\right)+M\Big(\sigma_{\rm gal,r}^2+\sigma_{\rm gal,t}^2+b_{\rm e}(T_0-2\Omega_0^2)\,R_{\rm c}^2\Big)~.
\end{equation}


\textbf{\textit{Total.}} Substituting
Eqs.~\ref{eq:virial_theorem_self_appendix} and \ref{eq:eext} into
Eq.~\ref{eq:separate_Virial_equation}, we obtain our final MVT for a
homogenous ellipsoidal cloud:
\begin{gather}
  \label{eq:modified_Virial_theorem_appendix}
  \begin{split}
    \frac{\ddot{I}}{2} = & \left[3M\sigma_{\rm sg,los}^2-3b_{\rm s}GM^2/R_{\rm c}\right]+E_{\rm ext} \\
    \approx & \left[3M\sigma_{\rm sg,los}^2-3b_{\rm s}GM^2/R_{\rm c}\right]+\Bigg[M\left(\sigma_{\rm gal,z}^2-b_{\rm e}\nu_0^2Z_{\rm c}^2\right) \\
  & ~~+M\Big(\sigma_{\rm gal,r}^2+\sigma_{\rm gal,t}^2+b_{\rm e}T_0\,(X_{\rm c}^2\cos^2\phi_{\rm PA}+Y_{\rm c}^2\sin^2\phi_{\rm PA})\\
    & ~~~~~~~~~~~-b_{\rm e}\Omega_0^2\,(X_{\rm c}^2+Y_{\rm c}^2)\Big)\Bigg]~.
  \end{split}
\end{gather}
For an axisymmetric cloud (i.e.\ $X_{\rm c}=Y_{\rm c}=R_{\rm c}$), we
then have
\begin{gather}
  \label{eq:modified_Virial_theorem_sphere_appendix}
  \begin{split}
    \frac{\ddot{I}}{2} \approx & \underbrace{\left[3M\sigma_{\rm sg,los}^2-3b_{\rm s}GM^2/R_{\rm c}\right]}_{\rm self~gravity} \\
    + & \left[\underbrace{M(\sigma_{\rm gal,z}^2-b_{\rm e}\nu_0^2Z_{\rm c}^2)}_{\rm external,~vertical~direction}+\underbrace{M\left(\sigma_{\rm gal,r}^2+\sigma_{\rm gal,t}^2+b_{\rm e}(T_0-2\Omega_0^2)\,R_{\rm c}^2\right)}_{\rm external,~plane}\right]~.
  \end{split}
\end{gather}


\subsection{Calculating $E_{\rm ext,z}$}
\label{app:eext_vertical}

According to Eq.~\ref{eq:split_eext_nrf}, the contribution of external
gravity to a cloud's energy budget in the vertical direction is
\begin{equation}
  \label{eq:e_ext_z}
  E_{\rm ext,z}=M\sigma_{\rm gal,z}^2+\int_{V}{\left(a_{\rm ext,z}d_{\rm z}\right)dm}~.
\end{equation}
As we have assumed the cloud's CoM to be in the galaxy mid-plane, the
component of the acceleration $\vec{a}_{\rm ext}$ in the vertical
direction can be approximated to
\begin{gather}
  \label{eq:a_ext_z_appendix}
  \begin{split}
    a_{\rm ext,z} & =-\frac{\partial\Phi_{\rm gal}(\vec{d})}{\partial z}\\
    & \approx\,-\frac{\partial\Phi_{\rm gal}(\vec{d})}{\partial z}\bigg|_{z=0}-\frac{\partial^2\Phi_{\rm gal}(\vec{d})}{\partial^2z}\bigg|_{z=0}\,d_{\rm z}\\
    & \approx\,-\frac{\partial^2\Phi_{\rm gal}(\vec{d})}{\partial^2z}\bigg|_{z=0}\,d_{\rm z}~.
  \end{split}
\end{gather}
For a thin gas disc,
\begin{equation}
  \label{eq:vertical_gravity_force_1}
  \frac{\partial^2\Phi_{\rm gal}(\vec{d})}{\partial^2z}\bigg|_{z=0}\approx4\pi G\rho_{\ast}(R,z=0)~,
\end{equation}
and we further assume
\begin{equation}
  \label{eq:vertical_gravity_force_2}
  \frac{\partial^2\Phi_{\rm gal}(\vec{d})}{\partial^2z}\bigg|_{z=0}\approx4\pi G\rho_{\ast,0}~,
\end{equation}
formally the total mass volume density, but we use here
$\rho_\ast(R,z=0)$, the stellar mass volume density in the mid-plane
of the disc, that can be reliably estimated from observations (here
our MGE model; see Appendix~\ref{app:stellar_density_calculation}),
and again $\rho_{\ast,0}\equiv\rho_{\ast}(R=R_0,z=0)$ is evaluated at
the cloud's CoM.
As expected from Poisson's equation,
Eq.~\ref{eq:vertical_gravity_force_1} only applies to a (thin) disc
where the variations in the gravitational potential are larger in the
vertical direction than in the plane (i.e.\
$\frac{\partial^2\Phi_{\rm
    gal}(\vec{d})}{\partial^2z}\gg\frac{\partial^2\Phi_{\rm
    gal}(\vec{d})}{\partial^2r}$ and
$\frac{\partial^2\Phi_{\rm
    gal}(\vec{d})}{\partial^2z}\gg\frac{\partial^2\Phi_{\rm
    gal}(\vec{d})}{\partial^2t}$; see also \citealt{koyama2009} and
\citealt{meidt2018}). Again as expected, given that $\rho_\ast$ is
positive, the gravitational potential of the galaxy along the $z$ axis
always has a confining effect on the cloud, i.e.\ a fluid element
moving away from the cloud's CoM will always experience a restoring
force in the $z$ direction back toward the galactic (i.e.\ mid-)
plane.

With these expressions (Eqs.~\ref{eq:a_ext_z_appendix} and
\ref{eq:vertical_gravity_force_2}), the volume integral in the second
term on the right-hand side of Eq.~\ref{eq:e_ext_z} simplifies to
$\int_{V}{d_{\rm z}^2\,dm}$. For an ellipsoidal cloud with semi-major
axis $Z_{\rm c}$ in the vertical direction (i.e.\ along the $z$ axis),
\begin{equation}
  \label{eq:square_distance_z_integration}
  \int_{V}{d_{\rm z}^2\,dm}=b_{\rm e}MZ_{\rm c}^2~,
\end{equation}
where $b_{\rm e}$ is the aforementioned geometrical factor that
quantifies the effects of the density inhomogeneities for the external
gravity term ($b_{\rm e}=\frac{1}{5}$ for a homogenous cloud).

Therefore, the total contribution of external gravity to the cloud's
energy budget in the vertical direction is
\begin{equation}
  \label{eq:eext_z_nrf}
  E_{\rm ext,z}\approx M\left(\sigma_{\rm gal,z}^2-b_{\rm e}\nu_0^2Z_{\rm c}^2\right)~,
\end{equation}
where as before $\nu_0^2\equiv4\pi G\rho_{\ast,0}$.


\subsection{Calculating $E_{\rm ext,plane}$ in the rotating frame}
\label{app:eext_rotating_frame}

In this section, we derive the contribution of external gravity to a
cloud's energy budget in the orbital plane $E_{\rm ext,plane}$, using
a frame of reference ($x^\prime$,$y^\prime$) that we will refer to as
the ``rotating frame''. This rotating frame is a local Cartesian
coordinate system centred at the cloud's CoM, that both orbits around
the galaxy centre with the cloud's CoM (with angular velocity
$\Omega_0$) and rotates on itself (with the same angular velocity
$\Omega_0$),
such that the $x^\prime$ axis always points in the direction of
increasing galactocentric radius and the $y^\prime$ axis always points
in the direction of orbital rotation at the cloud's CoM (see
Fig.~\ref{fig:rotating_frame}).

In the rotating frame, the contribution of external gravity to a
cloud's energy budget in the plane is (cf.\
Eq.~\ref{eq:split_eext_nrf})
\begin{gather}
  \begin{split}
    \label{eq:e_ext_plane_rf_expand}
    E_{\rm ext,plane} & =M\left(\sigma_{\rm gal,x^\prime}^{\prime2}+\sigma_{\rm gal,y^\prime}^{\prime2}\right)+\int_{V}{\left(a_{\rm ext,x^\prime}^{\prime}x^\prime+a_{\rm ext,y^\prime}^{\prime}y^\prime\right)dm} \\
    & =\underbrace{\left(M\sigma_{\rm gal,x^\prime}^{\prime2}+\int_{V}{(a_{\rm ext,x^\prime}^{\prime}x^\prime)\,dm}\right)}_{E_{\rm ext,x^\prime}}+\underbrace{\left(M\sigma_{\rm gal,y^\prime}^{\prime2}+\int_{V}{(a_{\rm ext,y^\prime}^{\prime}y^\prime)\,dm}\right)}_{E_{\rm ext,y^\prime}}~,
  \end{split}
\end{gather}
where $\sigma_{\rm gal,x^\prime}^{\prime2}$,
$a_{\rm ext,x^\prime}^\prime$ and $x^\prime$ are the components of
$\sigma_{\rm gal}^{\prime2}$, $\vec{a}_{\rm ext}^\prime$ and
$\vec{d}_{\rm plane}^\prime$ along the $\hat{x}^\prime$ direction,
respectively, similarly for $\sigma_{\rm gal,y^\prime}^{\prime2}$,
$a_{\rm ext, y^\prime}^\prime$ and $y^\prime$. Here,
$E_{\rm ext,x^\prime}$ and $E_{\rm ext,y^\prime}$ are the
contributions of external gravity to the cloud's energy budget in the
radial and the azimuthal direction, respectively.
 
In the rotating frame, the acceleration of a fluid element due to
galactic forces (i.e.\ the galactic gravitational potential) is
\begin{gather}
  \begin{split}
    \label{eq:total_acceleration_rf}
    \vec{a}_{\rm ext,plane}^\prime(\vec{d}_{\rm plane}^\prime) & =\,-\nabla\Phi_{\rm gal}(\vec{d}_{\rm plane}^\prime)\,+\,\Omega_0^2\vec{R}\,-\,2\vec{\Omega}_0\times\vec{v}_{\rm gal}^\prime\\ 
    &=\,-\,\Omega^2(R)\vec{R}\,+\,\Omega_0^2\vec{R}\,-\,2\vec{\Omega}_0\times\vec{v}_{\rm gal}^\prime~,
  \end{split}
\end{gather}
where
$\vec{v}_{\rm gal}^\prime\equiv\dot{\vec{d}}^\prime_{\rm plane}=v_{\rm
  gal,x^\prime}^\prime\hat{x}^\prime+v_{\rm
  gal,y^\prime}^\prime\hat{y}^\prime$ is the in-plane velocity of
gravitational motions induced by the external potential as measured in
the rotating frame. The last two terms on the RHS of
Eq.~\ref{eq:total_acceleration_rf} represent the centrifugal and the
Coriolis acceleration, respectively, as perceived in the rotating
frame.

We then expand $\vec{a}_{\rm ext,plane}^\prime$ from
Eq.~\ref{eq:total_acceleration_rf} in the radial ($\hat{x}^\prime$)
and azimuthal ($\hat{y}^\prime$) directions, and obtain
\begin{equation}
  \label{eq:ext_acceleration_plane_rf}
  \left\{
    \begin{array}{lr}
      a_{\rm ext,x^\prime}^\prime=\left(-\Omega^2(R)+\Omega_0^2\right)R\cos\phi+2\Omega_0v_{\rm gal,y^\prime}^\prime~,\\
      a_{\rm ext,y^\prime}^\prime=\left(-\Omega^2(R)+\Omega_0^2\right)R\sin\phi-2\Omega_0v_{\rm gal,x^\prime}^\prime~,\\
    \end{array}
  \right.
\end{equation}
where $\phi$ is the angle between $\vec{R}$ and $\vec{R}_0$ (see
Fig.~\ref{fig:rotating_frame}). If we assume the size of the cloud to
be much smaller than its galactocentric distance (i.e.\
$R_{\rm c}\ll R_0$), then $\cos\phi\approx1$ and
$\sin\phi\approx y^\prime/R$. The accelerations
$a_{\rm ext,x^\prime}^\prime$ and $a_{\rm ext,y^\prime}^\prime$ can
thus be approximated to
\begin{equation}
  \label{eq:aext_xy_rf}
  \left\{
    \begin{split}
      a_{\rm ext,x^\prime}^\prime & \approx\left(-\Omega^2(R)R+\Omega_0^2R\right)+2\Omega_0v_{\rm gal,y^\prime}^\prime \\
      & \approx\Bigg(-\Big(\Omega_0^2R_0+\frac{d(\Omega^2R)}{dR}\Bigr|_{R=R_0}(R-R_0)\Big) \\
      & ~~~~~~~+\Big(\Omega_0^2R_0+\frac{d(\Omega^2_0R)}{dR}\Bigr|_{R=R_0}(R-R_0)\Big)\Bigg)+2\Omega_0v_{\rm gal,y^\prime}^\prime \\
      & \approx\left(-\Big(\Omega_0^2R_0+(\Omega_0^2+R\frac{d\Omega^2}{dR}\Bigr|_{R=R_0})x^\prime\Big)+\left(\Omega_0^2R_0+\Omega_0^2x^\prime\right)\right) \\
      & ~~~~~~+2\Omega_0v_{\rm gal,y^\prime}^\prime \\
      & \approx-\left(R\frac{d\Omega^2}{dR}\Bigr|_{R=R_0}\right)x^\prime+2\Omega_0v_{\rm gal,y^\prime}^\prime \\
      & \approx T_0x^\prime+2\Omega_0v_{\rm gal,y^\prime}^\prime~,\\
      a_{\rm ext,y^\prime}^\prime & \approx\left(-\Omega^2(R)+\Omega_0^2\right)y^\prime-2\Omega_0v_{\rm gal,x^\prime}^\prime \\
      & \approx\left(-\Big(\Omega_0^2+\frac{d\Omega^2}{dR}\Bigr|_{R=R_0}(R-R_0)\Big)+\Omega_0^2\right)y^\prime-2\Omega_0v_{\rm gal,x^\prime}^\prime \\
      & \approx-\frac{d\Omega^2}{dR}\Bigr|_{R=R_0}x^{\prime}y^\prime-2\Omega_0v_{\rm gal,x^\prime}^\prime\\
      & \approx-2\Omega_0v_{\rm gal,x^\prime}^\prime~, 
    \end{split}
  \right.
\end{equation}
where we have assumed $(R-R_0)\approx x^\prime$ (as
$R_{\rm c}\ll R_0$) and expanded $a_{\rm ext,x^\prime}^\prime$ and
$a_{\rm ext,y^\prime}^\prime$ to first order in $x^\prime$ and
$y^\prime$ (i.e.\ $x^{\prime2}\approx0$, $y^{\prime2}\approx0$ and
$x^{\prime}y^\prime\approx0$). The term $T_0x^\prime$ represents the
tidal force (i.e.\ a combination of the external gravity and
centrifugal force), that is exclusively in the radial direction, while
the terms $2\Omega_0v_{\rm gal,y^\prime}^\prime$ and
$-2\Omega_0v_{\rm gal,x^\prime}^\prime$ represent the Coriolis force,
that is in both the radial and azimuthal directions.
 
It is worth noting that Eqs.~\ref{eq:aext_xy_rf} have solutions
\begin{equation}
  \label{eq:motion_solutions_appendix}
  \left\{
    \begin{array}{lr}
      x^{\prime}={\rm S}_1\sin(\kappa_0t+\varphi)+{\rm S}_2~,\\
      y^{\prime}=\frac{2\Omega_0}{\kappa_0}\,{\rm S}_1\cos(\kappa_0t+\varphi)-2A_0\,{\rm S}_2\,t+S_3~,\\
    \end{array}
  \right.
\end{equation}
where $\kappa_0$ is the epicyclic frequency evaluated at the cloud's
CoM
($\kappa_0^2\equiv\left(R\frac{d\Omega^2(R)}{dR}+4\Omega^2(R)\right)\vert_{R=R_0}$),
$A_0$ is Oort's constant $A$ quantifying shear evaluated at the
cloud's CoM
($A_0\equiv-\frac{R}{2}\frac{d\Omega(R)}{dR}\vert_{R=R_0}$), and
${\rm S}_1$, ${\rm S}_2$ and ${\rm S}_3$ (as well as the arbitrary
phase $\varphi$) are constants that depend on the given boundary
(e.g.\ initial) conditions.
Equations~\ref{eq:motion_solutions_appendix} show that the
gravitational motions associated with external gravity have two
contributions: epicyclic motions around the cloud's COM (i.e.\ the
``guiding centre''; see e.g.\ \citealt{meidt2018}), indicated by the
trigonometric terms ${\rm S}_1\sin(\kappa_0t+\varphi)$ and
$\frac{2\Omega_0}{\kappa_0}\,{\rm S}_1\cos(\kappa_0t+\varphi)$, and
linear shear motion, indicated by the $-2A_0\,{\rm S}_2\,t$ term
\citep[e.g.][]{gammie1991,tan2000,binney2020}. It is worth noting
that, in a model where all fluid elements of a cloud move on perfectly
circular orbits (around the galaxy centre) determined by the galactic
potential, the epicyclic amplitudes vanish and the gravitational
motions are completely dominated by the shear motions (see
Eq.~\ref{eq:motion_solutions_no_epicyclic}).


To calculate $E_{\rm ext,x^\prime}$ and $E_{\rm ext,y^\prime}$ (and
thus $E_{\rm ext,plane}$), we also need
$\sigma_{\rm gal,x^\prime}^{\prime2}$ and
$\sigma_{\rm gal,y^\prime}^{\prime2}$ measured in the rotating frame
(see Eq.~\ref{eq:e_ext_plane_rf_expand}). However, as
$\sigma_{\rm gal,x^\prime}^{\prime2}$ and
$\sigma_{\rm gal,y^\prime}^{\prime2}$ can not be obtained directly
from observations, we instead calculate them from the RMS velocities
of gravitational motions $\sigma_{\rm gal,r}^2$ and
$\sigma_{\rm gal,t}^2$ measured in an inertial frame centred at the
galaxy center, that
are related to the observed velocity dispersions
$\sigma_{\rm obs,los}$ and $\sigma_{\rm gs,los}$ through
Eq.~\ref{eq:sigma_obslos}.

To achieve this, we must first derive the velocity transformation
between the chosen inertial frame and our rotating frame. The velocity
of each fluid element due to gravitational motions in the inertial
frame is
$\vec{v}_{\rm
  gal,plane}\equiv\dot{\vec{R}}=\dot{\vec{R}}_0+\dot{\vec{d}}_{\rm
  plane}=\vec{\Omega}_0\times\vec{R}_0+\dot{\vec{d}}_{\rm plane}$,
where $\vec{d}_{\rm plane}$ is the in-plane position vector of the
fluid element with respect to the cloud's CoM in the inertial frame
(see Fig.~\ref{fig:rotating_frame}). The time derivative of
$\vec{d}_{\rm plane}$ is related to the time derivative of
$\vec{d}_{\rm plane}^\prime$ (i.e.\ $\vec{v}_{\rm gal,plane}^\prime$,
the velocity of the fluid element due to gravitational motions
measured in the rotating frame; see
Eq.~\ref{eq:total_acceleration_rf}) through the usual velocity
transformation between an inertial and a rotating frame, i.e.\
$\dot{\vec{d}}_{\rm plane}=\dot{\vec{d}}_{\rm
  plane}^\prime+\vec{\Omega}_0\times\vec{d}_{\rm plane}=\vec{v}_{\rm
  gal,plane}^\prime+\vec{\Omega}_0\times\vec{d}_{\rm plane}$. We thus
obtain
$\vec{v}_{\rm gal,plane}=\vec{\Omega}_0\times\vec{R}_0+\vec{v}_{\rm
  gal,plane}^\prime+\vec{\Omega}_0\times\vec{d}_{\rm plane} $, or
equivalently
\begin{equation}
  \vec{v}_{\rm gal,plane}^\prime=\vec{v}_{\rm gal,plane}-\vec{\Omega}_0\times\vec{R}_0-\vec{\Omega}_0\times\vec{d}_{\rm plane}~.
\end{equation}

Expanding $\vec{v}_{\rm gal,plane}^\prime$ in the radial
($\hat{x}^\prime$) and azimuthal ($\hat{y}^\prime$) directions, we
derive
\begin{equation}
  \label{eq:velocity_rf}
  \left\{
    \begin{array}{lr}
      v_{\rm gal,x^{\prime}}^{\prime}=v_{\rm gal,r}+\Omega_0d_{\rm t}~,\\
      v_{\rm gal,y^{\prime}}^{\prime}=v_{\rm gal,t}-\Omega_0R_0-\Omega_0d_{\rm r}~,
    \end{array}
  \right.
\end{equation}
where $v_{\rm gal,x^\prime}^{\prime}$ and
$v_{\rm gal,y^\prime}^{\prime}$ are the velocities of the fluid
element due to gravitational motions measured in the rotating frame
along the radial ($\hat{x}^\prime$) and the azimuthal
($\hat{y}^\prime$) direction, respectively, $v_{\rm gal,r}$ and
$v_{\rm gal t}$ are the corresponding velocities measured in the
inertial frame along the radial ($\hat{r}$ or $\vec{R}_0$) and the
azimuthal ($\hat{t}$) direction, respectively, and $d_{\rm r}$ and
$d_{\rm t}$ are the radial and the azimuthal component of
$\vec{d}_{\rm plane}$ measured in the inertial frame,
respectively. Using Eqs.~\ref{eq:velocity_rf}, we thus derive the mean
velocities $\overline{v^{\prime}}_{\rm gal,x^{\prime}}$ and
$\overline{v^{\prime}}_{\rm gal,y^{\prime}}$ of the fluid elements as
measured in the rotating frame:
\begin{equation}
  \label{eq:mean_xy_rf}
  \left\{
    \begin{split}
      \overline{v^{\prime}}_{\rm gal,x^{\prime}} & \equiv\frac{\int_V{v_{\rm gal,x^{\prime}}^{\prime}\,dm}}{M} \\
      & =\frac{\int_V{v_{\rm gal,r}\,dm}}{M}+\frac{\int_V{(\Omega_0d_{\rm t})\,dm}}{M} \\
      & =\overline{v}_{\rm gal,r}~,\\
      \overline{v^{\prime}}_{\rm gal,y^{\prime}} & \equiv\frac{\int_V{v_{\rm gal,y^{\prime}}^{\prime}\,dm}}{M} \\
      & =\frac{\int_V{v_{\rm
            gal,t}\,dm}}{M}-\frac{\int_V{(\Omega_0R_0)\,dm}}{M}-\frac{\int_V{(\Omega_0d_{\rm r})\,dm}}{M} \\
      & =\overline{v}_{\rm gal,t}-\Omega_0R_0~,\\
    \end{split}
  \right.
\end{equation}  
where $\overline{v}_{\rm gal,r}$ and $\overline{v}_{\rm gal,t}$ are
the mean velocities of the fluid elements as measured in the inertial
frame, and we have used
$\int_V{d_{\rm r}\,dm}=\int_V{d_{\rm t}\,dm}=0$ as a homogenous
ellipsoidal cloud has been assumed.
 
With Eqs.~\ref{eq:velocity_rf} and \ref{eq:mean_xy_rf}, the desired
RMS velocities of the fluid elements measured in the rotating frame
(i.e.\ $\sigma_{\rm gal,x^\prime}^{\prime2}$ and
$\sigma_{\rm gal,y^\prime}^{\prime2}$) can thus be related to those
measured in the inertial frame (i.e.\ $\sigma_{\rm gal,r}^2$ and
$\sigma_{\rm gal,t}^2$):
\begin{equation}
  \label{eq:sigma_xy_rf}
  \left\{
    \begin{split}
      \sigma_{\rm gal,x^\prime}^{\prime2} & \equiv\frac{\int_V{(v_{\rm gal,x^{\prime}}^{\prime}-\overline{v^{\prime}}_{\rm gal,x^{\prime}})^2\,dm}}{M} \\
      & =\frac{\int_V{\left((v_{\rm gal,r}-\overline{v}_{\rm gal,r})+\Omega_0d_{\rm t}\right)^2\,dm}}{M} \\
      & =\sigma_{\rm gal,r}^2+\frac{\Omega_0^2\int_V{d_{\rm t}^2\,dm}}{M}+\frac{2\Omega_0\int_V{(v_{\rm gal,r}-\overline{v}_{\rm gal,r})d_{\rm t}\,dm}}{M}~,\\
        \sigma_{\rm gal,y^\prime}^{\prime2} & \equiv\frac{\int_V{(v_{\rm gal,y^{\prime}}^{\prime}-\overline{v^{\prime}}_{\rm gal,y^{\prime}})^2\,dm}}{M} \\  
      & =\frac{\int_V{\left((v_{\rm gal,t}-\overline{v}_{\rm gal,t})-\Omega_0d_{\rm r}\right)^2\,dm}}{M} \\
      & =\sigma_{\rm gal,t}^2+\frac{\Omega_0^2\int_V{d_{\rm r}^2\,dm}}{M}-\frac{2\Omega_0\int_V{(v_{\rm gal,t}-\overline{v}_{\rm gal,t})d_{\rm r}\,dm}}{M}~,
    \end{split}
  \right.
\end{equation}
where we have used
$\sigma_{\rm gal,r}^2\equiv\frac{1}{M}\int_{V}{(v_{\rm
    gal,r}-\overline{v}_{\rm gal,r})^2dm}$ and
$\sigma_{\rm gal,t}^2\equiv\frac{1}{M}\int_{V}{(v_{\rm
    gal,t}-\overline{v}_{\rm gal,t})^2dm}$.

Substituting Eqs.~\ref{eq:aext_xy_rf}, \ref{eq:velocity_rf} and
\ref{eq:sigma_xy_rf} into Eq.~\ref{eq:e_ext_plane_rf_expand} yields
\begin{equation}
  \label{eq:eextx_eexty_rf}
  \left\{
    \begin{split}
      E_{\rm ext,x^\prime}= & \,M\sigma_{\rm gal,r}^2+(T_0-2\Omega_0^2)\int_V{d_{\rm r}^2\,dm}+\Omega_0^2\int_V{d_{\rm t}^2\,dm}
      \\
      + & \,2\Omega_0\int_V{\left(v_{\rm gal,r}d_{\rm t}+v_{\rm gal,t}d_{\rm r}\right)\,dm}~,\\
      E_{\rm ext,y^\prime}= & \,M\sigma_{\rm gal,t}^2+\Omega_0^2\int_V{d_{\rm r}^2\,dm}-2\Omega_0^2\int_V{d_{\rm t}^2\,dm} \\
      - & \,2\Omega_0\int_V{\left(v_{\rm gal,r}d_{\rm t}+v_{\rm gal,t}d_{\rm r}\right)\,dm}~,
    \end{split}
  \right.
\end{equation}
where we have adopted
$\int_V{x^{\prime2}\,dm}=\int_V{x^{\prime}d_{\rm r}\,dm}=\int_V{d_{\rm
    r}^2\,dm}$,
$\int_V{y^{\prime}d_{\rm t}\,dm}=\int_V{d_{\rm t}^2\,dm}$,
$\int_V{v_{\rm gal,t}x^{\prime}\,dm}=\int_V{v_{\rm gal,t}d_{\rm
    r}\,dm}$ and
$\int_V{v_{\rm gal,r}y^{\prime}\,dm}=\int_V{v_{\rm gal,r}d_{\rm
    t}\,dm}$ to simplify the notation, and again
$\int_V{x^{\prime}\,dm}=\int_V{d_{\rm r}\,dm}=\int_V{d_{\rm t}\,dm}=0$
as a homogenous ellipsoidal cloud has been assumed. The last term of
$E_{\rm ext,x^\prime}^\prime$ (resp.\ $E_{\rm ext,y^\prime}^\prime$)
represents the integration of the Coriolis force in the
$\hat{x}^\prime$ (resp.\ $\hat{y}^\prime$) direction.

We now calculate the terms $\int_V{d_{\rm r}^2\,dm}$ and
$\int_V{d_{\rm t}^2\,dm}$ of Eqs.~\ref{eq:eextx_eexty_rf} for a
homogenous ellipsoidal cloud with two semi-axes (semi-major axis
$X_{\rm c}$ and semi-minor axis $Y_{\rm c}$) located in the orbital
plane. For such a cloud, we have
\begin{equation}
  \label{equation:d_transformation}
  \left\{
    \begin{split}
      d_{\rm r} & =x_{\rm maj}\cos\phi_{\rm PA}-y_{\rm min}\sin\phi_{\rm PA}~, \\
      d_{\rm t} & =x_{\rm maj}\sin\phi_{\rm PA}+y_{\rm min}\cos\phi_{\rm PA}~,
    \end{split}
  \right.
\end{equation}
where $x_{\rm maj}$ and $y_{\rm min}$ are the components of
$\vec{d}_{\rm plane}$ along the major and the minor axis of the cloud,
respectively, and as before $\phi_{\rm PA}$ is the angle the
semi-major axis $X_{\rm c}$ makes with respect to the radial (i.e.\
$\hat{r}$ or $\vec{R}_0$) direction (see
Fig.~\ref{fig:rotating_frame}).

With Eqs.~\ref{equation:d_transformation}, we then have
\begin{equation}
  \label{eq:dr2_integral}
  \begin{split} 
    \int_{V}{d_{\rm r}^2\,dm} & =\int_{V}{\left(x_{\rm maj}\cos\phi_{\rm PA}-y_{\rm min}\sin\phi_{\rm PA}\right)^2\,dm} \\
    & =\int_{V}{\left(x_{\rm maj}^2\cos^2\phi_{\rm PA}+y_{\rm min}^2\sin^2\phi_{\rm PA}\right)\,dm} \\
    & =\cos^2\phi_{\rm PA}\int_{V}{x_{\rm maj}^2\,dm}+\sin^2\phi_{\rm PA}\int_{V}{y_{\rm min}^2\,dm} \\
    & =\cos^2\phi_{\rm PA}\int_{-X_{\rm c}}^{X_{\rm c}}{x_{\rm maj}^2\,\rho\pi Z_{\rm c}Y_{\rm c}\left(1-\frac{x_{\rm maj}^2}{X^2_{\rm c}}\right)dx_{\rm maj}} \\
    & ~~~~+\sin^2\phi_{\rm PA}\int_{-Y_{\rm c}}^{Y_{\rm c}}{y_{\rm min}^2\,\rho\pi Z_{\rm c}X_{\rm c}\left(1-\frac{y_{\rm min}^2}{Y^2_{\rm c}}\right)dy_{\rm min}} \\
    & =b_{\rm e}M\,(X^2_{\rm c}\cos^2\phi_{\rm PA}+Y^2_{\rm c}\sin^2\phi_{\rm PA})~,
  \end{split}
\end{equation}
and
\begin{equation}
  \label{eq:dt2_integral}
  \begin{split} 
    \int_{V}{d_{\rm t}^2\,dm} & =\int_{V}{\left(x_{\rm maj}\sin\phi_{\rm PA}+y_{\rm min}\cos\phi_{\rm PA}\right)^2\,dm} \\
    & =\int_{V}{\left(x_{\rm maj}^2\sin^2\phi_{\rm PA}+y_{\rm min}^2\cos^2\phi_{\rm PA}\right)\,dm} \\
    & =\sin^2\phi_{\rm PA}\int_{V}{x_{\rm maj}^2\,dm}+\cos^2\phi_{\rm PA}\int_{V}{y_{\rm min}^2\,dm} \\
    & =\sin^2\phi_{\rm PA}\int_{-X_{\rm c}}^{X_{\rm c}}{x_{\rm maj}^2\,\rho\pi Z_{\rm c}Y_{\rm c}\left(1-\frac{x_{\rm maj}^2}{X^2_{\rm c}}\right)dx_{\rm maj}} \\
    & ~~~~+\cos^2\phi_{\rm PA}\int_{-Y_{\rm c}}^{Y_{\rm c}}{y_{\rm min}^2\,\rho\pi Z_{\rm c}X_{\rm c}\left(1-\frac{y_{\rm min}^2}{Y^2_{\rm c}}\right)dy_{\rm min}} \\
    & =b_{\rm e}M\,(X^2_{\rm c}\sin^2\phi_{\rm PA}+Y^2_{\rm c}\cos^2\phi_{\rm PA})~,\\  
  \end{split}
\end{equation}
where $b_{\rm e}$ is the usual geometrical factor quantifying the
effects of density inhomogeneities for the external gravity term
($b_{\rm e}=\frac{1}{5}$ for a homogenous cloud), we have used
$\int_{V}{x_{\rm maj}y_{\rm min}\,dm}=0$ as a homogenous ellipsoidal
cloud has been assumed, and
$ dm=\rho dV=\rho\pi Z_{\rm c}Y_{\rm c}(1-x_{\rm maj}^2/X^2_{\rm
  c})dx_{\rm maj}=\rho\pi Z_{\rm c}X_{\rm c}(1-y_{\rm min}^2/Y^2_{\rm
  c})dy_{\rm min}$.

Finally, substituting Eqs.~\ref{eq:dr2_integral} and
\ref{eq:dt2_integral} into Eqs.~\ref{eq:eextx_eexty_rf}, we obtain
\begin{equation}
  \label{eq:eextx_eexty_rf_nonspherical}
  \left\{
    \begin{split}
      E_{\rm ext,x^\prime} & =M\sigma_{\rm gal,r}^2+b_{\rm e}M(T_0-2\Omega_0^2)\,(X_{\rm c}^2\cos^2\phi_{\rm PA}+Y_{\rm c}^2\sin^2\phi_{\rm PA}) \\
      & ~~~~~~~~~~~~~~~~~\,+b_{\rm e}M\Omega_0^2\,(X_{\rm c}^2\sin^2\phi_{\rm PA}+Y_{\rm c}^2\cos^2\phi_{\rm PA})\\
      & ~~~~~~~~~~~~~~~~~\,+2\Omega_0\int_{V}{(v_{\rm gal,r}d_{\rm t}+v_{\rm gal,t}d_{\rm r})\,dm}~,\\
      E_{\rm ext,y^\prime} & =M\sigma_{\rm gal,t}^2+b_{\rm e}M\Omega_0^2\,(X_{\rm c}^2\cos^2\phi_{\rm PA}+Y_{\rm c}^2\sin^2\phi_{\rm PA}) \\
      & ~~~~~~~~~~~~~~~~~\,-2b_{\rm e}M\Omega_0^2\,(X_{\rm c}^2\sin^2\phi_{\rm PA}+Y_{\rm c}^2\cos^2\phi_{\rm PA}) \\
      & ~~~~~~~~~~~~~~~~~\,-2\Omega_0\int_{V}{(v_{\rm gal,r}d_{\rm t}+v_{\rm gal,t}d_{\rm r})\,dm}~,
    \end{split}
  \right.
\end{equation}
and thus
\begin{gather}
  \label{eq:eext_plane_rf}
  \begin{split}
    E_{\rm ext,plane} & =E_{\rm ext,x^\prime}+E_{\rm ext,y^\prime} \\
    & =M\Big(\sigma_{\rm gal,r}^2+\sigma_{\rm gal,t}^2+b_{\rm e}T_0\,(X_{\rm c}^2\cos^2\phi_{\rm PA}+Y_{\rm c}^2\sin^2\phi_{\rm PA})\\
    & ~~~~~~~~~~-b_{\rm e}\Omega_0^2\,(X_{\rm c}^2+Y_{\rm c}^2)\Big)~,
  \end{split}
\end{gather}
where
$(\sigma_{\rm gal,r}^2+\sigma_{\rm gal,t}^2)=\sigma_{\rm gal,plane}^2$
is the RMS velocity of in-plane gravitational motions caused by the
external potential as measured in the inertial frame.

For an axisymmetric cloud ($X_{\rm c}=Y_{\rm c}=R_{\rm c}$), we thus
have
\begin{equation}
  \label{eq:eextx_eexty_rf_sphere}
  \left\{
    \begin{split}
      E_{\rm ext,x^\prime} & =M\sigma_{\rm gal,r}^2+b_{\rm e}M(T_0-\Omega_0^2)R_{\rm c}^2 \\
      & ~~+2\Omega_0\int_{V}{(v_{\rm gal,r}d_{\rm t}+v_{\rm gal,t}d_{\rm r})\,dm}~,\\
      E_{\rm ext,y^\prime} & =M\sigma_{\rm gal,t}^2-b_{\rm e}M\Omega_0^2R_{\rm c}^2 \\
      & ~~~-2\Omega_0\int_{V}{(v_{\rm gal,r}d_{\rm t}+v_{\rm gal,t}d_{\rm r})\,dm}~,
    \end{split}
  \right.
\end{equation}
and
\begin{gather}
  \label{eq:eext_plane_rf_sphere}
  \begin{split}
    E_{\rm ext,plane} & =E_{\rm ext,x^\prime}+E_{\rm ext,y^\prime} \\
    & =M\Big(\sigma_{\rm gal,r}^2+\sigma_{\rm gal,t}^2+b_{\rm e}(T_0-2\Omega_0^2)R_{\rm c}^2\Big)~.
  \end{split}
\end{gather}


\section{Effective Virial Parameter}
\label{app:effective_virial_parameter}

Overall, our modified Virial theorem can be written simply as (see
Eq.~\ref{eq:modified_Virial_theorem_appendix})
\begin{equation}
 \begin{split}
  \label{equation:mvt_appendixB}
  \frac{\ddot{I}}{2} & =\left(3M\sigma_{\rm sg,los}^2-3b_{\rm s}GM^2/R_{\rm c}\right)+E_{\rm ext}\\
  & =\frac{3b_{\rm s}GM^2}{R_{\rm c}}\left(\frac{\sigma_{\rm sg,los}^2R_{\rm c}}{b_{\rm s}GM}+\frac{E_{\rm ext}}{3b_{\rm s}GM^2/R_{\rm c}}-1\right)~,
\end{split}
\end{equation}
where $E_{\rm ext}$ is the contribution of external gravity to a
cloud's energy budget (see Eqs.~\ref{eq:eext} and
\ref{eq:eext_sphere}). We define
\begin{equation}
  \label{eq:beta_appendix}
  \beta\equiv\frac{E_{\rm ext}}{3b_{\rm s}GM^2/R_{\rm c}}~,
\end{equation}
the ratio between the contribution of external gravity and the
(absolute value of the) cloud's self-gravitational energy
($\vert U_{\rm sg}\vert=3b_{\rm s}GM^2/R_{\rm c}$), so that
Eq.~\ref{equation:mvt_appendixB} can be written as
\begin{equation}
  \frac{\ddot{I}}{2}=\frac{3b_{\rm s}GM^2}{R_{\rm c}}\left(\alpha_{\rm sg,vir}+\beta-1\right)~,
\end{equation}
where
\begin{equation}
  \label{eq:alpha_sg,vir_appendix}
  \alpha_{\rm sg,vir}\equiv\frac{\sigma_{\rm sg,los}^2R_{\rm c}}{b_{\rm s}GM}
\end{equation}
is the traditional Virial parameter regulated by self-gravity only
(see Eq.~\ref{eq:alpha_vir}).

This naturally leads us to define an effective Virial parameter
\begin{equation}
  \label{eq:alpha_eff,vir_appendix}
  \alpha_{\rm eff,vir}\equiv\alpha_{\rm sg,vir}+\beta
\end{equation}
such that
\begin{equation}
  \label{eq:modified_Virial_theorem_alpha_eff_appendix}
  \frac{\ddot{I}}{2}=\frac{3b_{\rm s}GM^2}{R_{\rm c}}\left(\alpha_{\rm eff,vir}-1\right)~.
\end{equation}
Thus, just like the standard Virial parameter, this effective Virial
parameter informs on the dynamical stability of a cloud. If
$\alpha_{\rm eff,vir}\approx1$, the cloud is gravitationally bound and
in Virial equilibrium even in the presence of the external (i.e.\
galactic) gravitational potential. If $\alpha_{\rm eff,vir}\gg1$, the
cloud is unlikely to be bound (i.e.\ it is transient unless confined
by other forces). If $\alpha_{\rm eff,vir}\lesssim1$, the molecular
cloud is likely to collapse. For clouds that are (marginally)
gravitationally bound, we again require
$\alpha_{\rm eff,vir}\le\alpha_{\rm vir,crit}=2$ \citep{kauffmann2013,
  kauffmann2017}, or equivalently $\beta\le1$ if an internal Virial
equilibrium is established by self-gravity (i.e.\ if
$\alpha_{\rm sg,vir}\approx1$; see Eq.~\ref{eq:alpha_eff,vir}).

Equivalently, from Eq.~\ref{eq:alpha_vir}, we can define an
effective velocity dispersion
\begin{equation}
  \label{eq:effective_velocity_dispersion_2_appendix}
  \sigma_{\rm eff,los}^2=\alpha_{\rm eff,vir}\,b_{\rm s}GM/R_{\rm c}~,
\end{equation}
and thus our modified Virial equation
(Eq.~\ref{equation:mvt_appendixB}) can be simplified to
\begin{equation}
  \label{equation:mvt_appendixB_2}
  \frac{\ddot{I}}{2}\approx\left(3M\sigma_{\rm eff,los}^2-3b_{\rm s}GM^2/R_{\rm c}\right)~.
\end{equation}
The parameters $\alpha_{\rm eff,vir}$ (via
Eq.~\ref{eq:modified_Virial_theorem_alpha_eff_appendix}) or equivalently
$\sigma_{\rm eff,los}$ (via Eq.~\ref{equation:mvt_appendixB_2})
thus embody our MVT and offer a straightforward method to test the
gravitational boundedness of a cloud in the presence of an external
(i.e.\ galactic) gravitational field.

Having said that, a major challenge to calculate the effective virial
parameter $\alpha_{\rm eff,vir}$ (and $\beta$) or the effective
velocity dispersion $\sigma_{\rm eff,los}$ is to determine the
in-plane ($\sigma_{\rm gal,r}$ and $\sigma_{\rm gal,t}$) and vertical
($\sigma_{\rm gal,z}$) RMS velocities of gravitational motions induced
by the external potential in an inertial frame (see Eqs.~\ref{eq:eext}
and \ref{eq:eext_sphere}). By making increasingly stringent
assumptions, we however show below that it is possible to evaluate
those quantities from observables alone.

If we assume the cloud to be in vertical equilibrium, i.e.\
$E_{\rm ext,z}=M(\sigma_{\rm gal,z}^2-b_{\rm e}\nu_0^2Z_{\rm
  c}^2)\approx0$, we have
\begin{equation}
\beta \equiv\frac{E_{\rm ext}}{3b_{\rm s}GM^2/R_{\rm c}}
 \approx \frac{E_{\rm ext,plane}}{3b_{\rm s}GM^2/R_{\rm c}} ~.
\end{equation}
In this case, we only need to derive the in-plane RMS velocities of
gravitational motions (i.e.\
$\sigma_{\rm gal,plane}^2=\sigma_{\rm gal,r}^2+\sigma_{\rm
  gal,t}^2$).

In the following, we will estimate $\sigma_{\rm gal,r}$ and
$\sigma_{\rm gal,t}$ (and thereby $\beta$, $\alpha_{\rm eff,vir}$ and
$\sigma_{\rm eff,los}$) using two different methods: one using
observations, the other using a shear model.

\textbf{\textit{Observations.}}  Although $\sigma_{\rm gal,r}$ and
$\sigma_{\rm gal,t}$ can not be measured directly from observations,
it is nevertheless possible to glean some information about them from
the observables $\sigma_{\rm obs,los}$ and $\sigma_{\rm
  gs,los}$. Indeed, the observed velocity dispersion of a cloud
$\sigma_{\rm obs,los}$ can be expressed as
\begin{equation}
  \label{eq:vrms_obs_los_appendix}
  \sigma_{\rm obs,los}^2\approx\sigma_{\rm sg,los}^2\,+\,\left(\sigma_{\rm gal,r}^2\sin^2\theta+\sigma_{\rm gal,t}^2\cos^2\theta\right)\sin^2i\,+\,\sigma_{\rm gal,z}^2\cos^2i~,
\end{equation}
where $i$ is the inclination of the galactic disc with respect to the
line of sight, and $\theta$ is the (deprojected) azimuthal angle of
the cloud's CoM with respect to the kinematic major axis of the disc
(see Eq.~32 of \citealt{meidt2018}).

Assuming that the vertical gravitational motions can be treated as
random motions that balance the weight of the disc (i.e.\ no bulk
motion in the vertical direction), analogously to turbulent motions
due to self-gravity, the only bulk motions will originate from
in-plane gravitational motions. The gradient-subtracted velocity
dispersion $\sigma_{\rm gs,los}$ can therefore be written as
\begin{equation}
  \label{eq:vrms_gs_los_appendix}
  \sigma_{\rm gs,los}^2\approx\sigma_{\rm sg,los}^2\,+\,\sigma_{\rm gal,z}^2\cos^2i~.
\end{equation}
Our gradient-subtracted velocity dispersion $\sigma_{\rm gs,los}$ thus
removed the second term (in-plane bulk gravitational motions) but kept
the first term (turbulent self-gravitational motions) and last term
(vertical random gravitational motions) on the RHS of
Eq.~\ref{eq:vrms_obs_los_appendix}.

If we assume the gas motions induced by the galactic potential to be
isotropic in the plane (i.e.\
$\sigma_{\rm gal,r}=\sigma_{\rm gal,t}$), the RMS velocities of the
in-plane gravitational motions due to external gravity can easily be
derived by combining Eqs.~\ref{eq:vrms_obs_los_appendix} and
\ref{eq:vrms_gs_los_appendix}:
\begin{equation}
  \label{eq:in_plane_rms_velocity_appendix}
  \sigma_{\rm gal,r}^2=\sigma_{\rm gal,t}^2\approx\frac{\sigma_{\rm obs,los}^2-\sigma_{\rm gs,los}^2}{\sin^2i}~.
\end{equation}
Substituting Eq.~\ref{eq:in_plane_rms_velocity_appendix} into
Eq.~\ref{eq:eext_plane_rf}, we then obtain
\begin{gather}
\label{eq:eobs_plane}
  \begin{split}
    E_{\rm ext} & \approx E_{\rm ext,plane} \\ 
    & \approx M\,\Bigg[\frac{2(\sigma_{\rm obs,los}^2-\sigma_{\rm gs,los}^2)}{\sin^2i}\\
    & ~~~~~~~~~~~+b_{\rm e}T_0\,(X_{\rm c}^2\cos^2\phi_{\rm PA}+Y_{\rm c}^2\sin^2\phi_{\rm PA})-b_{\rm e}\Omega_0^2\,(X_{\rm c}^2+Y_{\rm c}^2)\Bigg]~,
  \end{split}
\end{gather}
and thus (see Eqs.~\ref{eq:beta_appendix},
\ref{eq:alpha_sg,vir_appendix}, \ref{eq:alpha_eff,vir_appendix} and
\ref{eq:effective_velocity_dispersion_2_appendix})
\begin{gather}
  \label{eq:alpha_vir_obs}
  \begin{split}
    \alpha_{\rm eff,vir} & \approx\frac{\sigma_{\rm gs,los}^2R_{\rm c}}{b_{\rm s}GM}+\frac{R_{\rm c}}{3b_{\rm s}GM}\Bigg[\frac{2(\sigma_{\rm obs,los}^2-\sigma_{\rm gs,los}^2)}{\sin^2i}\\
    & ~~~~+b_{\rm e}T_0\,(X_{\rm c}^2\cos^2\phi_{\rm PA}+Y_{\rm c}^2\sin^2\phi_{\rm PA})-b_{\rm e}\Omega_0^2\,(X_{\rm c}^2+Y_{\rm c}^2)\Bigg]
  \end{split}
\end{gather}
and 
\begin{gather}
  \begin{split}
    \label{eq:vrms_eff_obs}
    \sigma_{\rm eff,los}^2 & \approx\sigma_{\rm gs,los}^2+\frac{1}{3}\Bigg[\frac{2(\sigma_{\rm obs,los}^2-\sigma_{\rm gs,los}^2)}{\sin^2i}\\
    & ~~~~+b_{\rm e}T_0\,(X_{\rm c}^2\cos^2\phi_{\rm PA}+Y_{\rm c}^2\sin^2\phi_{\rm PA})-b_{\rm e}\Omega_0^2\,(X_{\rm c}^2+Y_{\rm c}^2)\Bigg]~,
  \end{split}
\end{gather}
where we have used
$\sigma_{\rm gs,los}^2\approx\sigma_{\rm sg,los}^2$.
 
For an axisymmetric cloud (i.e.\ $X_{\rm c}=Y_{\rm c}=R_{\rm c}$), we
have
\begin{gather}
  \label{eq:eobs_plane_sphere}
  \begin{split}
    E_{\rm ext} & \approx E_{\rm ext,plane} \\
    & \approx M\,\Bigg[\frac{2(\sigma_{\rm obs,los}^2-\sigma_{\rm gs,los}^2)}{\sin^2i}+b_{\rm e}(T_0-2\Omega_0^2)R_{\rm c}^2\Bigg]~,
  \end{split}
\end{gather}
and thus
\begin{gather}
  \label{eq:alpha_vir_obs_sphere}
  \begin{split}
    \alpha_{\rm eff,vir} & \approx\frac{\sigma_{\rm gs,los}^2R_{\rm c}}{b_{\rm s}GM}+\frac{R_{\rm c}}{3b_{\rm s}GM}\Bigg[\frac{2(\sigma_{\rm obs,los}^2-\sigma_{\rm gs,los}^2)}{\sin^2i}+b_{\rm e}(T_0-2\Omega_0^2)R_{\rm c}^2\Bigg]
  \end{split}
\end{gather}
and 
\begin{gather}
  \begin{split}
    \label{eq:vrms_eff_obs_sphere}
    \sigma_{\rm eff,los}^2 & \approx\sigma_{\rm gs,los}^2+\frac{1}{3}\Bigg[\frac{2(\sigma_{\rm obs,los}^2-\sigma_{\rm gs,los}^2)}{\sin^2i}+b_{\rm e}(T_0-2\Omega_0^2)R_{\rm c}^2\Bigg]~.
  \end{split}
\end{gather}

\textbf{\textit{Shear model.}} For the model we will refer to as our
``shear model'', we assume that all fluid elements of a cloud populate
perfectly circular orbits (around the galaxy centre) determined by the
galactic potential. In this case, the radial and azimuthal velocities
of each fluid element measured in the inertial frame (centred at the
galaxy centre) can be written as
\begin{equation}
  \label{eq:gravitational_velocity_appendixB}
  \left\{
    \begin{split} 
      v_{\rm gal,r}^{\rm mod} & =-\Omega(R)R\sin\phi \\
      & \approx-\Omega(R)d_{\rm t} \\
      & \approx-\left(\Omega_0+\frac{d\Omega(R)}{dR}\biggr|_{R=R_0}(R-R_0)\right)d_{\rm t} \\
      & \approx-\left(\Omega_0+\frac{d\Omega(R)}{dR}\biggr|_{R=R_0}d_{\rm r}\right)d_{\rm t} \\
      & \approx-\Omega_0d_{\rm t}~,\\
      v_{\rm gal,t}^{\rm mod} & =\Omega(R)R\cos\phi \\
      & \approx\Omega(R)R \\
      & \approx\Omega_0R_0+\frac{d(\Omega(R)R)}{dR}\biggr|_{R=R_0}(R-R_0) \\
      & \approx\Omega_0R_0+\left(\Omega_0+R\frac{d\Omega(R)}{dR}\biggr|_{R=R_0}\right)d_{\rm r} \\
      & \approx\Omega_0R_0+(\Omega_0-2A_0)d_{\rm r}~,
    \end{split}
  \right.
\end{equation}
where as before $\phi$ is the angle between $\vec{R}$ and $\vec{R}_0$
(see Fig.~\ref{fig:rotating_frame}), we have assumed
$\cos\phi\approx1$, $\sin\phi\approx d_{\rm t}/R$ and
$R-R_0\approx d_{\rm r}$ (as $R_{\rm c}\ll R_0$), and we have expanded
$v_{\rm gal,r}$ and $v_{\rm gal,t}$ to first order about $R_0$ (i.e.\
$d_{\rm r}d_{\rm t}\approx0$).

Using the velocity transformation between the rotating frame and the
inertial frame (Eq.~\ref{eq:velocity_rf}), we can derive the
gravitational motion of each fluid element due to the external
potential as measured in the rotating frame:
\begin{equation}
  \label{eq:gravitational_velocity_rotatingFrame_appendixB}
  \left\{
    \begin{split}
      v_{\rm gal,x^{\prime}}^{\prime{\rm mod}} & =0~,\\
      v_{\rm gal,y^{\prime}}^{\prime{\rm mod}} & =-2A_0d_{\rm r}~,
    \end{split}
  \right.
\end{equation}
that have solutions
\begin{equation}
  \label{eq:motion_solutions_no_epicyclic}
  \left\{
    \begin{split}
      x^{\prime{\rm mod}} & ={\rm S}_2~,\\
      y^{\prime{\rm mod}} & =-2A_0\,{\rm S}_2\,t+{\rm S}_3~,\\
    \end{split}
  \right.
\end{equation}
where ${\rm S}_2$ and ${\rm S}_3$ are constants that depend on the
given boundary (e.g.\ initial) conditions. As expected, if all fluid
elements of a cloud move on perfectly circular orbits (around the
galaxy centre) determined by the galactic potential, the gravitational
motions of the cloud are completely dominated by shear motions and the
epicyclic amplitudes vanish (cf.\
Eq.~\ref{eq:motion_solutions_shear_2}). We thus refer to this model as
the ``shear model''.

Considering Eqs.~\ref{eq:gravitational_velocity_appendixB}, the mean
velocities of the fluid elements of a cloud along the radial
($\hat{r}$ or $\vec{R}_0$) and azimuthal ($\hat{t}$) directions as
measured in the inertial frame are
\begin{equation}
  \label{eq:mean_gravitational_velocity_appendixB}
  \left\{
    \begin{split}
      \overline{v}_{\rm gal,r}^{\rm mod} & \equiv\frac{1}{M}\int_{V}{v_{\rm gal,r}^{\rm mod}\,dm} \\
      & \approx\frac{1}{M}\int_{V}{(-\Omega_0d_{\rm t})\,dm}~,\\
      & \approx0~,\\
      \overline{v}_{\rm gal,t}^{\rm mod} & \equiv\frac{1}{M}\int_{V}{v_{\rm gal,t}^{\rm mod}\,dm}\\
      & \approx\frac{1}{M}\int_{V}{\left(\Omega_0R_0+(\Omega_0-2A_0)d_{\rm r}\right)\,dm} \\
      & \approx\Omega_0R_0~,
    \end{split}
  \right.
\end{equation}
where as before we have used
$\int_{V}{d_{\rm r}\,dm}=\int_{V}{d_{\rm t}\,dm}=0$ as a homogenous
ellipsoidal cloud has been assumed. The RMS velocities of
gravitational motions cause by an external potential as measured in
the inertial frame are thus
\begin{equation}
  \label{eq:vrms_gravitational_velocity_appendixB}
  \left\{
    \begin{split}
      (\sigma_{\rm gal,r}^{\rm mod})^2 & \equiv\frac{1}{M}\int_{V}{(v_{\rm gal,r}^{\rm mod}-\overline{v}_{\rm gal,r}^{\rm mod})^2dm} \\
      & \approx\frac{1}{M}\int_{V}{(-\Omega_0d_{\rm t})^2dm}~,\\
      & \approx\Omega_0^2b_{\rm e}\,(X_{\rm c}^2\sin^2\phi_{\rm PA}+Y_{\rm c}^2\cos^2\phi_{\rm PA})~,\\
      (\sigma_{\rm gal,t}^{\rm mod})^2 & \equiv\frac{1}{M}\int_{V}{(v_{\rm gal,t}^{\rm mod}-\overline{v}_{\rm gal,t}^{\rm mod})^2dm} \\
      & \approx\frac{1}{M}\int_{V}{\big((\Omega_0-2A_0)d_{\rm r}\big)^2dm},\\
      & \approx(\Omega_0-2A_0)^2b_{\rm e}\,(X_{\rm c}^2\cos^2\phi_{\rm PA}+Y_{\rm c}^2\sin^2\phi_{\rm PA})~,\\
    \end{split}
  \right.
\end{equation}
where Eqs.~\ref{eq:dr2_integral} and \ref{eq:dt2_integral} have been
used for the integrals $\int_{\rm V}{d_{\rm r}^2\,dm}$ and
$\int_{\rm V}{d_{\rm t}^2\,dm}$ for a homogenous ellipsoidal cloud.

Substituting Eq.~\ref{eq:vrms_gravitational_velocity_appendixB} into
Eq.~\ref{eq:eext_plane_rf}, we obtain
\begin{equation}
  \label{eq:eModel_plane}
  \begin{split}
    E_{\rm ext}^{\rm mod} & \approx E_{\rm ext,plane}^{\rm mod} \\
    & \approx 4A_0^2b_{\rm e}M\,(X_{\rm c}^2\cos^2\phi_{\rm PA}+Y_{\rm c}^2\sin^2\phi_{\rm PA})~,
  \end{split}
\end{equation}
and thus (see Eqs.~\ref{eq:beta_appendix},
\ref{eq:alpha_eff,vir_appendix} and
\ref{eq:effective_velocity_dispersion_2_appendix})
\begin{equation}
  \label{eq:alpha_vir_Model}
  \alpha_{\rm eff,vir}^{\rm mod}=\alpha_{\rm sg,vir}+\frac{4A_0^2b_{\rm e}\,(X_{\rm c}^2\cos^2\phi_{\rm PA}+Y_{\rm c}^2\sin^2\phi_{\rm PA})}{3b_{\rm s}GM/R_{\rm c}}
\end{equation}
and
\begin{equation}
  \label{eq:vrms_eff_Model}
  (\sigma_{\rm eff,los}^{\rm mod})^2=\sigma_{\rm sg,los}^2+\frac{4A_0^2b_{\rm e}\,(X_{\rm c}^2\cos^2\phi_{\rm PA}+Y_{\rm c}^2\sin^2\phi_{\rm PA})}{3}~.
\end{equation} 

For an axisymmetric cloud (i.e.\ $X_{\rm c}=Y_{\rm c}=R_{\rm c}$), we
then have
\begin{equation}
  \label{eq:eModel_plane_sphere}
  \begin{split}
    E_{\rm ext}^{\rm mod} & \approx E_{\rm ext,plane}^{\rm mod} \\
    & \approx 4A_0^2b_{\rm e}MR_{\rm c}^2~,
  \end{split}
\end{equation}
and thus
\begin{equation}
  \label{eq:alpha_vir_Model_sphere}
  \alpha_{\rm eff,vir}^{\rm mod}=\alpha_{\rm sg,vir}+\frac{4A_0^2b_{\rm e}R_{\rm c}^2}{3b_{\rm s}GM/R_{\rm c}}
\end{equation}
and
\begin{equation}
  \label{eq:vrms_eff_Model_sphere}
  (\sigma_{\rm eff,los}^{\rm mod})^2=\sigma_{\rm sg,los}^2+\frac{4A_0^2b_{\rm e}R_{\rm c}^2}{3}~.
\end{equation} 

Finally, if we assume an internal Virial equilibrium has been
established by self-gravity (i.e.\
$3M\sigma_{\rm sg,los}^2-3b_{\rm s}GM^2/R_{\rm c}\approx0$) and the
cloud is in vertical equibrium (i.e.\ $E_{\rm ext, z}\approx0$), our
shear model predicts that
\begin{equation}
  \label{eq:solve_virial_equation}
  \begin{split}
    \frac{\ddot{I}^{\,\rm mod}}{2} & \approx E_{\rm ext,plane}^{\rm mod} \\
    & \approx4A_0^2b_{\rm e}M(R_{\rm c}^{\rm mod})^2 \\
    & =2A_0^2I^{\rm mod}~,
  \end{split}
\end{equation}
where $I^{\rm mod}=2b_{\rm e}M(R_{\rm c}^{\rm mod})^2$ for a spherical
cloud. As $A_0>0$ for all clouds, we easily obtain
$I^{\rm mod}\propto e^{2A_0t}$ and thus
$R_{\rm c}^{\rm mod}\propto e^{A_0t}$ and
$\Sigma_{\rm gas}^{\rm mod}\propto e^{-2A_0t}$ for a spherical cloud
with a constant mass. This suggests that, if all fluid elements of a
cloud move on perfectly circular orbits (around the galaxy centre)
determined by the galactic potential, a cloud experiencing strong
shear will grow larger and larger and become less and less bound over
a timescale $\sim1/2A_0$, which we name the ``shear timescale''.


\section{Stellar Density Calculation}
\label{app:stellar_density_calculation}

For a number of calculations, we must know the local stellar mass
density at the position of each cloud ($\rho_{\ast,0}$). For this, we
adopt the multi-Gaussian expansion (MGE) formalism of
\citet{emsellem1994} and \citet{cappellari2002}, and specifically the
existing model of NGC4429 from \citet{davis2018}, constrained from
dynamical modelling of the same molecular gas data as used here.

In short, the luminous matter distribution was first parametrised
using a MGE model of the stellar light, constructed by applying the
{\texttt{MGE\_FIT\_SECTORS}} package of \citet{cappellari2002} to a
{\it HST} Wide-Field Planetary Camera~2 (WFPC2) F606W image combined
with an $r^{\prime}$-band image from the Sloan Digital Sky Survey
(SDSS; \citealt{adelmanmccarthy2008}). Each Gaussian component $j$ of
the model has an observed surface brightness $I_j$, standard deviation
(width) $\sigma_j$ and axial ratio $q_j$. The best-fitting MGE model
is tabulated in Table~1 of \citet{davis2018} and is shown visually in
their Fig.~7. Using this MGE parametrisation, the surface brightness
distribution of the galaxy can be accurately reproduced (see Eq.~12 of
\citealt{cappellari2008}). The next step is to obtain the intrinsic
luminosity density by deprojecting the surface brightness model, which
the MGE parametrisation allows to do trivially under the assumption of
(oblate) axisymmetry and a known inclination (see Eq.~13 of
\citealt{cappellari2008}). The stellar density of the cloud is then
derived by multiplying the deprojected MGE luminosity density with the
(spatially-variable) stellar mass-to-light ratio $\Psi(R_{\rm gal})$
derived from the \citet{davis2018} dynamical modelling:
\begin{equation}
  \rho_\ast(R_{\rm gal})=\Psi(R_{\rm gal})\sum^N_j\frac{I_jq_j}{\sqrt{2\pi}\,\sigma_jq^\prime_j}\,\exp\left(-\frac{R^2_{\rm gal}}{2\sigma^2_j}\right)~,
\end{equation}
where we have assumed that all the clouds are in the equatorial plane
($z=0$), $R_{\rm gal}$ is the galactocentric distance (radius in the
plane of the disc) of the cloud as usual, $q^\prime_j$ is the
intrinsic axial ratio of the $j^{\rm th}$ Gaussian component,
\begin{equation}
  q^\prime_j=\frac{\sqrt{q^2_j-\cos^2i}}{\sin i}~,
\end{equation}
(note that \citealt{cappellari2008} instead uses $q$ and $q^\prime$
for the intrinsic and the observed axial ratio, respectively), and the
sum is taken over the $N$ Gaussian components. The derived central
stellar mass density of each cloud is listed in
Table~\ref{tab:shear_parameters} and ranges from $6$ to
$60$~M$_\odot$~pc$^{-3}$.


\bsp	
\label{lastpage}
\end{document}